\documentclass[a4paper,12pt]{article}
\usepackage{graphicx}
\usepackage{epsfig}
\usepackage[cp1251]{inputenc}
\usepackage[english]{babel}
\usepackage{amssymb,latexsym,amsmath}
\usepackage{amsmath,enumerate, amsfonts, amssymb,amsthm}
\usepackage{cite}

\newcommand{\be}[1]{\begin{equation}\label{#1}}
\newcommand{\ee}{\end{equation}}
\newcommand{\ba}[1]{\begin{eqnarray}\label{#1}}
\newcommand{\ea}{\end{eqnarray}}
\newcommand{\rf}[1]{(\ref{#1})}
\newcommand{\nn}{\nonumber}
\newcommand{\ov}{\overline}

\newcommand{\opensquare}{\mbox{$\rlap{$\sqcap$}\sqcup$}}

\newcommand{\const}{\mbox{\rm const}\,}

\newcommand{\sign}{ \mbox{\rm sign}\,}

\font\msbm=msbm10
\def\RR{\hbox{\msbm R}}

\def\ZZ{\hbox{\msbm Z}}

\def\beq{\begin{equation}}
\def\eeq{\end{equation}}
\def\ber{\begin{eqnarray}}
\def\eer{\end{eqnarray}}

\def\lsim{\
  \lower-1.5pt\vbox{\hbox{\rlap{$<$}\lower5.3pt\vbox{\hbox{$\sim$}}}}\ }
\def\gsim{\
  \lower-1.5pt\vbox{\hbox{\rlap{$>$}\lower5.3pt\vbox{\hbox{$\sim$}}}}\ }

\def\n {\noindent}

\def \lleq {\
  \lower-1.5pt\vbox{\hbox{\rlap{$<$}\lower5.3pt\vbox{\hbox{$\sim$}}}}\ }
\def \ggeq {\
    \lower-1.5pt\vbox{\hbox{\rlap{$>$}\lower5.3pt\vbox{\hbox{$\sim$}}}}\ }

\def\lsim{\
  \lower-1.5pt\vbox{\hbox{\rlap{$<$}\lower5.3pt\vbox{\hbox{$\sim$}}}}\ }
  \def\gsim{\
    \lower-1.5pt\vbox{\hbox{\rlap{$>$}\lower5.3pt\vbox{\hbox{$\sim$}}}}\ }

\numberwithin{equation}{section}
\begin{document}
 \normalsize\thispagestyle{empty}

\centerline {THE I.I. MECHNIKOV ODESSA NATIONAL UNIVERSITY}
\centerline {The manuscript}
\medskip\medskip\medskip\medskip\medskip\medskip\medskip\medskip\medskip
\centerline {Saidov Tamerlan Adamovich}

\medskip\medskip\medskip\medskip\medskip\medskip\medskip\medskip\medskip\medskip
\rightline {UDC 524.83: 531.51: 530.145}
\medskip\medskip\medskip\medskip\medskip\medskip\medskip\medskip\medskip

\centerline {THE COMPACTIFICATION PROBLEMS OF ADDITIONAL DIMENSIONS }
\centerline {IN MULTIDIMENSIONAL COSMOLOGICAL THEORIES}

\medskip\medskip\medskip\medskip\medskip\medskip\medskip\medskip\medskip\medskip\medskip\medskip\medskip

\centerline {01.04.02 - theoretical physics}
\medskip\medskip\medskip\medskip\medskip\medskip\medskip\medskip\medskip\medskip\medskip\medskip\medskip

\centerline {The PhD thesis for}
\centerline {physical and mathematical science}

\medskip\medskip\medskip\medskip\medskip\medskip\medskip\medskip\medskip\medskip\medskip\medskip\medskip

\rightline {Scientific adviser $\quad\quad\quad\qquad\qquad\;\;$}
\rightline {Zhuk Alexander Ivanovich $\quad\qquad\;\:$}
\rightline {Dr. of phys.-math. sci., Professor $\;$}

\medskip\medskip\medskip\medskip\medskip\medskip\medskip\medskip\medskip\medskip\medskip\medskip\medskip
\medskip\medskip\medskip\medskip\medskip\medskip\medskip\medskip\medskip\medskip\medskip\medskip\medskip

\centerline {Odessa -- 2011}

\newpage
\normalsize\thispagestyle{empty}

\medskip\medskip\medskip\medskip\medskip\medskip\medskip\medskip\medskip
\medskip\medskip\medskip\medskip\medskip\medskip\medskip\medskip\medskip
\[\]\[\]\[\]
\begin{flushright}
You can't look forward with head down.
\end{flushright}

\begin{flushright}
A.A. Cron
\end{flushright}

\newpage
\setcounter{page}{2}
 \renewcommand{\refname}{\center{LITERATURE}}
 \renewcommand{\contentsname}{\begin{center}$\;$ 
CONTENT \\ \vspace{2cm}
 \end{center}}
 \normalsize\thispagestyle{empty} $\;$  \large \pagestyle{myheadings}

 \large
 \tableofcontents


\newpage
 \section*{\center{LIST OF ABBREVIATIONS}}
 \numberwithin{equation}{section}
  \addcontentsline{toc}{section}{LIST OF ABBREVIATIONS}
  \medskip\medskip\medskip\medskip
\begin{description}
\item[ADD-model] -- ADditional Dimensions model (or large extra dimensions model);
\item[AdS] -- Anti de Sitter;
\item[BBN] -- Big Bang Nucleosynthesis;
\item[$\Lambda$CDM] -- Cold Dark Matter with Lambda-term;
\item[CMB] -- Cosmic Microwave Background;
\item[ETG] -- Extended Theory of Gravitation;
\item[GR] -- General Relativity;
\item[KK] -- Kaluza-Klein;
\item[MCM] -- Multidimensional Cosmological Model.
\end{description}


 \setcounter{section}{0}
 \setcounter{equation}{0}
 \newpage
 \pagestyle{myheadings}
 \section*{\center{INTRODUCTION}}
 \numberwithin{equation}{section}
 \addcontentsline{toc}{section}{INTRODUCTION}

 \medskip\medskip
Latest observation of Ia type supernovas and CMB yields to following composition of the Universe: 4\% of baryonic matter,
20\% of dark matter and 76\% of dark energy. The term "dark matter" is implemented for the unknown matter, which has an ability for clustering, but was not yet detected in lab-conditions. The "dark energy" is the energy not been detected yet, but also unable for clustering, as common energy does.

Most probably, the dark energy is responsible for the value of the cosmological constant. Recent experiments indicate
the value of the cosmological constant to be too small if originated only by vacuum energy of common matter.
This brings some difficulties for developing of corresponding theories. A few ways for interpretation of presents for the cosmological constant are known. Some of those are dynamical alternatives
of dark energy.

In this Thesis the attention is focused on  $f(\bar{R})$ theory.
This kind of gravitational theory is the result of the lagrangian generalization in the Hilbert-Einstein
action. On the base of this theory a study will be undertaken on the problem of the effective cosmological constant, accelerated expansion of the Universe and the compactification (non-observation) of additional dimensions.
\medskip

\textbf{Topicality of the subject}.
\medskip

Multidimensionality of our Universe is one of the most intriguing
assumption in modern physics. It follows naturally from theories
unifying different fundamental interactions with gravity, e.g.
M/string theory \cite{pol-wit,pol-wita}. The idea has received a great deal
of renewed attention over the last few years. However, it also brings a row of additional questions.

According to observations the internal space should be static or
nearly static at least from the time of primordial
nucleosynthesis, otherwise the fundamental physical constants
would vary. This means that at the present evolutionary stage of
the Universe there are two possibilities: slow variation or compactification of internal space scale parameters \cite{WCOS}.

in many recent studies the problem of extra dimensions stabilization was studied
for so-called ADD (see e.g., Refs.
\cite{sub-mill2,d2,sub-mill3,CGHW,Geddes,demir,NSST,PS}). Under these approaches a massive
scalar fields (gravitons or radions) of external space-time can be presented as conformal excitations.

In above mentioned works it was assumed that multidimensional action to be linear with respect to curvature. Although as follows from string theory, the gravity action needs to be extended to nonlinear one. In order to investigate effects of nonlinearity, in this Thesis a multidimensional Lagrangian will be studied, having the form $L = f(\bar{R})$, where  $f(\bar{R})$ is an arbitrary smooth function of the scalar curvature.

\medskip

\textbf{Connection of the Thesis with scientific programs,}

\textbf{plans and projects}.

\medskip
The study under this Thesis was undertaken as part of the budget project of Odessa National Mechnikov University
\#519 "Compactification of space-time in quantum cosmology".
\medskip

 \textbf{The target and the task of the study}.
\medskip

The task of this study is an investigation of diverse non-linear cosmological models with respect to condition of additional dimensions compactification and
an analysis of additional dimensions effect on the evolution of the Universe.

For that, several surveys have to be implemented:
an analysis of non-linearity influence on the effective potential in the equivalent theory; capability of extrema for these potentials as result of additional dimensions non-observability condition. Stabilization and compactification of additional dimensions and their consistency with observations.

Hence, the attention is accumulated on the possibility to achieve models, in which external space (our 4D space-time) behaves as being observed Universe and internal space is stabilized and compactified on the Plank scales. These kind of models allow to build a bridge between observations, multidimensionality and evolution of the Universe, proving by this the principal possibility for existence of additional dimensions.
\medskip

\textbf{The scientific novelty of obtained results}.
\medskip

The multidimensional models of type $\bar{R}^{2}+\bar{R}^{4}$  for pure geometric action and of $\bar{R}^{-1}$ with forms are studied for stabilization and compactification of additional dimensions.

It is shown, that for non-linearity of type $\bar{R}^{-1}$ in multidimensional case with forms, the stabilization of internal spaces is independent on signatures of internal space curvature, multidimensional and effective cosmological constants. Moreover, effective cosmological constant may satisfy to observed densities of dark energy by applying fine tuning. First time for this kind of model, the capability for stable compactification of additional dimensions and accelerated expansion of the Universe is shown. When in case of pure geometrical model of $\bar{R}^{-1}$ this two phenomena simultaneously are not obtainable.

The analysis of inflation is undertaken for linear, quadratic, quartic models with forms and for $\bar{R}^{2}+\bar{R}^{4}$ model with pure gravitational action.

A new type of inflation is found for $\bar{R}^{2}+\bar{R}^{4}$ model, called as bouncing inflation.
For this type of inflation there is no need for original potential
to have a minimum or to check the slowroll
conditions. A necessary condition is the existence
of the branching points.
It is shown that this inflation takes place both in
the Einstein and Brans-Dicke frames.
\medskip

 \textbf{Practical value of obtained results}.
\medskip

The obtained results have fundamental and theoretical value, giving possibility for assay of the Universe evolution, taking into account presence of additional dimensions for diverse non-linear models either pure gravitational or with forms. The parameters values obtained in the Thesis may become an object for further study and justification in experiments.
\medskip

\medskip

 \textbf{Approbation of the Thesis results}.
 \medskip

The main results of the study, present in the Thesis, were reported on following conferences:

1. International conference "QUARQS 2008", May 23-26, Ser.-Pasad,
Russia;

2. The 8-th G. Gamow Odessa International Astronomical Summer
School, Astronomy and beyond: Astrophysics, Radio astronomy,
Cosmology and Astrobiology, August 1-5, 2006, Odessa, Ukraine;

3. The 6-th G. Gamow Odessa International Astronomical Summer
School,  Astronomy and beyond: Astrophysics, Radio astronomy,
Cosmology and Astrobiology, August 1-5, 2006, Odessa, Ukraine;

4. VI International Conference: Relativistic Astrophysics,
Gravitation and Cosmology, May 24-26 2006, Kyiv, Ukraine;

5. Kyiv High Energy Astrophysics Semester (by ISDC, CERN), April 17
- May 12, 2006, Kiyv, Ukraine;

6. III International Symposium Fundamental Problems in Modern
Quantum Theories and Experiments, September 2005,
Kiyv-Sevastopol, Ukraine.

\medskip

 \textbf{Publications}.
 \medskip

The main results of the study, present in the Thesis, were explained in following publications:

1. Saidov T. 1/R multidimensional gravity with form-fields: Stabilization of extra dimensions, cosmic acceleration, and domain walls / Saidov T.,  Zhuk. A. // Phys. Rev. D. -- 2007. -- V. 75. -- 084037, 10 p.

2. Saidov T. AdS Nonlinear Curvature-Squared and
Curvature-Quartic Multidimensional (D=8) Gravitational Models with
Stabilized Extra Dimensions / Saidov T.,  Zhuk. A. // Gravitation and Cosmology. -- 2006. -- V. 12. --  P. 253--261.

3. Saidov T. A non linear multidimensional gravitational
model $R+R^{-1}$ with form fields and stabilized extra dimensions / Saidov T.,  Zhuk. A. //
Astronomical \& Astrophysical Transactions. -- 2006. -- V. 25.  -- P. 447--453.

4. Saidov T. Problem of inflation in nonlinear multidimensional cosmological models / Saidov T.,  Zhuk. A. //
Phys. Rev. D. -- 2009. -- V. 79. -- 024025,  18 p.

5. Saidov T. Bouncing inflation in a nonlinear $R^2+R^4$ gravitational model / Saidov T.,  Zhuk. A. //
Phys. Rev. D. -- 2010. -- V. 81. -- 124002, 13 p.


\newpage
 \addcontentsline{toc}{section}{CHAPTER 1. LITERATURE OVERVIEW}
 \section*{\center{CHAPTER 1 \\ $\;$ \\
LITERATURE OVERVIEW}}
 \setcounter{section}{1}
 \setcounter{equation}{0}
\label{sec first}

\medskip\medskip
The currently observed accelerated expansion of the Universe
suggests that cosmic flow dynamics is dominated by some unknown
form of dark energy characterized by a large negative pressure\cite{FRR2}.
Most recent observations of CMB and Ia supernovas indicates that Universe consists of:
76\% of dark energy, 20\% of dark matter and 4\% of baryonic matter  \cite{WMAP2006,Riess:2004nr,Astier:2005qq,Eisenstein:2005su}.
Many approaches are known for distinguishing of dark matter and dark energy. Usually it is assumed that dark energy despite of dark matter is incapable for clustering and satisfies for Strong Energy Condition\cite{FRR1,wald}.
Due to dark energy dominates over matter, it is responsible for late-time acceleration of the Universe (i.e. determines the value of effective cosmological constant), following after early-time acceleration as predicted with inflation paradigm \cite{Guth:1980zm,infl1,infl2}.
In between these to stages, the period of decelerated expansion is needed to provide eras of radiation domination for BBN and matter domination for formation of Universe structure

As general approach, the standard Einstein's General Relativity
(GR) is used as paradigm (foundation)  for developing of new theories, called as
Extended Theories of Gravity (ETG). This allows to keep successful
features of GR and apply extensions and corrections to Einstein's theory.
Usually by adding higher order curvature invariants and/or
minimally or non-minimally coupled scalar fields to the dynamics;
these corrections emerge from  the effective action of quantum
gravity \cite{FRR2,odintsov2011}.

Among many, the simplest theory capable more or less to describe experimental data is so-called $\Lambda$CDM model.
It gives appropriate qualitative picture of the observed Universe, but do not explain the inflation. Also, in different models well known problem of the cosmological constant magnitude arises being higher in 120 orders of magnitude then observed one \cite{CarLCDM}.

As next step for modification of GR, one may apply Mach's
principle, which states that the local inertial frame is determined by
the average motion of distant astronomical objects \cite{bondi}.
This means  that
the gravitational coupling could be determined by the
distant distribution of matter, and it can be scale-dependent
and related to some scalar field. As a consequence,  the concept
of ''inertia'' and the Equivalence Principle have to be revised.
Brans-Dicke theory \cite{BDn} constituted the first consistent
and complete theory alternative to Einstein's GR. Brans-Dicke
theory incorporates  a variable gravitational coupling strength
whose dynamics are  governed by a scalar field non-minimally
coupled to the geometry, which implements  Mach's principle
in the gravitational theory \cite{FRR2,BDn,cimento,sciama}.

Nonlinear ETG models may arise either due to quantum fluctuations of matter fields including gravity \cite{BirrDav}, or as a result of compactification
of extra spatial dimensions \cite{NO1}. Compared, e.g., to others higher-order gravity theories,
$f(\bar{R})$ theories are free of ghosts and of Ostrogradski instabilities \cite{Woodard}.
Recently, it was realized that these models can also explain the late time acceleration
of the Universe. This fact resulted in a new wave of papers devoted to this topic (see e.g., recent reviews \cite{reviews,reviewsa,FRR2}).

Another intriguing assumption in modern physics is the multidimensionality of our Universe.
It follows from theories which unify
different fundamental interactions with gravity, such as M or
string theory \cite{pol-wit,pol-wita}, and which have their most consistent
formulation in spacetimes with more than four dimensions. Thus,
multidimensional cosmological models have received a great deal of
attention over the last years.

Stabilization of additional dimensions near their present day
values (dilaton/geometrical moduli stabilization) is one of the
main problems for any multidimensional theory because a dynamical
behavior of the internal spaces results in a variation of the
fundamental physical constants. Observations show that internal
spaces should be static or nearly static at least from the time of
recombination (in some papers arguments are given in favor of the
assumption that variations of the fundamental constants are absent
from the time of primordial nucleosynthesis \cite{Kolb}). In other
words, from this time the compactification scale of the internal
space should either be stabilized and trapped at the minimum of
some effective potential, or it should be slowly varying (similar
to the slowly varying cosmological constant in the quintessence
scenario). In both cases, small fluctuations over stabilized or
slowly varying compactification scales (conformal
scales/geometrical moduli) are possible.

Stabilization of extra dimensions (moduli stabilization) in models
with large extra dimensions (ADD-type models) has been considered
in a number of papers (see e.g., Refs. \cite{sub-mill2,PS}). In the corresponding approaches, a product topology of the
$(4+D^{\prime })-$dimensional bulk spacetime was constructed from
Einstein spaces with scale (warp) factors depending only on the
coordinates of the external $4-$dimensional component. As a
consequence, the conformal excitations had the form of massive
scalar fields living in the external spacetime. Within the
framework of multidimensional cosmological models (MCM) such
excitations were investigated in
\cite{GZ(PRD1997),GZ(PRD2000)}
where they were called gravitational excitons. Later, since the
ADD
compactification approach these geometrical moduli excitations are
known as radions \cite{sub-mill2,sub-mill3}.

Most of the aforementioned  papers are devoted to the
stabilization of large extra dimensions in theories with a linear
multidimensional gravitational action. String theory suggests that
the usual linear Einstein-Hilbert action should be extended with
higher order nonlinear curvature terms. In the papers
\cite{GMZ(PRDa),GZBR} a simplified model is considered with
multidimensional Lagrangian of the form $L = f(\bar{R})$, where $f(\bar{R})$
is an arbitrary smooth function of the scalar curvature. Without
connection to stabilization of the extra-dimensions, such models
($4-$dimensional as well as multidimensional ones) were considered
e.g. in Refs. \cite{Kerner,Kernera1,Kernera2,Kernera3,BC,Kernera5,Kernera6,Kernera7,Kernera8}.
There, it was shown that the nonlinear models are equivalent to
models with linear gravitational action plus a minimally coupled
scalar field with self-interaction potential. Similar approach was
elaborated in Refs. \cite{NO} where the main attention was paid to
a possibility of the late time acceleration of the Universe due to
the nonlinearity of the model.

The most simple, and, consequently, the most studied models are
polynomials of $\bar{R}$:  $f(\bar{R})=\sum_{n=0}^k C_n \bar{R}^n \,$ $(k >1)$, e.g., quadratic $\bar{R}+\bar{R}^2$ and quartic $\bar{R}+\bar{R}^4$ ones. Active investigation of these models, which started in 80-th years of the last century \cite{80-th,80-tha1,Maeda}, continues up to now \cite{Ketov,Ketova}. Obviously, the correction terms (to the Einstein action) with $n>1$ give the main contribution in the case of large $\bar{R}$, e.g., in the early stages of the Universe evolution. As it was shown first in \cite{Star1} for the quadratic model, such modification of gravity results in early inflation. From the other hand,
function $f(\bar{R})$ may also contain negative degrees of $\bar{R}$. For example, the simplest model is $\bar{R}+\bar{R}^{-1}$. In this case the correction term plays the main role for small $\bar{R}$, e.g., at the late
stage of the Universe evolution (see e.g. \cite{GZBR,ZhukSaidov} and numerous references therein). Such modification of gravity may result in the late-time acceleration of our Universe \cite{Carrolletal}.
Nonlinear models with polynomial as well as $\bar{R}^{-1}$-type correction terms have also been generalized to the multidimensional case (see e.g.,  \cite{GZBR,ZhukSaidov,Ellis,GMZ1,GMZ2,ZhukSaidov2,Bronnikov,SZPRD2009}).

In this Thesis several models of $f(\bar{R})$-theory are analyzed for pure gravitational case in chapter 2, and with forms in chapter 3, as well as inflation possibility is studied in chapter 4.

 \newpage
 \addcontentsline{toc}{section}{CHAPTER 2. ASYMPTOTICALLY AdS NONLINEAR \\ GRAVITATIONAL MODELS}
 \section*{\center{CHAPTER 2\\ $\;$ \\
ASYMPTOTICALLY AdS NONLINEAR GRAVITATIONAL MODELS}}
 \setcounter{section}{2}

In this chapter the non-linear gravitational models with curvatures of type $\bar{R}^{-1}$ and
$\bar{R}^{2}+\bar{R}^{4}$ are studied. It is shown that for particular values of parameters, the stabilization and compactification of additional dimensions is achievable with negative constant curvature of internal space. In this case the $4-$dimensional effective cosmological constant becomes negative. As result homogenous and isotropic external time-space appears to be $\mbox{AdS}_4$. The correlation between $D-$dimensional and $4-$dimensional fundamental masses imposes restrictions on the parameters in the models.

 \addcontentsline{toc}{subsection}{2.1. General setup}
 \setcounter{subsection}{1}
 \medskip\medskip\medskip\medskip
 \textbf{2.1. General setup}
\medskip\medskip\medskip\medskip

Let us consider a $D= (4+D^{\prime})-$dimensional nonlinear pure
gravitational theory with action functional
\be{1.1a} S = \frac {1}{2\kappa^2_D}\int_M d^Dx \sqrt{|\bar g|}
f(\bar R)\; , \ee
where $f(\bar R)$ is an arbitrary smooth function with mass
dimension $\mathcal{O}(m^2)$ \ ($m$ has the unit of mass) of a
scalar curvature $\bar R = R[\bar g]$ constructed from the
$D-$dimensional metric $\bar g_{ab}\; (a,b = 1,\ldots,D)$.
$D^{\prime}$ is the number of extra dimensions and $\kappa^2_D $
denotes the $D-$dimensional gravitational constant which is
 connected with the fundamental mass scale
$M_{*(4+D^{\prime})}$ and the surface area $S_{D-1}=2\pi
^{(D-1)/2}/\Gamma [(D-1)/2]$ of a unit sphere in $D-1$ dimensions
by the relation \cite{add1}
\be{1.2a1}  \kappa^2_D = 2S_{D-1} /
M_{*(4+D^{\prime})}^{2+D^{\prime}}.\ee
Before endowing the metric of the  pure gravity theory \rf{1.1a}
with explicit structure, let us recall that this $\bar R-$nonlinear
theory is equivalent to a theory which is linear in another scalar
curvature $R$ but which contains an additional self-interacting
scalar field. According to standard techniques
\cite{Kerner,Kernera1,Kernera2,Kernera3,BC,Kernera5,Kernera6,Kernera7,Kernera8,Maeda}, the corresponding $R-$linear theory has the
action functional:
\be{1.6a} S = \frac{1}{2\kappa^2_D} \int_M d^D x \sqrt{|g|} \left[
R[g] - g^{ab} \phi_{,a} \phi_{,b} - 2 U(\phi )\right]\; , \ee
where
\be{1.7a} f'(\bar R) = \frac {df}{d \bar R} := e^{A \phi} > 0\;
,\quad A := \sqrt{\frac{D-2}{D-1}}\;  ,\ee
and where the self-interaction potential $U(\phi )$ of the scalar
field $\phi$ is given by
\ba{1.8-1} U(\phi ) &=&
 \frac12 \left(f'\right)^{-D/(D-2)} \left[\; \bar R f' - f\right]\;
 ,\nn\\
&=& \frac12 e^{- B \phi} \left[\bar R (\phi )e^{A \phi } -
f\left( \bar R (\phi )\right) \right], B := \frac
{D}{\sqrt{(D-2)(D-1)}} .\label{1.8a1} \ea
 This scalar field
$\phi$ carries the nonlinearity degrees of freedom in $\bar R$ of
the original theory, and for brevity will be called the nonlinearity
field. The metrics $g_{ab}$, $\bar g_{ab}$   of the two theories
\rf{1.1a} and \rf{1.6a} are conformally connected by the relations
\be{1.9} g_{ab} = \Omega^2 \bar g_{ab} = \left[ f'(\bar
R)\right]^{2/(D-2)}\bar g_{ab}\;.  \ee

Next, let us assume that the D-dimensional bulk space-time $M$
undergoes a spontaneous compactification
to a warped product manifold
\be{1.18} M = M_0 \times M_1 \times \ldots \times M_n \ee
with  metric
\be{1.19} \bar g=\bar g_{ab}(X)dX^a\otimes dX^b=\bar
g^{(0)}+\sum_{i=1}^ne^{2\bar {\beta} ^i(x)}g^{(i)}\; . \ee
The coordinates on the $(D_0=d_0+1)-$dimensional manifold $M_0 $
(usually interpreted as our observable $(D_0=4)-$dimensional
Universe) are denoted by $x$ and the corresponding metric by
\be{1.20} \bar g^{(0)}=\bar g_{\mu \nu }^{(0)}(x)dx^\mu \otimes
dx^\nu\; . \ee
For simplicity, the internal factor manifolds $M_i$ are chosen as
$d_i-$dimensional Einstein spaces with metrics
$g^{(i)}=g^{(i)}_{m_in_i}(y_i)dy_i^{m_i}\otimes dy_i^{n_i},$ so
that the relations
\be{1.21} R_{m_in_i}\left[ g^{(i)}\right] =\lambda
^ig_{m_in_i}^{(i)},\qquad m_i,n_i=1,\ldots ,d_i \ee
and
\be{1.22} R\left[ g^{(i)}\right] =\lambda ^id_i\equiv R_i
\;  \ee hold.
The specific metric ansatz \rf{1.19} leads to a scalar curvature
$\bar R$ which depends only on the coordinates $x$ of the external
space: $\bar R[\bar g] = \bar R(x)$. Correspondingly, also the
nonlinearity field $\phi$ depends on $x$ only: $\phi = \phi (x)$.

Passing from the $\bar R-$nonlinear theory \rf{1.1a} to the
equivalent $R-$linear theory \rf{1.6a} the metric \rf{1.19}
undergoes the conformal transformation $\bar g \mapsto g$ [see
relation \rf{1.9}]
\be{1.23} g = \Omega^2 \bar g = \left( e^{A \phi
}\right)^{2/(D-2)} \bar g\: := g^{(0)}+\sum_{i=1}^ne^{2
\beta^i(x)}g^{(i)}\;  \ee
with
\be{1.24} g^{(0)}_{\mu \nu} := \left( e^{A \phi}\right)^{2/(D-2)}
\bar g^{(0)}_{\mu \nu}\; , \quad
 \beta^i := \bar {\beta} ^i + \frac{A}{D-2} \phi\; . \ee

 \addcontentsline{toc}{subsection}{2.2. Stabilization of internal dimensions}
 \setcounter{subsection}{1}
 \medskip\medskip\medskip\medskip
 \textbf{2.2. Stabilization of internal dimensions}
 \medskip\medskip\medskip\medskip

The main subject of subsequent considerations will be the
stabilization of the internal space components. A strong argument
in favor of stabilized or almost stabilized internal space scale
factors $\bar {\beta} ^i(x)$, at the present evolution stage of
the Universe, is given by the intimate relation between variations
of these scale factors and those of the fine-structure constant
$\alpha$ \cite{GSZ}. The strong restrictions on
$\alpha-$variations in the currently observable part of the
Universe \cite{alpha-var,alpha-vara1,alpha-vara2,alpha-vara3} imply a correspondingly strong
restriction on these scale factor variations \cite{GSZ}. For this
reason, the derivation of criteria
ensuring a freezing stabilization of the scale factors will be performed below.

In Ref. \cite{GZ(PRD2000)} it was shown that for models with a
warped product structure \rf{1.19} of the bulk spacetime $M$ and a
minimally coupled scalar field living on this spacetime, the
stabilization of the internal space components requires a
simultaneous freezing of the scalar field. Here a
similar situation with simultaneous freezing stabilization of the
scale factors ${\beta} ^i(x)$ and the nonlinearity field
$\phi(x)$ is expected. According to \rf{1.24}, this will also imply a
stabilization of the scale factors $\bar {\beta} ^i(x)$ of the
original nonlinear model.

In general, the model will allow for several stable scale factor
configurations (minima in the landscape over the space of volume
moduli). Let us choose one of them\footnote{Although the toy model
ansatz \rf{1.1a} is highly oversimplified and far from a realistic
model, one can roughly think of the chosen minimum, e.g., as that
one which is expected to correspond to current evolution stage of
our observable Universe.}, denote the corresponding scale factors
as $\beta^i_0$, and work further on with the deviations
\be{1.25} \hat \beta^i (x)= \beta^{i}(x) - \beta^{i}_0\ee
as the dynamical fields. After dimensional reduction of the action
functional \rf{1.6a} one passes from the intermediate Brans-Dicke
frame to the Einstein frame via a conformal transformation
\begin{equation}
\label{1.26}g_{\mu \nu }^{(0)}=\hat{\Omega} ^2\hat g_{\mu \nu
}^{(0)} =\left( \prod^{n}_{i=1} e^{d_i\hat \beta ^i}
\right)^{-2/(D_0-2)} \hat g_{\mu \nu }^{(0)}\,
\end{equation}
with respect to the scale factor deviations $\hat \beta^i (x)$
\cite{GMZ(PRDa),GZBR}. As result the following action is achieved
\ba{1.27}
 \nn S = \frac 1{2\kappa
_{D_0}^2}\int\limits_{M_0}d^{D_0}x\sqrt{|\hat g^{(0)}|}\left\{
\hat R\left[ \hat g^{(0)}\right]\right. &-&\left.\bar G_{ij}\hat g^{(0)\mu \nu}\partial _\mu\hat \beta ^i\,\partial _\nu \hat \beta ^j \right. \\&-& \left. \hat
g^{(0)\mu \nu }\partial _\mu\phi \,\partial _\nu\phi-2U_{eff}\right\},
\ea
which contains the scale factor offsets $\beta^i_0$ through the
total internal space volume
\be{1.29} V_{D'} \equiv V_I\times v_0\equiv
\prod^{n}_{i=1}\int_{M_i}d^{d_i}y \sqrt{|g^{(i)}|}\times
\prod^{n}_{i=1}e^{d_i\beta^{i}_0}\ee
in the definition of the effective gravitational constant $\kappa
_{D_0}^2$ of the dimensionally reduced theory
\be{1.28}\kappa _{(D_0=4)}^2=\kappa_D ^2/V_{D'}= 8\pi/M^2_{4}\quad
\Longrightarrow \quad M_{4}^2 = \frac{4\pi}{S_{D-1}}V_{D'}
M_{*(4+D^{\prime})}^{2+D^{\prime}}.\ee
Obviously, at the present evolution stage of the Universe, the
internal space components should have a total volume which would
yield a four-dimensional mass scale of order of the Planck mass
$M_{(4)} = M_{Pl}$. The tensor components of the midisuperspace
metric (target space metric on $\RR _T^n$) reads: $\bar
G_{ij}=d_{i}\delta_{ij}+d_{i}d_{j}/(D_{0}-2)$, where $
i,j=(1,\ldots ,n)$, see \cite{IMZ,RZ}. The effective potential has
the explicit form
\begin{equation}
\label{1.31}U_{eff} ( \hat \beta , \phi ) ={\left(
\prod_{i=1}^ne^{d_i\hat \beta ^i}\right) }^{-\frac 2{D_0-2}}\left[
-\frac 12\sum_{i=1}^n\hat R_{i}e^{-2\hat \beta ^i}+ U(\phi )
\right] \, ,
\end{equation}
where abbreviated
\be{1.31a} \hat R_{i} := R_i
\exp{(-2\beta^{i}_{0})}. \ee

A freezing stabilization of the internal spaces will be achieved
if the effective potential has at least one minimum with respect
to the fields $\hat \beta^i(x)$. Assuming, without loss of
generality, that one of the  minima  is located at $\beta^{i} =
\beta^{i}_{0} \Rightarrow \hat \beta^{i} = 0$, the extremum
condition reads:
\be{1.32} \left.\frac{\partial U_{eff}}{\partial \hat
\beta^{i}}\right|_{\hat \beta =0} =0 \Longrightarrow \hat R_{i} =
\frac{d_i}{D_0-2}\left( -\sum_{j=1}^n \hat R_{j} +2 U(\phi)
\right)\, . \ee
From its structure (a constant on the l.h.s. and a dynamical
function of $\phi (x)$ on the r.h.s) it follows that a
stabilization of the internal space scale factors can only occur
when the nonlinearity field $\phi (x)$ is stabilized as well. In
the freezing scenario this will require a minimum with respect to
$\phi$:
\be{1.33} \left. \frac{\partial U(\phi )}{\partial \phi
}\right|_{\phi_{0}} = 0\; \Longleftrightarrow\;  \left.
\frac{\partial U_{eff}}{\partial \phi}\right|_{\phi_{0}}  = 0\, .
\ee
Hence, a stabilization problem is arrived, some of whose general
aspects have been analyzed already in Refs.
\cite{GZ(PRD1997),GZ(PRD2000),GMZ(PRDa),GZBR}. For brevity let us summarize the
corresponding essentials as they will be needed for more detailed
discussions in the next consideration.
\begin{enumerate}
\item \label{c1a1} Eq. \rf{1.32} implies that the scalar curvatures
$\hat R_i$ and with them the compactification scales
$e^{\beta^i_{0}}$ [see relation \rf{1.31a}] of the internal space
components are finely tuned
\be{1.34} \frac{\hat R_i}{d_i} = \frac{\hat R_j}{d_j}\, , \quad
i,j = 1,\ldots ,n \, . \ee
\item \label{c2a1} The masses of the normal mode excitations of the
internal space scale factors (gravitational excitons/radions) and
of the nonlinearity field $\phi$ near the minimum position are
given as \cite{GZ(PRD2000)}:
\ba{1.35} m_{1}^2 &=& \dots \; = m_{n}^2 = \, -\frac{4}{D-2}
U(\phi_{0})\,= -2\frac{\hat R_{i}}
{d_i} > 0\, , \label{1.35a} \\
&\phantom{-} & \nn\\
m_{\phi }^2 &:=& \left. \frac{d^2 U(\phi )}{d \phi^2}
\right|_{\phi_{0}}>0\, .\label{1.35b} \ea
\item \label{c3} The value of the effective potential at the
minimum plays the role of an effective 4D cosmological constant of
the external (our) spacetime $M_0$:
\be{1.36} \Lambda_{{eff}} :=\left. U_{eff}\vphantom{\int} \right|_
{\genfrac{}{}{0pt}{1}{\hat \beta^i =0,}{\phi = \phi_{0}}}\; =\,
\frac{D_0-2}{D-2} U(\phi_{0})\, =\, \frac{D_0-2}{2}\frac{\hat
R_{i}}{d_i}\, . \ee
\item \label{c4} Relation \rf{1.36} implies \be{1.36c4} \sign
\Lambda_{eff} =\sign U(\phi_{0}) =\sign R_{i}\, . \ee Together
with condition \rf{1.35} this shows that in a pure geometrical
model stable configurations can only exist for internal spaces
with negative curvature\footnote{Negative constant curvature
spaces $M_i$ are compact if they have a quotient structure: $M_i =
H^{d_i}/\Gamma_i$, where $H^{d_i}$ and $\Gamma_i$ are hyperbolic
spaces and their discrete isometry group, respectively.}:
\be{1.36c4-4}R_{i} <0\qquad (i=1,\ldots ,n)\, . \ee
Additionally, the effective cosmological constant $\Lambda_{eff}$
as well as the minimum of the potential $U(\phi )$ should be
negative too:
\be{1.36c4a} \Lambda_{eff} <0, \qquad U(\phi_{0})<0\, . \ee
\end{enumerate}
Plugging the potential $U(\phi)$ from Eq. \rf{1.8a1} into the
minimum conditions \rf{1.33}, \rf{1.35b}  yields with the help of
$\partial_\phi \bar R=A f'/f''$ the conditions
\ba{1.11}
\left.\frac{d U}{d\phi}\right|_{\phi_{0}} &=&
\left.\frac{A}{2(D-2)}\left(f'\right)^{-D/(D-2)}h\right|_{\phi_{0}}
= 0,\nn\\ \nn\\h&:=&Df-2\bar R f', \quad \Longrightarrow \quad \
h(\phi_0) = 0\, ,\label{1.11a}\\\nn\\
\left.\frac{d^2 U}{d\phi^2}\right|_{\phi_{0}} &=& \frac12 A
e^{(A-B)\phi_{0}} \left[ \partial_\phi \bar R + (A-B)\bar R
\right]_{\phi_{0}}\nn\\\nn\\&=& \left.\frac{1}{2(D-1)}
 \left(f'\right)^{-2/(D-2)} \frac{1}{f''}\partial_{\bar R}
 h \right|_{\phi_{0}}>0\, ,\label{1.11b}
\ea
where the last inequality can be reshaped into the suitable form
\be{1.11c} \left.f''\partial_{\bar R}
h\right|_{\phi_{0}}=f''\left[(D-2)f'-2\bar R f''\right]_{\phi_0}>0
\; . \ee
Furthermore, from Eq. \rf{1.11a} follows
\be{1.12-1}
U(\phi_0)=\frac{D-2}{2D}\left(f'\right)^{-\frac{2}{D-2}}\bar
R(\phi_0) \ee
so that \rf{1.36c4a} leads to the additional restriction
\be{1.13-1}
\bar R(\phi_0)<0
\ee
at the extremum.

Thus, to avoid the effective four-dimensional fundamental constant variation, it is necessary
to provide the mechanism of the internal spaces stabilization. In these models, the scale factors of the internal spaces play the role of
additional scalar fields (geometrical moduli/gravexcitons \cite{GZ(PRD1997)}).
To achieve their stabilization, an effective potential should have  minima with respect to all scalar fields (gravexcitons and scalaron).
Previous analysis (see e.g. \cite{GZ(PRD2000)}) shows that for a model of the form \rf{1.6a} the stabilization is possible only for the
case of negative minimum of the potential $U(\phi )$.
According to the definition
\rf{1.7a}, the positive branch of
$f'(\bar{R})>0$ will be considered.
Although the negative $f'<0$ branch can be considered as well (see
e.g. Refs. \cite{Maeda,GZBR,GMZ2}). However, negative values of $f'(\bar R)$
result in negative effective gravitational "constant" $G_{eff}=\kappa^2_D/f'$. Thus $f'$
should be positive for the graviton to carry positive kinetic energy (see e.g., \cite{FRR2}).
From action \rf{1.6a} the equation of motion of scalaron field $\phi$ can be obtained in the form:
\be{1.11ab}
\opensquare \phi -\frac{\partial U}{\partial \phi} =0\, .
\ee

Now, let us analyze the internal space
stabilization conditions \rf{1.34} - \rf{1.36c4a} and \rf{1.11} -
\rf{1.13-1} on their compatibility with particular scalar
curvature nonlinearity $f(\bar R)$.

 \addcontentsline{toc}{subsection}{2.3. $R^{-1}$ model}
 \setcounter{subsection}{1}
 \medskip\medskip\medskip\medskip
 \textbf{2.3. $R^{-1}$ model}
 \medskip\medskip\medskip\medskip

Recently it has been shown in Refs. \cite{1/R-1,1/R-1a1,1/R-1a2,1/R-1a3,1/R-1a4,1/R-1a5,1/R-1a6,1/R-1a7,1/R-1a8} that cosmological
models with a nonlinear scalar curvature term of the type $\bar
R^{-1}$ can provide a possible explanation of the observed late-time
acceleration of our Universe within a pure gravity setup. The
equivalent linearized model contains an effective potential with a
positive branch which can simulate a transient inflation-like
behavior in the sense of an effective dark energy. The corresponding
considerations have been performed mainly in four
dimensions\footnotetext[8]{A discussion of pro and contra of a
higher dimensional origin of $\bar R^{-1}$ terms can be found in
Ref. \cite{NO1}.}. Here these analyses are extended  to higher
dimensional models, assuming that the scalar curvature nonlinearity is of the same
form in all dimensions. Starting from a nonlinear coupling of the
type:
\be{2.1a1} f(\bar R) = \bar R - \mu / \bar R\, , \quad \mu >0 \,
,\ee
in front of the $\bar R^{-1}-$term, the minus sign  is chosen,
because otherwise the potential $U(\phi )$ will have no extremum.

With the help of definition \rf{1.7a}, the scalar
curvature $\bar R$ can be expressed in terms of the nonlinearity field $\phi$ and
obtain two real-valued solution branches
\be{2.2a1} \bar R_\pm = \pm \sqrt{\mu} \left( e^{A\phi} -
1\right)^{-1/2}\, ,\quad \Longrightarrow \quad \phi >0
\ee
of the quadratic equation $f'(\bar R)=e^{A\phi}$.
The corresponding potentials
\be{2.3a1} U_\pm(\phi) = \pm \sqrt{\mu}\, e^{-B\phi }\,
\sqrt{e^{A\phi }-1}\, \ee
have extrema for curvatures [see Eq. \rf{1.11}]
\ba{2.4a1} \bar R_{0,\pm} &=&\pm \sqrt{\mu}\sqrt{\frac{D+2}{D-2}}\nn\\
e^{A\phi_0} &=& \frac{2B}{2B-A} = \frac{2D}{D+2}
> 1 \quad \mbox{for} \quad D \geq 3\, \ea
and take for these curvatures the values
\be{2.5a}U_\pm(\phi_0) = \pm
\sqrt{\mu}\sqrt{\frac{D-2}{D+2}}\; e^{-B\phi_0} = \pm
\sqrt{\mu}\sqrt{\frac{D-2}{D+2}}\left( \frac{2D}{D+2}
\right)^{-D/(D-2)}\, .\ee
The stability defining second derivatives [Eq. \rf{1.11b}] at
the extrema \rf{2.4a1},
\ba{2.6a} \left.  \partial^2_\phi U_\pm\right|_{\phi_{0}} &=&
\mp \sqrt{\mu} \frac{D}{D-1}\sqrt{\frac{D+2}{D-2}}\;
e^{B\phi_{0}}\nn\\ &=& \mp \sqrt{\mu}
\frac{D}{D-1}\sqrt{\frac{D+2}{D-2}} \left( \frac{2D}{D+2}
\right)^{-D/(D-2)}\, ,\ea
show that only the negative curvature branch $\bar R_-$ yields a
minimum with stable internal space components. The positive branch
has a maximum with $U_+(\phi_0)>0$. According to \rf{1.36c4} it
can provide an effective dark energy contribution with
$\Lambda_{eff}>0$, but due to its tachyonic behavior with
$\partial_\phi^2U(\phi_0)<0$ it cannot give stably frozen internal
dimensions. This means that the simplest extension of the
four-dimensional purely geometrical $\bar R^{-1}$ setup of Refs.
\cite{1/R-1,1/R-1a1,1/R-1a2,1/R-1a3,1/R-1a4,1/R-1a5,1/R-1a6,1/R-1a7,1/R-1a8} to higher dimensions is incompatible with a freezing
stabilization of the extra dimensions. A possible circumvention of
this behavior could consist in the  existence of different
nonlinearity types $f_i(\bar R_i)$ in different factor spaces
$M_i$ so that their dynamics can decouple one from the other. This
could allow for a freezing of the scale factors of the internal
spaces even in the case of a late-time acceleration with
$\Lambda_{eff}>0$. Another circumvention could consist in a
mechanism which prevents the dynamics of the internal spaces from
causing strong variations of the fine-structure constant $\alpha$.
The question of whether one of these schemes could work within a
physically realistic setup remains to be clarified.

Finally, in the minimum of the effective potential
$U_{eff}(\phi,\beta^i)$, which is provided by the negative
curvature branch $\bar R(\phi)$, one finds excitation masses for
the gravexcitons/radions and the nonlinearity field (see Eqs.
\rf{1.35}, \rf{2.5a} and \rf{2.6a}) of order
\be{2.7a1} m_1 = \ldots = m_n \sim m_{\phi} \sim \mu^{1/4}\, .\ee
For the four-dimensional effective cosmological constant
$\Lambda_{eff}$ defined in \rf{1.36} one obtains in accordance
with Eq. \rf{2.5a} \ $\Lambda_{eff} \sim - \sqrt{\mu}$.

\medskip\medskip\medskip\medskip
\textbf{Summary}

 As stability
condition the existence of a minimum of the effective
potential of the dimensionally reduced theory is assumed, so that a late-time
attractor of the system could be expected with freezing
stabilization of the extra-dimensional scale factors and the
nonlinearity field. It was shown in Refs. \cite{GMZ1,GMZ2}, that
for purely geometrical setups this is only possible for negative
scalar curvatures, $(\bar R<0)$, independently of the concrete
form of the function $f(\bar R)$.

Four-dimensional purely gravitational models with $\bar R^{-1}$
curvature contributions have been proposed recently as possible
explanation of the observed late-time acceleration (dark energy)
of the Universe \cite{1/R-1,1/R-1a1,1/R-1a2,1/R-1a3,1/R-1a4,1/R-1a5,1/R-1a6,1/R-1a7,1/R-1a8}. It is shown that higher dimensional models with the
same $\bar R^{-1}$ scalar curvature nonlinearity reproduce (after
dimensional reduction) the two solution branches of the
four-dimensional models. But due to their oversimplified structure
these models cannot simultaneously provide a late-time
acceleration of the external four-dimensional spacetime and a
stabilization of the internal space. A late-time acceleration is
only possible for one of the solution branches
--- for that which yields  a positive maximum of the potential
$U(\phi)$ of the nonlinearity field. A stabilization of the
internal spaces requires a negative minimum of $U(\phi)$ as it can
be induced by the other solution branch.

 \addcontentsline{toc}{subsection}{2.4. $R^{2}+R^{4}$ model}
 \setcounter{subsection}{1}
 \medskip\medskip\medskip\medskip
 \textbf{2.4. $R^{2}+R^{4}$ model}
 \medskip\medskip\medskip\medskip

\begin{figure}
  \center
    \includegraphics[width=4in,height=3in]{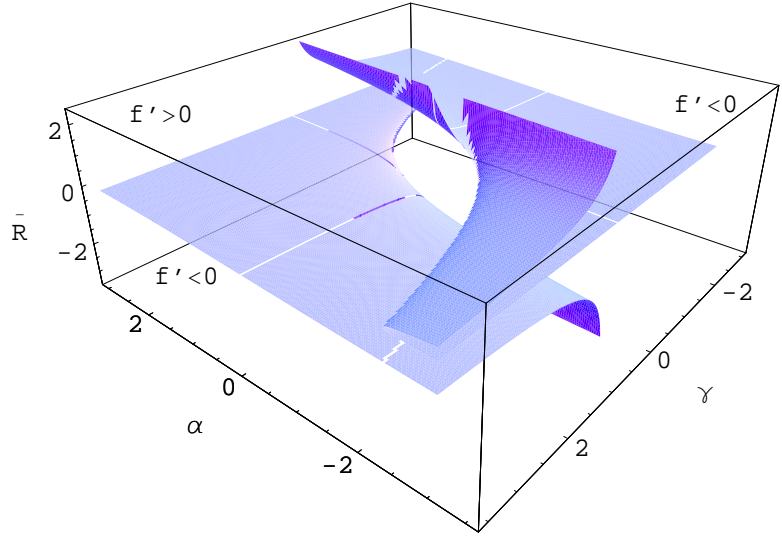}\\
\caption {The surfaces $f'=0$ as a functions
$\bar R = \bar R(\alpha , \gamma)$ for the model \rf{4.1a}.\label{f=0}}
\end{figure}

In this section let us analyze a model with
curvature-quadratic and curvature-quartic correction terms of the
type
\be{4.1a}
f(\bar{R})=\bar{R}+\alpha\bar{R}^{2}+\gamma\bar{R}^{4}-2\Lambda_{D}\,
. \ee
According to eq. \rf{1.7a}:
\be{4.2a}
f' = e^{A\phi} = 1 +2\alpha \bar{R} + 4\gamma
\bar{R}^3\; .
\ee
The definition \rf{1.7a} $f'=\exp(A\phi)$ clearly indicates that the positive branch $f'>0$ is chosen. For the given model \rf{4.1a}, the surfaces $f'=0$
as a functions
$\bar R = \bar R(\alpha , \gamma)$ are given in Fig. \ref{f=0}. As it easily follows from Eq. \rf{4.2a}, points where all three values $\bar R, \gamma$ and
$\alpha$ are positive correspond to the region $f' >0$. Thus, this picture shows that there is one simply connected region $f'>0$ and two disconnected regions $f'<0$.

Eq. \rf{4.2a} can be rewritten equivalently in the form:
\be{s1a} \bar R^3+\frac{\alpha}{2\gamma}\bar R-\frac
1{4\gamma}X=0,\ee
\ba{4.4a1} X \equiv e^{A\phi} - 1\, ,\quad &&-\infty <\phi
<+\infty\nn\\ \Longrightarrow &&-1<X<+\infty\, . \ea
Eq. \rf{s1a} has three solutions $\bar R_{1,2,3}$, where one or three of
them are real-valued. Let
\be{s4}
q:=\frac{\alpha}{6\gamma},\quad r:=\frac 1{8\gamma}X.
\ee
The sign of the discriminant
 \be{s5}
 Q:=r^2+q^3
 \ee
 defines the
number of real solutions:
\ba{s6}
Q>0&\qquad
\Longrightarrow \qquad & \Im \bar R_1=0,\quad \Im \bar R_{2,3}\neq 0\, ,\nn\\
Q=0&\qquad \Longrightarrow \qquad & \Im \bar R_i=0 \ \forall i,
\quad \bar R_1=\bar R_2\, ,\nn \\
Q<0&\qquad \Longrightarrow \qquad & \Im \bar R_i=0 \ \forall i\, .
\ea
Physical scalar curvatures correspond to real solutions $\bar
R_i(X)$.
It is the most convenient to consider $\bar R_i=\bar R_i(X)$ as
solution family depending on the two additional parameters
$(\alpha,\gamma)$: $sign(\alpha)=sign(\gamma)\;\Longrightarrow
Q>0$, $sign(\alpha)\neq sign(\gamma)\;\Longrightarrow Q\gtrless
0$.

For $Q>0$ the single real solution $\bar R_1$ is given
as
\be{s7a}\bar
R_1=\left[r+Q^{1/2}\right]^{1/3}+\left[r-Q^{1/2}\right]^{1/3} := z_1+z_2\, ,
\ee
where $z_{1,2}$ is defined in the form:
\ba{4.5a1}
&&z_{1,2}^3=p\, e^{\pm\theta}\, , \quad p^2 = r^2-Q=-q^3\, ,\nn\\&&\cosh (\theta ) = \frac{r}{\sqrt{-q^3}}\, .
\ea
Taking into account eq. \rf{s4}, the function $X$ reads
\be{4.6a1}
X(\theta ) = 8\gamma \sqrt{-q^3}\cosh (\theta)\, .
\ee

 The three real solutions $\bar R_{1,2,3}(X)$ for $Q<0$ are
given as
\ba{s8}
\bar R_1&=&s_1+s_2,\nn\\
\bar R_2&=&\frac 12 (-1+i\sqrt 3)s_1+\frac 12 (-1-i\sqrt 3)s_2\nn
\\&=&e^{i\frac{2\pi}3}s_1+e^{-i\frac{2\pi}3}s_2,\nn\\
\bar R_3&=&\frac 12 (-1-i\sqrt 3)s_1+\frac 12 (-1+i\sqrt 3)s_2\nn
\\&=&e^{-i\frac{2\pi}3}s_1+e^{i\frac{2\pi}3}s_2,
 \ea
  where one can
fix the Riemann sheet of $Q^{1/2}$ by setting in the definitions
of $s_{1,2}$
 \be{s9} s_{1,2}:=\left[r\pm i|Q|^{1/2}\right]^{1/3}.
\ee
 A simple Mathematica calculation gives for Vieta's relations
from \rf{s8}
\ba{s10a1}
\bar R_1+\bar R_2+\bar R_3&=&0,\nn\\
\bar R_1\bar R_2+\bar R_1\bar R_3+\bar R_2\bar R_3&=&-3s_1s_2=3q,\nn\\
\bar R_1\bar R_2\bar R_3&=&s_1^3+s_2^3=2r.
\ea
 In order to work
with explicitly real-valued $\bar R_i$ let us rewrite $s_{1,2}$ from
\rf{s9} as follows
\ba{s11} s_{1,2}&=&|b|^{1/3}e^{\pm i\vartheta/3},\nn\\
|b|^2&=&r^2+|Q|=r^2-Q=-q^3, \nn\\
\cos(\vartheta)&=&\frac{r}{|b|}=\frac {r}{\sqrt{-q^3}}.
\ea
 and get via
\rf{s8}
\ba{s12}
\bar R_1&=&s_1+s_2=2|b|^{1/3}\cos (\vartheta/3),\\
\bar R_2&=&e^{i\frac{2\pi}3}s_1+e^{-i\frac{2\pi}3}s_2=2|b|^{1/3}\cos \left(\vartheta/3+2\pi/3\right),\nn\\
\bar
R_3&=&e^{-i\frac{2\pi}3}s_1+e^{i\frac{2\pi}3}s_2=2|b|^{1/3}\cos
\left(\vartheta/3-2\pi/3\right)\nn
\ea
or
\ba{s13}
\bar R{_k}&=&2|b|^{1/3}\cos \left(\frac{\vartheta+2\pi
k}3\right)\nn\\&=&2\sqrt{-q}\cos \left(\frac{\vartheta+2\pi k}3\right),\;
k=-1,0,1.\quad
\ea
   In order to understand the qualitative
behavior of these three real-valued solutions as part of the
global solution picture let us first note that, according to \rf{s1a},
one may interpret $X$ as single-valued function
\be{s14}
X(\bar R)=4\gamma \bar R^3+2\alpha \bar R
\ee
 and look
what is happening when $(\alpha,\gamma)$ are changed. Obviously, the
inverse function $\bar R(X)$ has three real-valued branches when
$X(\bar R)$ is not a monotonic function but instead has a minimum
and a maximum, i.e. when
\be{s15}
\partial_{\bar R}X:= X'=12 \gamma \bar R^2+2\alpha=0\;
\Longrightarrow \; \bar R^2=-\frac{\alpha}{6\gamma}
\ee
has two real solutions $\bar R_\pm=\pm \sqrt{-\alpha/(6\gamma)}$
and corresponding extrema\footnote{It is worth of noting that $ f''(\bar R_\pm) =X'(\bar R_\pm) =0$.}
\be{s15a}
X(\bar R_\pm)=\frac 43 \alpha \bar R_\pm\, .
\ee
It should
hold $\sign (\alpha)\neq \sign (\gamma)$ in this case, so that one
find
\ba{s15b}
\gamma>0,\alpha<0:&\;& X_{max}=X(\bar R_-), \
X_{min}=X(\bar
R_+)\nn\\
\gamma<0,\alpha>0:&\; & X_{max}=X(\bar R_+), \ X_{min}=X(\bar
R_-)\,.\nn\\
\ea
 The transition from the three-real-solution regime to
the one-real-solution regime occurs when maximum and minimum
coalesce at the inflection point
\be{s16}
\bar R_+=\bar R_-=0\; \Longrightarrow \;
\alpha=0, \ \gamma \neq 0\, .
\ee
 (The
non-degenerate case $\gamma\neq 0$ is considered here. Models with $\gamma=0$ are
degenerated ones and are characterized by quadratic scalar
curvature terms only.) Due to $-1\le X\le +\infty$
the limit $X\to +\infty$ where in leading approximation may be considered:
\be{s17}
4\gamma \bar R^3\approx X\to +\infty \ee so that \be{s18}
\bar R(X\to \infty)\to \sign(\gamma)\times \infty\, .
\ee
Leaving
the restriction $X\ge -1$ for a moment aside, it was found that
for product $\alpha\gamma<0$
there exist three real solution branches $\overline{\mathcal{R}}_{1,2,3}$:
\ba{s19} \gamma>0:&\; & -\infty \le \overline{\mathcal{R}}_1\le \bar R_-, \quad -\infty \le X \le X_{max},\nn\\
&& \bar R_-\le \overline{\mathcal{R}}_2 \le \bar R_+, \quad X_{max}\ge X \ge X_{min},\nn\\
&& \bar R_+\le \overline{\mathcal{R}}_3\le +\infty, \quad X_{min}\le X \le +\infty, \nn\\
\gamma<0:&\; & -\infty \le \overline{\mathcal{R}}_1 \le \bar R_-, \quad +\infty\ge X\ge X_{min},\nn\\
&&\bar R_-\le \overline{\mathcal{R}}_2 \le \bar R_+, \quad X_{min}\le X \le X_{max},\nn\\
&&\bar R_+\le \overline{\mathcal{R}}_3 \le +\infty, \quad X_{max}\ge X \ge
-\infty.\nn\\\ea
It remains for each of these branches to check which
of the solutions $\bar R_k$ from \rf{s13} can be fitted into this
scheme. Finally, one will have to set the additional restriction
$X\ge -1$ on the whole picture.

To define the conditions for minima  of the effective potential
$U_{eff}$, first the extremum positions of the potential
$U(\phi )$ is to be obtained. The extremum condition \rf{1.11a} for the given particular
model \rf{4.1a} reads:
\begin{equation}\label{extRa1}
\gamma\bar R_{(0)i}^{4}\left(\frac{D}{2}-4\right)+\alpha\bar
R_{(0)i}^{2}\left(\frac{D}{2}-2\right)+\bar
R_{(0)i}\left(\frac{D}{2}-1\right)-D\Lambda_{D} =0\, ,
\end{equation}

 \addcontentsline{toc}{subsection}{2.5. Case $D=8$ - analytical solution }
 \setcounter{subsection}{1}
 \medskip\medskip\medskip\medskip
 \textbf{2.5. Case $D=8$ - analytical solution}
 \medskip\medskip\medskip\medskip

In this part let us investigate the case of positive $Q(\phi)$ that
is equivalent to the condition
\be{s10}
Q(\phi)>0 \quad \Rightarrow \quad \sign \alpha =\sign \gamma\; .
\ee
The case $\sign \alpha \neq \sign \gamma \:$ that corresponds to
different signatures of the discriminant $Q$ will be considered in chapter 4.

To define the conditions for minima  of the effective potential
$U_{eff}$, first let us obtain the extremum positions of the potential
$U(\phi )$. The extremum condition \rf{1.11a} for the given particular
model \rf{4.1a} reads:
\begin{equation}\label{extR}
\bar R_{(0)1}^{4}\gamma\left(\frac{D}{2}-4\right)+\bar
R_{(0)1}^{2}\alpha\left(\frac{D}{2}-2\right)+\bar
R_{(0)1}\left(\frac{D}{2}-1\right)-D\Lambda_{D} =0\, ,
\end{equation}
where subscript "1" indicates seeking the extremum positions
for the solution \rf{s7a}. Eq. \rf{extR} clearly shows that $D=8$
is the critical dimension for the model \rf{4.1a} in full agreement
with the result of the appendix 1 (see \rf{a8-1}). In what follows,
that this critical case is investigated here. For $D=8$ eq. \rf{extR} is
reduced to a quadratic one
\begin{equation}\label{extR8}
\bar R_{(0)1}^{2}+\frac{3}{2\alpha}\bar
R_{(0)1}-\frac{4\Lambda_8}{\alpha}=0\;;\quad\Lambda_8\equiv\Lambda_{D=8}
\end{equation}
with the following two roots:
\begin{equation}\label{R0}
\bar
R_{(0)1}^{(\pm)}=-\frac{3}{4\alpha}\pm\sqrt{\left(\frac{3}
{4\alpha}\right)^{2}+\frac{4}{\alpha}\Lambda_8}\quad .
\end{equation}
These roots are real if parameters $\alpha$ and $\Lambda_8$
satisfy the following condition:
\begin{equation}\label{cona}
\left(\frac{3}{4\alpha}\right)^{2}+\frac{4}{\alpha}\Lambda_8\geq0\;
.
\end{equation}
If $\sign(\alpha)=\sign(\Lambda_8)$, then condition (\ref{cona}) is
automatically executed, else
\begin{equation}\label{LL}
|\Lambda_8|\leq\frac{9}{64|\alpha|}\; , \quad \sign(\alpha)\ne
\sign(\Lambda_8)\; .
\end{equation}

To insure that roots (\ref{R0}) correspond to a minimum value of
$U(\phi )$, they should satisfy the condition (\ref{1.11c}):
\be{usmin} f''\left[(D-2)f'-2\bar R f''\right]_{\phi_0}>0 \:
\Longleftrightarrow f''[3+4\alpha \bar R ]_{\phi_0}>0\;, \ee where
\be{qwer1} f''=2\alpha+12\gamma\bar R^2\;. \ee

Because for $Q>0$ eq. \rf{s7a} is the single real solution of the
cubic eq. \rf{extR}, then $\bar{R} = \bar{R}_1(\phi )$ is a
monotonic function of $\phi $
Thus, the derivative
$\partial_{\phi}\bar R_1=Af'/f''\;$ does not change its sign.
Keeping in mind that the $f'>0$ branch is considered, the function
$\bar{R}_1(\phi)$ is a monotone increasing one for $f''>0$. As
apparent form eq. \rf{s7a}, for increasing $\bar{R}_1$ one should
take $\gamma >0$. In a similar manner, the function
$\bar{R}_1(\phi)$ is a monotone decreasing one for $f''
,\gamma<0$.
Thus, for the minimum position $\bar{R}_{(0)1}$, inequality
(\ref{usmin}) leads to the following conditions (reminding that
according to eq. \rf{1.13-1} the minimum position $\bar{R}_{(0)1}$
should be negative and according to eq. \rf{s10} $\sign \alpha =
\sign \gamma $):

\textbf{I.} $\;\; f'' ,\gamma, \alpha >0\;:$
\be{g>0}3+4\alpha\bar
R^{\pm}_{(0)1}>0 \Longleftrightarrow
|R^{\pm}_{(0)1}|<\frac{3}{4\alpha }\;. \ee

\textbf{II.} $\;\; f'' ,\gamma , \alpha <0\;:$
\be{g<0}3+4\alpha\bar R^{\pm}_{(0)1}<0 \Longleftrightarrow
-|R^{\pm}_{(0)1}|>\frac{3}{4|\alpha|}\;. \ee
Obviously, inequality \rf{g<0} is impossible and one arrives to the
conclusion that the minimum of the effective potential $U_{eff}$
is absent if $\sign \alpha = \sign \gamma = -1$.

Additionally, it can be easily seen that in the case
\be{sign}
\sign \alpha = \sign \gamma = \sign \Lambda_D = +1
\ee
the effective potential $U_{eff}$ has no minima also. This
statement follows from the form of the potential $U(\phi )$ for
the model \rf{4.1a}. According to eq. \rf{1.8a1}, $U(\phi )$ reads:
\be{4.8b}
U(\phi ) = (1/2)e^{-B\phi
}\left(\alpha\bar{R}^2+3\gamma\bar{R}^4+2\Lambda_D\right)\, .
\ee
Thus, this potential is always positive for parameters satisfying
\rf{sign} and one arrives to the contradiction with the minimum
condition \rf{1.36c4a}. Therefore, the investigation carried above
indicates that the internal space stable compactification is
possible only if the parameters satisfy the following sign
relation:
\be{signs}
\alpha >0, \gamma>0, \Lambda_8<0\, .
\ee
Let us investigate this case in more detail.
%
%
%
%
%
For this choice of signs of the parameters, it can be easily seen
that both extremum values $\bar R^{(\pm)}_{(0)1}$ from eq. \rf{R0}
satisfy the condition \rf{1.13-1}:  $\bar R^{(\pm)}_{(0)1}<0$.
However, the expression
\begin{equation}\label{n88}
f'\left(\bar R^{(\pm)}_{(0)1}\right) = 1 + 2\alpha
\bar{R}^{(\pm)}_{(0)1} + 4\gamma \bar
{R}^{(\pm)3}_{(0)1} =-\frac{1}{2} \pm \sqrt{\frac{9}{4}-
16\alpha|\Lambda_8|}-4\gamma\left|\bar R^{(\pm)}_{(0)1}\right|^{3}
\end{equation}
shows that only $\bar R^{(+)}_{(0)1}$ can belong to $f'>0$ branch.
To make $f'\left(\bar R^{(+)}_{(0)1}\right)$ positive, parameter
$\gamma$ should satisfy the condition
\begin{equation}\label{Lcond}
  \gamma<\frac{-\frac{1}{2}+\sqrt{\frac{9}{4}-
16\alpha|\Lambda_8|}}{4\left|\bar R^{(+)}_{(0)1}\right|^{3}}\quad.
\end{equation}
As apparent from this equation, parameter $\gamma $ remains
positive if $\Lambda_8$ belongs to the interval
\be{lambda}
\Lambda_8\in\left(-\frac{1}{8 \alpha},0\right)\, .
\ee
For this values of $\Lambda_8$, the condition \rf{LL} is
automatically satisfied. It should be noted, that for positive $\alpha$
and negative $\bar R^{(+)}_{(0)1}$ the condition \rf{g>0} is also
satisfied. Taking into account the interval \rf{lambda}, the
corresponding allowed interval for $\gamma $
reads\footnote{Similar interval for the allowed values of $\gamma$
was also found in \cite{GZBR} for the curvature-quartic model.}
\be{gm}
\gamma\in\left(0,\frac{1}{4\left|\bar
R^{(+)}_{(0)1}\right|^{3}}\right)\;.
\ee

Thus, for any positive value of $\alpha$, Eqs. \rf{lambda} and
\rf{gm} define allowed intervals for parameters  $\Lambda_8$ and
$\gamma$ which ensure the existence of a global minimum of the
effective potential $U_{eff}$. Here, the required
stable compactification of the internal space is obtained. The position of the
minimum $(\beta^1_0,\phi_0 )$ and its value can be easily found
(via the root $\bar{R}^{(+)}_{(0)1}$) with the help of Eq.
\rf{4.2a}. The Figs.\rf{fig221}, \rf{fig222}
demonstrate such minimum for a particular choice of the
parameters: $\alpha=1,\gamma=1,\Lambda_8=-0.1$. Moreover, the limit $\Lambda_8 \rightarrow 0$
corresponds to $\bar{R}^{(+)}_{(0)1} \rightarrow 0$ which results
in the decompactification of the internal space $\beta^1_0
\rightarrow \infty$.

 \medskip\medskip\medskip\medskip
\textbf{Summary}

In this section the model with curvature-quadratic and
curvature-quartic correction terms of the type \rf{4.1a} was analyzed and it was shown
that the stable compactification of the internal space takes place
for the sign relation \rf{signs}. Moreover, the parameters of the
model should belong to the allowed intervals (regions of
stability) \rf{lambda} and \rf{gm}. The former one can be
rewritten in the form
\be{con1}
\Lambda_8 =\frac{\xi}{8\alpha}\, , \quad \xi\in\left(-1,0\right)\,
.
\ee
Thus, for the root $\bar{R}^{(+)}_{(0)1}$ and parameter $\gamma$
one obtains respectively
\be{con2}
\bar{R}^{(+)}_{(0)1} = \frac{\eta}{\alpha}\, , \qquad
\eta\equiv\frac{1}{4}\left(-3+\sqrt{9+8\xi}\right)<0
\ee
and
\be{con3}
\gamma=\frac{\zeta\alpha^{3}}{4|\eta|^{3}}\; ,\quad
\zeta\in\left(0,1\right)\; .
\ee
Eq. \rf{con2} shows that $\bar{R}^{(+)}_{(0)1} \in
\left(-\frac{1}{2\alpha},0\right)$.

It is of interest to estimate the masses of the gravitational
excitons \rf{1.35a} and of the nonlinearity field $\phi$
\rf{1.35b} as well as the effective cosmological constant
\rf{1.36}. From Eqs. \rf{con1}-\rf{con3} follows that
$\bar{R}^{(+)}_{(0)1} \sim \Lambda_8 \sim
-\alpha^{-1}\:,\;\gamma\sim \alpha^{3} \Longrightarrow f'(\phi_0)
\sim \mathcal{O}(1)\; , f''(\phi_0) \sim \alpha\; ,
 U(\phi_0) \sim -\alpha^{-1}$. Then, the corresponding estimates
 read:
\be{con4}
-\Lambda_{eff}\sim m^{2}_{1}\sim m^{2}_{\phi}\sim \alpha^{-1}\, .
\ee
From other hand (see Eqs. \rf{1.31a} and \rf{1.35a})
\be{con5}
U(\phi_0) \sim \exp (-2\beta^1_0) = b_{(0)1}^{-2}\, .
\ee
So, if the scale factor of the stabilized internal space is of the
order of the Fermi length: $b_{(0)1} \sim L_F \sim 10^{-17}$cm,
then $\alpha \sim L_F^2$ and for the effective cosmological
constant and masses one obtains: $-\Lambda_{eff}\sim m^{2}_{1}\sim
m^{2}_{\phi}\sim 1\mbox{TeV}^2$.

The internal space stable
compactification analysis was performed in the case $Q(\phi )>0 \Rightarrow
\sign \alpha = \sign \gamma $. In forthcoming chapters the extend of
this investigation to the case of negative $Q(\phi )$ is performed, where the
function $\bar{R}(\phi )$ has three real-valued branches.

For the visualization reasons, let us consider the model with one
internal space and critical dimension $D=8$ (as usual, for the
external spacetime $D_0=4$). Then, the effective potential
\rf{1.31} reads:
\be{b1}
U_{eff}(\hat{
\beta^1},\phi)=e^{-4\hat{\beta^1}}\left[-\frac{1}{2}
\hat{R}_{1}e^{-2\hat{\beta^1}}+U(\phi)\right]\; .
\ee
To draw this effective potential, let us define $\hat{R}_{1}$ via
$U(\phi_0)$ in Eq. \rf{1.35a}. In its turn, $U(\phi_0)$ is defined
in Eq. \rf{1.12-1} where $\bar{R}(\phi_0) =
\bar{R}^{(+)}_{(0)1}$ and $f'\left(\bar R^{(+)}_{(0)1}\right)$ can
be found from \rf{n88}. In Figs.\rf{fig221}, \rf{fig222} the generic form of the
$U_{eff}$ is illustrated by a model with parameters
$\alpha=1,\gamma=1,\Lambda_{8}=-0.1$ from the stability regions
\rf{lambda} and \rf{gm}.

\begin{figure}[htbp]
\centerline{\includegraphics[width=2.5in,height=2.5in]{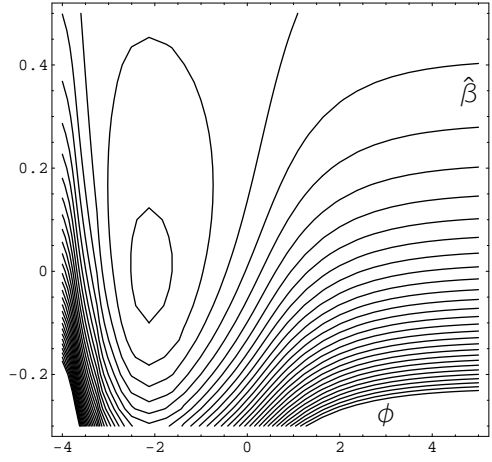}}
\caption{Typical contour plot of the effective potential
$U_{eff}(\hat{
\beta^1},\phi)$ given in Eq. \rf{b1}
with parameters $\alpha=1,\gamma=1,\Lambda_{8}=-0.1$. $U_{eff}$
reaches the global minimum at $(\hat{\beta}^1 =0, \phi\approx
-2.45)$.
\label{fig221}}
\end{figure}

\begin{figure}[htbp]
\centerline{\includegraphics[width=3in,height=3in]{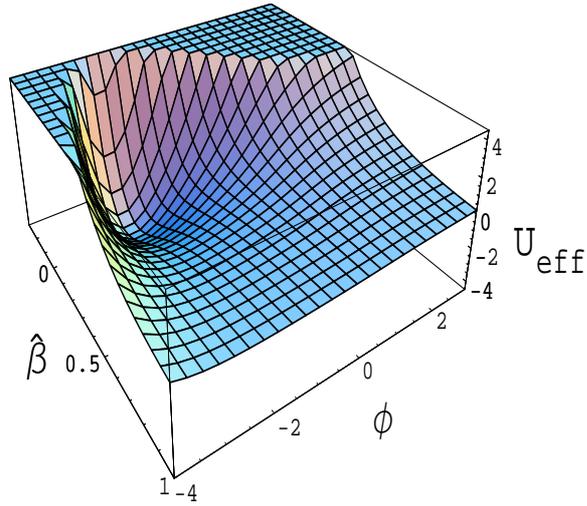}}
\caption{Typical form of the effective potential $U_{eff}(\hat{
\beta^1},\phi)$ given in Eq. \rf{b1}
with parameters $\alpha=1,\gamma=1,\Lambda_{8}=-0.1$.
\label{fig222}}
\end{figure}

 \newpage
 \addcontentsline{toc}{section}{CHAPTER 3. MULTIDIMENSIONAL NONLINEAR \\ MODELS WITH FORMS}
 \section*{\center{CHAPTER 3 \\ $\;$ \\
MULTIDIMENSIONAL NONLINEAR MODELS WITH FORMS}}
 \setcounter{section}{3}
 \setcounter{equation}{0}
\medskip\medskip

In the present chapter, the nonlinear gravitational
multidimensional cosmological model is considered with action of the type
$\bar{R}+1/\bar{R}$ with form-fields as a matter source, including a
bare cosmological term as an additional parameter of the theory.
It is assumed that the corresponding higher-dimensional space-time
manifold undergos a spontaneous compactification to a manifold
with warped product structure of the external and internal spaces.
Each of spaces has its own scale factor. A model without
form-fields and bare cosmological constant was considered in previous chapter,
where the internal space freezing stabilization was
achieved due to negative minimum of the effective potential. Thus,
such model is asymptotically AdS without accelerating behavior of
our Universe. It is well known that inclusion of usual matter can
uplift potential to the positive values \cite{GMZ2}. One of
the main task of present investigations is to observe such
uplifting due to the form-fields. Indeed, it is demonstrated that for
certain parameter regions the late-time acceleration scenario in
present model becomes reachable. However, it is not simple uplifting
of the negative minimum of the theory (\ref{1.1a}) to the positive
values. The presence of the form-fields results in much more rich
structure of the effective potential then for (\ref{1.1a}). Here,
 additional branches are obtained with extremum points, and one of such
extremum corresponds to the positive minimum of the effective
potential. This minimum plays the role of the positive
cosmological constant. With the corresponding fine tuning of the
parameters, it can provide the late-time accelerating expansion of
the Universe. Moreover, it is  shown that for this branch of the
effective potential there is also a saddle point. Thus, domain walls
are obtained, which separate regions with different vacua. It is
demonstrated that these domain walls do not undergo inflation
because the effective potential is not flat enough around the
saddle point.

It is also worth of noting that the effective potential in this
reduced model has a branchpoint. It gives very interesting
possibility to investigate transitions from one branch to another
by analogy with catastrophe theory or similar to phase transitions
in statistical theory. This idea needs more detail investigation.

The chapter is structured as follows. First, a brief description of multidimensional models with scalar
curvature nonlinearity $f(\bar{R})$ and the form-fields as a matter
source is given. Then, dimensional reduction is performed and
effective four-dimensional action with effective potential is obtained.
General formulas from this section are applied to the specific
model $f(\bar{R})=\bar{R}-\mu /{R}-2\Lambda_{D}$.
Then minimum conditions of the effective potential are obtained. These
conditions are analyzed for the cases of zero and positive effective
cosmological constants respectively. Furthermore, it is demonstrated
that the positive minimum of the
effective potential plays the role of the positive cosmological
constant and can provide the late-time accelerating expansion.
Additionally, this minimum is accompanied by a saddle point. It
results in non-inflating domain walls in the Universe.


 \addcontentsline{toc}{subsection}{3.1. General equations}
 \setcounter{subsection}{1}
 \medskip\medskip\medskip\medskip
 \textbf{3.1. General equations}
\medskip\medskip\medskip\medskip

Let us consider a $D= (D_0+D^{\prime})-$dimensional nonlinear
gravitational theory with action functional
\ba{1.1} S&=&\frac{1}{2
\kappa^{2}_{D}}\int_{M}d^{D}x\sqrt{|\bar{g}|}
f(\bar{R})\nn\\&-&\frac{1}{2}\int_{M}d^{D}x\sqrt{|\bar{g}|}\sum^{n}_{i=1}
\frac{1}{d_{i}!}\left(F^{(i)}\right) ^{2}\, ,\ea
where $f(\bar R)$ is an arbitrary smooth function of a scalar
curvature $\bar R := R[\bar g]$ constructed from the
$D-$dimensional metric $\bar g_{ab}\; (a,b = 1,\ldots,D)$.
$D^{\prime}$ is the number of extra dimensions. $\kappa^2_D $
denotes the $D-$dimensional gravitational constant. In action
(\ref{1.1}), a form field (flux) $F$ has block-orthogonal
structure consisting of $n$ blocks. Each of these blocks is
described by its own antisymmetric tensor field $F^{(i)}
(i=1,\ldots ,n)$ of rank $d_{i}$ ($d_{i}$-form field strength).
Additionally, let us assume that for the sum of the ranks holds
$\sum^{n}_{i=1}d_{i}=D'$.

Following Refs. \cite{GZBR,GMZ(ASS)}, it can be shown that the
nonlinear gravitational theory \rf{1.1} is equivalent to a linear
theory $R = R[g]$ with conformally transformed metric
\be{1.2} g_{ab} = \Omega^2 \bar g_{ab} = \left[ f'(\bar
R)\right]^{2/(D-2)}\bar g_{ab}\;  \ee
and an additional minimal scalar field $\phi=\ln[f'(\bar R)]/A$
coupled with fluxes. The scalar field $\phi$ is the result and the
carrier of the curvature nonlinearity of the original theory.
Thus, for brevity, referring to the field $\phi$ as
nonlinearity scalar field, a self-interaction potential $U(\phi )$
of the scalar field $\phi$ reads
\be{1.3} U(\phi ) = \frac12 e^{- B \phi} \left[\; \bar R (\phi
)e^{A \phi } - f\left( \bar R (\phi )\right) \right]\; , \ee
where
\be{1.4} A = \sqrt{\frac{D-2}{D-1}}\, , \quad B = \frac
{D}{\sqrt{(D-2)(D-1)}}\, . \ee

Furthermore, let us assume that the multidimensional space-time
manifold undergoes a spontaneous compactification \be{1.5} M
\longrightarrow M = M_0 \times M_1 \times \ldots \times M_n \ee in
accordance with the block-orthogonal structure of the field
strength $F$, and that the form fields $F^{(i)}$, each nested in
its own $d_i-$dimensional factor space $M_i\, (i=1,\ldots ,n)$,
respect a generalized Freund-Rubin ansatz \cite{FR}. Here,
($D_0=4$)-dimensional space-time $M_0$ is treated as our external
Universe with metric $g^{(0)}(x)$.

This allows performing of a dimensional reduction of the model
along the lines of Refs.
\cite{GZ(PRD1997),GMZ(PRDa),GZ(PRD2000),RZ}.
The factor spaces $M_i$ are then Einstein spaces with metrics
$g^{(i)} \equiv e^{2\beta^i(x)} \gamma^{(i)}$ which depend only
through the warp factors $a_i(x) :=e^{\beta^i(x)}$ on the
coordinates $x$ of the external space-time $M_0$. For the
corresponding scalar curvatures holds $R\left[ \gamma^{(i)}\right]
=\lambda ^id_i\equiv r_i $ (in the case of the constant curvature
spaces $\lambda^i =k_i(d_i-1),\, \, k_i = 0,\pm 1$). The warped
product of Einstein spaces leads to a scalar curvature $\bar{R}$
which depends only on the coordinate $x$ of the $D_0-$dimensional
external space-time $M_0$: $\bar{R}[\bar{g}] = \bar{R}(x)$. This
implies that the nonlinearity field $\phi$ is also a function only
of $x$: $\phi = \phi (x)$. Additionally, it can be easily seen
\cite{GZBR} that the generalized Freund-Rubin ansatz results in
the following expression for the form-fields: $
\left(F^{(i)}\right)^2=f_i^2/a_i^{2d_i} $ where $ f^i=const $.

In general, the model will allow for several stable scale factor
configurations (minima in the landscape over the space of volume
moduli). Let us choose one of them (which is expected to correspond to
current evolution stage of our observable Universe), denote the
corresponding scale factors as $\beta^i_0$, and work further on
with the deviations $\hat \beta^i (x)= \beta^{i}(x) -
\beta^{i}_0$.

Without loss of generality\footnote{The difference between a
general model with $n>1$ internal spaces and the particular one
with $n=1$ consists in an additional diagonalization of the
geometrical moduli excitations. \label{n=1}}, one shall consider a
model with only one $d_1$-dimensional internal space. After
dimensional reduction and subsequent conformal transformation to
the Einstein frame the action functional \rf{1.1}
reads\footnote{The equivalency between original higher dimensional
and effective dimensionally reduced models was investigated in a
number of papers (see e.g.\cite{GZaa,Ivashchuk,Ivashchuka1}). The origin of this
equivalence results from high symmetry of considered models (i.e.
because of specific metric ansatz which is defined on the manifold
consisting of direct product of the Einstein spaces).}
\ba{1.6} S&=&\frac 1{2\kappa _0^2}\int\limits_{M_0}d^{D_0}x\sqrt{|
\tilde{g}^{(0)}|}\left\{ R\left[  \tilde{g}^{(0)}\right] -
\tilde{g}^{(0) \mu \nu}
\partial_{\mu}\varphi \partial_{\nu} \varphi \right.\nn\\ &-& \left. \tilde{g}^{(0) \mu \nu}
\partial_{\mu}\phi \partial_{\nu} \phi -2U_{eff} (\varphi ,\phi )
\right\} \, , \ea
where $\varphi := -\sqrt{d_1(D-2)/(D_0-2)}\, \hat \beta^1$ and
$\kappa^2_0 := \kappa^2_D/V_{d_1}$ denotes the $D_0-$ dimensional
(four-dimensional) gravitational constant. $V_{d_1} \sim
\exp{(d_1\beta_{0}^{1})}$ is the volume of the internal space at
the present time.

A stable compactification of the internal space $M_1$ is  ensured
when its scale factor $\varphi$ is frozen at the minimum of the
effective potential
\be{1.7} U_{eff}=e^{b\varphi}\left[-\frac 12 R_1
e^{a\varphi}+U(\phi)+h e^{c\phi}e^{ad_1 \varphi} \right]\, , \ee
where $R_1 := r_1 \exp{(-2\beta^1_0)}$ defines the curvature of
the internal space at the present time and contribution of the
form-field into the effective action is described by
$h:=\kappa^2_D\,f^2_1 \exp{(-2d_1\beta^1_0)}
>0$. For brevity let us introduce notations
\ba{1.8} a:&=&2\sqrt{\frac{D_0-2}{d_1(D-2)}}\; ,\;
b:\;=2\sqrt{\frac{d_1}{(D-2)(D_0-2)}}\; ,\nn\\ c:&=&
\frac{2d_1-D}{\sqrt{(D-1)(D-2)}}\; . \ea

 \addcontentsline{toc}{subsection}{3.2. $R^{-1}$ model with forms}
 \setcounter{subsection}{1}
 \medskip\medskip\medskip\medskip
 \textbf{3.2. $R^{-1}$ model with forms}
  \medskip\medskip\medskip\medskip

In this section  the conditions of the compactification are analyzed
for a model with
\be{2.1a} f(\bar{R})=\bar{R}-\frac{\mu}{\bar{R}}-2\Lambda_{D}\;.
\ee
Then from the relation $f^{\prime}(\bar R) = \exp{(A\phi)}$ one
obtains
\be{R} \bar{R}=q\sqrt{\frac{|\mu |}{s(e^{A\phi}-1)}}\;,\quad
q=\pm1\;, \, s =\sign (\mu)\; . \ee
Thus, the ranges of variation of $\phi$ are $\phi\in(-\infty,0)$
for $\mu <0\; (s=-1)$ and $\phi\in(0,+\infty)$ for $\mu >0\;
(s=+1)$.

It is worth of noting that the limit $\phi \to \pm 0\; (f^{\prime}
\to 1 )$ corresponds to the transition to a linear theory: $f(\bar
R) \to \bar R -2\Lambda_D$ and $R\to \bar R$. This is general
feature of all nonlinear models $f(\bar R)$. For example, in the
case \rf{2.1a} one obtains  $f(\bar R) = \bar R \left(2 - \exp (A\phi
)\right) - 2\Lambda_D \ \longrightarrow \bar R -2\Lambda_D$ for
$\phi \to 0$. From other hand, for particular model \rf{2.1a},
eq. \rf{R} shows that the point $\phi = 0$ maps into infinity
$\bar R,R = \pm \infty$. Thus, in this sense, one shall refer to
the point $\phi =0$ as singularity.

For the model \rf{2.1a}, potential \rf{1.3} $U(\phi)$ reads
\begin{equation}\label{U}
U(\phi)=\frac{1}{2}e^{-B\phi}\left(2qs\sqrt{|\mu|}\sqrt{se^{A\phi}-s}+
2\Lambda_{D}\right)\, .
\end{equation}

It is well known (see e.g.
\cite{GMZ2,GMZ(PRDa),GZ(PRD2000)}) that in
order to ensure a stabilization and asymptotical freezing of the
internal space $M_{1}$, the effective potential \rf{1.7} should
have a minimum with respect to both scalar fields $\varphi$ and
$\phi$. Let us remind that the minimum position is chosen with respect
to $\varphi$ at $\varphi =0$. Additionally, the eigenvalues of the
mass matrix of the coupled $(\varphi,\phi)$-field system, i.e. the
Hessian of the effective potential at the minimum position,
\begin{equation}\label{Hessian}
    J:=\left.\left(
\begin{array}{cc}
  \partial^{2}_{\varphi\varphi}U_{eff} &  \partial^{2}_{\varphi\phi}U_{eff} \\\\
   \partial^{2}_{\phi\varphi}U_{eff} &  \partial^{2}_{\phi\phi}U_{eff} \\
\end{array}
\right)\right|_{extr}
\end{equation}
should be positive definite (this condition ensures the
positiveness of the mass squared of scalar field excitations).
According to the Silvester criterion this is equivalent to the
condition:
\begin{equation}\label{con}
J_{11}>0\;,\quad J_{22}>0\;,\quad \mbox{det}(J)>0\quad.
\end{equation}

It is convenient in further consideration to introduce the
following notations:
\begin{equation}\label{xc}
\phi_{0}:=\phi|_{extr}\,,\, X:=\sqrt{se^{A\phi_{0}}-s}>0\,
\rightarrow \, X_{(s=-1)}<1\, .
\end{equation}
Then potentials $U(\phi )$, $U_{eff}(\varphi ,\phi
)$ and derivatives of the $U_{eff}$ at an extremum (possible
minimum) position ($\varphi=0,\phi_{0}$) can be rewritten as follows:
\be{U0} U_{0}\equiv
U|_{extr}=\left(1+sX^{2}\right)^{-B/A}\left(qs\sqrt{|\mu|}X+
\Lambda_{D}\right), \ee
\be{Ueff0}
U_{eff}|_{extr}=-\frac{1}{2}R_{1}+U_{0}(X)+h\left(1+sX^{2}\right)^{c/A}\;,
\ee
\ba{vpUeff}
\partial_{\varphi}U_{eff}|_{extr}&=&-\frac{a+b}{2}R_{1}+bU_{0}(X)\nn\\&+&(ad_{1}+
b)h\left(1+sX^{2}\right)^{c/A}=0,\;\quad \ea
\ba{pUeff}
\partial_{\phi}U_{eff}|_{extr}&=&
ch\left(1+sX^{2}\right)^{c/A}-BU_{0}(X)\nn\\&+&\frac{qA\sqrt{|\mu|}}{2X}
\left(1+sX^{2}\right)^{1-B/A}=0,\qquad \ea
\ba{vpvpUeff}
\partial^{2}_{\varphi\varphi}U_{eff}|_{extr}&=&-\frac{(a+b)^2}{2}R_{1}+
b^{2}U_{0}(X)\nn\\&+&(ad_{1}+b)^{2}h\left(1+sX^{2}\right)^{c/A},\qquad
\ea
\be{pvpUeff}
\partial^{2}_{\varphi\phi}U_{eff}|_{extr}=chad_{1}\left(1+sX^{2}\right)^{c/A}\;,
\ee
\ba{ppUeff}
\partial^{2}_{\phi\phi}U_{eff}|_{extr}&=&ch\left(c-A+2B\right)
\left(1+sX^{2}\right)^{c/A}\nn\\&+&B(A-B)U_{0}(X)\nn \\
&-&
\frac{qs\sqrt{|\mu|}A^{2}}{4X^{3}}\left(1+sX^{2}\right)^{2-B/A}.
\:\qquad \ea
The most natural strategy for extracting detailed information
about the location of stability region of parameters in which
compactification is possible would consist in solving
(\ref{pUeff}) for $X$ with subsequent back-substitution of the
found roots into the inequalities (\ref{con}) and the equation
(\ref{vpUeff}). To get the main features  of the model under
consideration, it is sufficient to investigate two particular
nontrivial situations. Both of these cases are easy to handle
analytically.

 \addcontentsline{toc}{subsection}{3.3. Case $\Lambda_{eff}=0$}
 \setcounter{subsection}{1}
 \medskip\medskip\medskip\medskip
 \textbf{3.3. Case $\Lambda_{eff}=0$}
  \medskip\medskip\medskip\medskip

It can be easily seen from eqs. \rf{Ueff0} and \rf{vpUeff} that
condition $\Lambda_{eff}=U_{eff}|_{extr}=0$ results in relations
\be{R1}
R_{1}=2d_{1}h\left(1+sX^{2}\right)^{c/A}=\frac{2d_{1}}{d_{1}-1}U_{0}(X)\;,\quad
d_{1}\geq2\; , \ee
which enables to get from eq. (\ref{pUeff}) quadratic equation
for $X$
\be{4.1a1} (d_1+1)X^2 + qsd_1zX - s(d_1-1) = 0\, ,\quad z\equiv
2\Lambda_{D}/\sqrt{|\mu|} \ee
with the following solutions:
\ba{s1}
X_{p}&=&qs\frac{d_{1}}{2(d_{1}+1)}\left(-z+p\:\sqrt{z^2+4s\frac{d^{2}_{1}
-1}{d_{1}^{2}}}\right)\; ,\nn\\ p&=&\pm 1\,.\ea
In the case $s=-1\;$ parameter $z$ should satisfy condition $
|z|\geq z_{0}\equiv2\sqrt{d_{1}^{2}-1}/d_{1}$ $<2$ and for $z=z_0$
two solutions $X_{p}$ degenerate into one: $X_{p} \equiv X_0 =
-qs\sqrt{(d_1-1)/(d_1+1)}$.

Because of conditions $h\geq0$ and $e^{A\phi_{0}} = 1+sX^2
>0$, the relations \rf{R1} show that  parameters $R_{1}$ and
$U_{0}(X)$ should be non-negative: $R_{1}\geq0 , \;
U_{0}(X)\geq0$. Obviously, only one of the solutions \rf{s1}
corresponds to a minimum of the effective potential. With respect
to this solution let us define parameters in the relation \rf{R1}.
Therefore, one musts distinguish now which of $X_{p}$ corresponds to
the minimum of $U_{eff}$. Let us investigate solutions \rf{s1} for
the purpose of their satisfactions to conditions $e^{A\phi_{0}}> 0
,\, U_{0}(X)\geq 0$ and $X_{p} \geq 0$.

\textbf{The condition $e^{A\phi_{0}}=1+sX_{p}^{2}>0$:}

Simple analysis shows that solutions $X_{p}$ satisfy  this
inequality for the following combinations of parameters:
\begin{equation}\label{Xs}
\begin{tabular}{|c|c|}
\hline $\begin{array}{c}\mu >0\\(s=+1)\end{array}$ & $\; \, p=\pm1
\; :
\; z\in(-\infty,+\infty)\quad\qquad\qquad$\\
\hline $\begin{array}{c}\mu<0\\(s=-1)\end{array}$ &
$\begin{array}{c}p=+1
\;:\; z\in(-2,-z_{0})\cup(z_{0},+\infty) \\
 p=-1 \;:\; z\in(-\infty,-z_{0})\cup(z_{0},2)\quad\:\\ z=z_{0}\qquad\qquad\qquad\qquad\qquad\quad\qquad\qquad\end{array}$ \\
\hline
\end{tabular}
\end{equation}

\textbf{The condition $U_{0}(X)\geq0$:}

As appears from eq. \rf{U0}, this condition takes place if $X_{p}$
satisfies inequality $2qsX_{p}+z\geq 0$ which leads to the
conditions:
\begin{equation}\label{val}
\begin{tabular}{|c|c|}
  \hline
      $\begin{array}{c}\mu >0\\(s=+1)\end{array}$ &
      $\begin{array}{c}\; \; p=+1\;:\;z\in(-\infty,+\infty)\;\end{array}$ \\
  \hline
   $\begin{array}{c}\mu<0\\(s=-1)\end{array}$ &
   $\begin{array}{c}p=+1\;  :\; z\in(z_{0},+\infty)\;\;\;\\
   p=-1\;  :\; z\in(z_{0},2]\qquad\:\\z=z_{0}\qquad\qquad\qquad\qquad\qquad\end{array}$ \\
  \hline
\end{tabular}
\end{equation}
\textbf{The condition $X_{p}>0$:}

This condition is satisfied for the combinations:
\begin{equation}\label{X<>}
\begin{tabular}{|c|c|}
  \hline
      $\begin{array}{c}\mu >0\\(s=+1)\end{array}$ &
      $\begin{array}{c}q=+1:\left\{\begin{array}{c}
      z<0:\quad X_{+}>0 \quad X_{-}<0\\
      z>0:\quad X_{+}>0 \quad X_{-}<0\end{array}\right.\\
      q=-1:\left\{\begin{array}{c}
      z<0:\quad X_{+}<0 \quad X_{-}>0\\
      z>0:\quad X_{+}<0 \quad X_{-}>0\end{array}\right.\end{array}$ \\
  \hline
   $\begin{array}{c}\mu <0\\(s=-1)\end{array}$ &
   $\begin{array}{c}q=+1:\left\{\begin{array}{c}
      z<0:\quad X_{+}<0 \quad X_{-}<0\\
      z>0:\quad X_{+}>0 \quad X_{-}>0\end{array}\right.\\
      q=-1:\left\{\begin{array}{c}
      z<0:\quad X_{+}>0 \quad X_{-}>0\\
      z>0:\quad X_{+}<0 \quad X_{-}<0\end{array}\right.\end{array}$ \\
  \hline
\end{tabular}
\end{equation}

The comparison of \rf{Xs}, \rf{val} and \rf{X<>} shows that they
are simultaneously satisfied only for the following combinations:
 \begin{equation}\label{XF}
 \begin{tabular}{|c|c|}
  \hline
      $\begin{array}{c}\mu >0\\(s=+1)\end{array}$ &
      $\begin{array}{c}\; \; \; \; p=+1\, :\;\;q=+1\;:\;z\in(-\infty,+\infty)\end{array}\;$ \\
  \hline
   $\begin{array}{c}\mu <0\\(s=-1)\end{array}$ &
   $\begin{array}{c}p=+1\;:\;\;q=+1\;:\;z\in(z_{0},+\infty)\\
\;\;p=-1\;:\;\;q=+1\;:\;z\in(z_{0},2)\qquad\:\\\;q=+1\;:\;z=z_{0}\qquad\qquad\qquad\qquad\qquad\end{array}$ \\
  \hline
\end{tabular}
\end{equation}

Additionally, the extremum solutions $X_{p}$ should correspond to
the minimum of $U_{eff}$. The inequalities (\ref{con}) are the
sufficient and necessary  conditions for that. Let us analyze them in
the case of four-dimensional external space $D_0=4$. Taking into
account definitions \rf{1.4}, \rf{1.8}, \rf{vpvpUeff}-\rf{ppUeff}
and relations \rf{R1}, for $J_{11}, J_{22}$ and $J_{21}$ one gets
respectively:
\be{Lj11} J_{11}=\frac{8}{d_{1}+2}U_{0}(X_{p})\, , \ee
 \begin{equation}\label{Lj22}
 \begin{split}
  J_{22}=&-\sqrt{|\mu|}\left[\frac{6d_{1}(2qsX_{p}+z)\left(1+sX_{p}^{2}\right)^{-\frac{d_{1}+4}{d_{1}+2}}}{(d_{1}-1)(d_{1}+2)(d_{1}+3)}
  \right.
\\ &+\left.\frac{qs(d_{1}+2)}{4X^{3}_{p}(d_{1}+3)}
\left(1+sX_{p}^{2}\right)^{\frac{d_{1}}{d_{1}+2}}\right] \, ,
\end{split}
\end{equation}
\ba{Lj21}
J_{21}=&-&\sqrt{|\mu|}\left[\frac{(d_{1}-4)(2qsX_{p}+z)}{(d_{1}-1)(d_{1}+2)}\right.\nn\\
&\times &\left. \sqrt{\frac{2}{d_{1}(d_{1}+3)}}\left(1+sX_{p}^{2}
\right)^{-\frac{d_{1}+4}{d_{1}+2}}\right]\, . \quad\ea
It is supposed in these equations that each of $X_{p}$ can define
zero minimum of $U_{eff}$. In what follows, one shall check this
assumption for every $X_{p}$ with corresponding combinations of
signs of the parameters $s$ and $q$ in accordance with the table
\rf{XF}.

According to the Silvester criterion \rf{con}, $J_{11}$ should be
positive. Thus eqs. \rf{R1} and \rf{Lj11} result in the following
conclusions: the potential $U_0$ should be positive $U_0>0$, the
internal space should have positive curvature $R_1>0$ (hence,
$d_1>1$) and its stabilization (with zero minimum
$\Lambda_{eff}=0$) takes place only in the present of form-field
($h>0$). Transition from the non-negativity condition $U_0\geq 0$
to the positivity one $U_0>0$ corresponds to the only substitution
in \rf{val} $(z_{0},2] \rightarrow (z_{0},2)$  for the case
$s=-1,\; p=-1$. Exactly this interval $(z_{0},2)$ appears in
concluding table \rf{XF}. Therefore, $J_{11}$ is positive for all
$X_{p}$ from the table \rf{XF}.

\begin{figure}[htbp]
\centerline{\includegraphics[width=3in,height=2in]{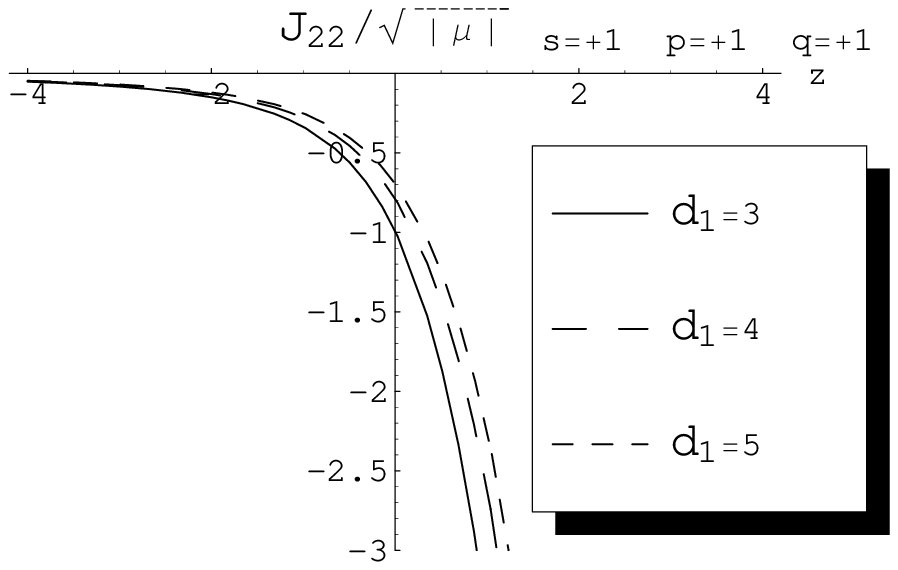}
\includegraphics[width=3in,height=2in]{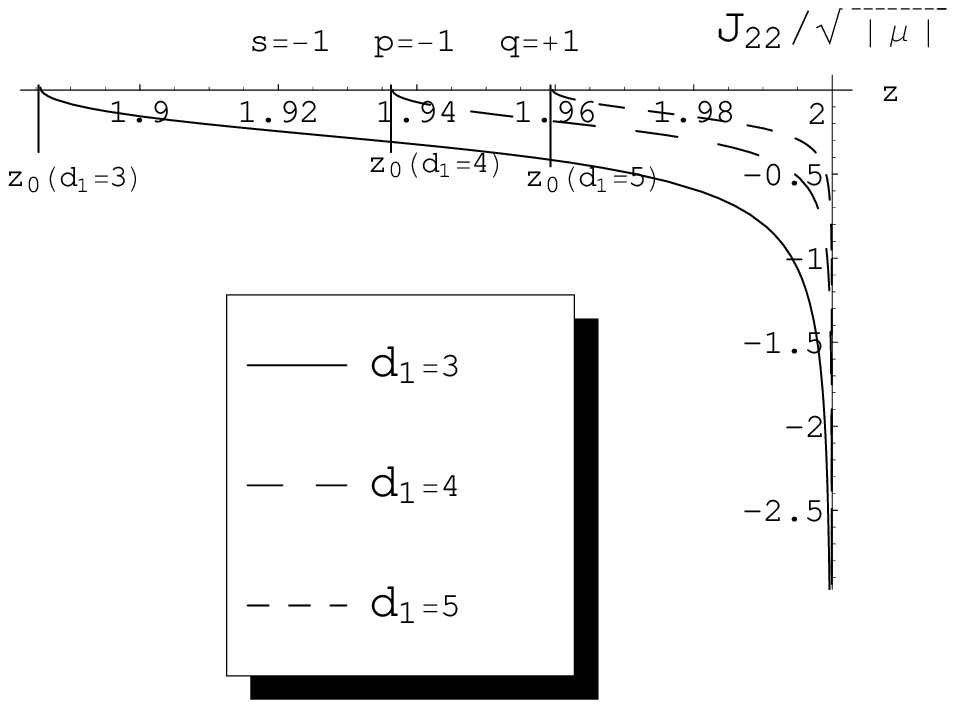}}
 \caption{Typical form of $J_{22}/\sqrt{|\mu|}$ (eq. \rf{Lj22}) for parameters
 $s=+1,\; q=+1,\; p=+1$ (left) and $s=-1,\; q=+1,\; p=-1$ (right). For both of these
 combinations of the parameters, $J_{22}<0$. \label{Lj22-1}}
\end{figure}

\begin{figure}[htbp]
\centerline{\includegraphics[width=3in,height=2in]{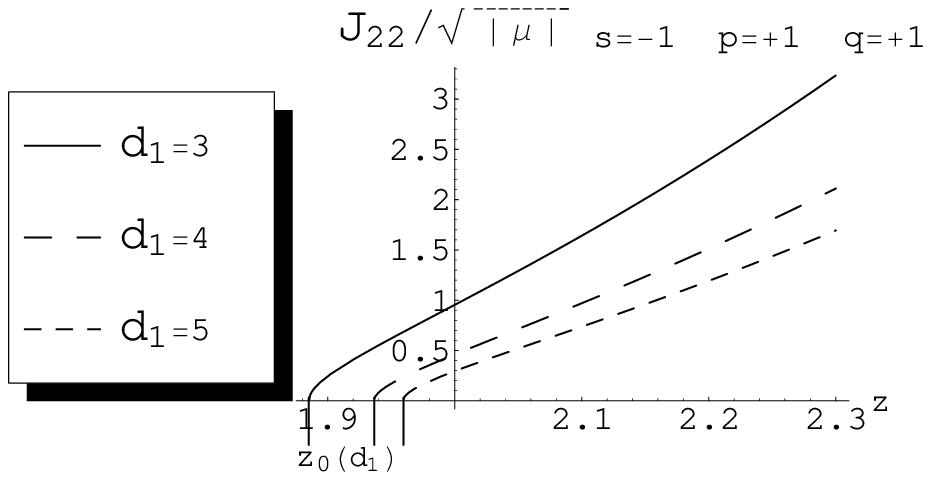}
\includegraphics[width=3in,height=2in]{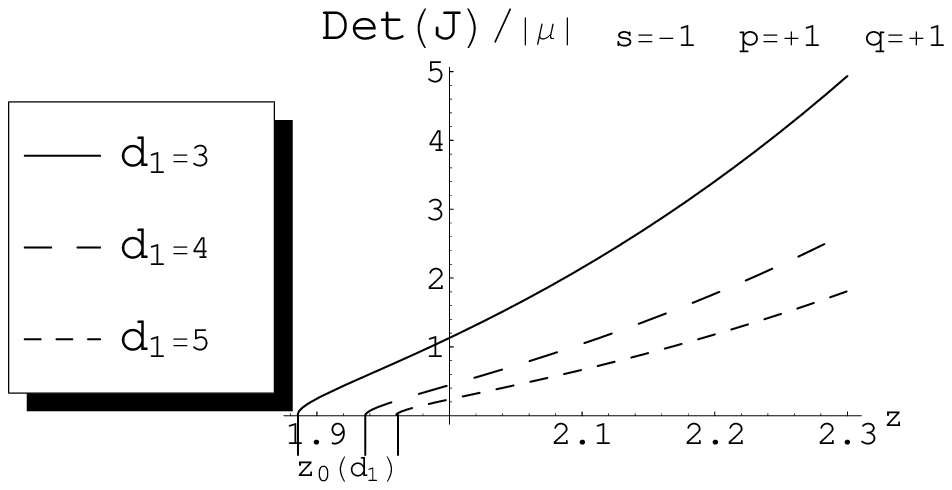}}
 \caption{Typical form of $J_{22}/\sqrt{|\mu|}$ (left) and
 $\mbox{det}(J)/|\mu|$ (right) (see eqs. \rf{Lj11}-\rf{Lj21}) for parameters
  $s=-1,\; q=+1,\; p=+1$. For this combination of the parameters, $J_{22}>0$
  and $\mbox{det}J>0$. \label{Lj22-2}}
\end{figure}

Concerning expressions $J_{22}$ and
$\mbox{det}(J)=J_{11}J_{22}-J_{12}^2$,\, graphical plotting (see
Fig.\ref{Lj22-1} and Fig.\ref{Lj22-2}) demonstrates that they are
negative for $s=+1,\; p=+1,\; q=+1$ and $s=-1,\; p=-1,\; q=+1$ but
positive in the case $s=-1,\; p=+1,\; q=+1$. For this latter
combination $z \in (z_0,\infty)$. The case $s=-1,\; q=+1$ and $z =
z_0$ should be investigated separately. Here, $X_{p}\equiv
X_{0}=\sqrt{(d_{1}-1)/(d_{1}+1)}$ and for $J_{22}$ and $J_{21}$ one
obtains:
\ba{J022}
J_{22}=&-&\sqrt{|\mu|}X^{-3}_{0}\left(1+sX_{0}^{2}\right)^{-\frac{d_{1}+4}{d_{1}+2}}
\nn\\
&\times
&\left(\frac{12(d_{1}-1)-(d_{1}+2)^{2}}{(d_{1}+1)^{2}(d_{1}+2)(d_{1}+3)}\right),
\quad\ea
\ba{J021}
J_{21}=&-&\sqrt{|\mu|}X^{-3}_{0}\left(1+sX_{0}^{2}\right)^{-\frac{d_{1}+4}{d_{1}+2}}
\nn\\ &\times
&\left(\frac{2\sqrt{2}(d_{1}-4)(d_{1}-1)}{(d_{1}+1)^{2}(d_{1}+2)\sqrt{d_1(d_{1}+3)}}\right).
\qquad \ea
It can be easily seen from eqs. (\ref{J022}) and \rf{J021} that
$J_{22}>0$ for $d_{1}\neq 4$ and   $J_{22} = J_{21}=0$ for
$d_{1}=4$. Additionally, $\mbox{det}(J)>0$ for $d_{1}\neq 4$.

Thus, it can finally concluded that zero minimum of the effective
potential $U_{eff}$ takes place either for $s=-1,\; q=+1,\; z\in
(z_{0},\infty), \forall \, d_{1}>1$ (position of this minimum is
defined by solution \rf{s1} with $p=+1$) or for $ s=-1,\; q=+1,\;
z=z_{0}, \forall \, d_{1}\neq 4$. Concerning the signs of
parameters, one obtains that $\mu <0$ and $\Lambda_D>0$.

 \addcontentsline{toc}{subsection}{3.4. Decoupling of excitations: $d_{1}=D_{0}$}
 \setcounter{subsection}{1}
 \medskip\medskip\medskip\medskip
 \textbf{3.4. Decoupling of excitations: $d_{1}=D_{0}$}
  \medskip\medskip\medskip\medskip

It can be easily seen from eq. \rf{1.8} that in the case
$d_{1}=D_{0}$ parameter $c=0$ that leads to condition $
\partial^{2}_{\varphi\phi}U_{eff}|_{extr} =0$ (see eq.
\rf{pvpUeff}. Thus, the Hessian \rf{Hessian} is diagonalized. It
means that the excitations of the fields $\varphi$ and $\phi$ near
the extremum position are decoupled from each other\footnote{In
the vicinity of a minimum of the effective potential, squared
masses of these excitations are $m^2_{\varphi} =J_{11}$ and
$m^2_{\phi}=J_{22}$.}.

Dropping the $h$ term in eq. \rf{pUeff} (because of $c=0$) and
taking into account eq. \rf{U0}, one obtains quadratic equation for
$X$
\be{5.1a} (D+2)X^2 + DqszX - (D-2)s =0\, , \ee
which for $D_0=d_1$ exactly coincides with eq. \rf{4.1a}. Thus, in
spite of the fact that the condition
$\Lambda_{eff} =0$ is not applied directly, one obtains in the case $d_1=D_0$ precisely the
same solutions \rf{s1}. However, parameters $R_1, U_0$ and $h$
satisfy now relations different from \rf{R1}. For example, for the
most physically interesting case $D_0=d_1=4$, eqs. \rf{Ueff0} and \rf{vpUeff}
 result in the following relations:
\be{RR} R_{1}=4\left[\frac{1}{3}U_{0}(X)+h\right] \; , \quad
\Lambda_{eff}(X)=\frac{1}{3}U_{0}(X)-h\, . \ee Nonzero components
of the Hessian read
\ba{Dj11} J_{11}&=&\frac{2}{3}\left[9h-U_{0}(X)\right]\:,\:
\\J_{22}&=&
    -\sqrt{|\mu|}\left[\frac{4(z+2qsX_{p})}{21(1+sX_{p}^{2})^{4/3}}+\frac{3qs(1+sX_{p}^{2})^{2/3}}{14X_{p}^{3}}\right].
    \nn
\ea
Now, let us to look for a positive minimum of the
effective potential. It means that $\Lambda_{eff} >0$, $J_{11}>0$
and $J_{22}>0$. From the positivity of $J_{11}$ and
$\Lambda_{eff}$ follows respectively\footnote{It is interesting
to note that (in the case $D_0=d_1=4$) relations \rf{5.1a},
$J_{11}$ in \rf{Dj11} and inequalities \rf{J11>0}, \rf{posit}
coincide with the analogous expressions in paper \cite{GMZ2}
with quadratic nonlinear model. This is not surprising because
they do not depend on the form of nonlinearity $f(\bar R)$ (and,
consequently, on the form of $U(\phi )$). However, the expressions
for $J_{22}$ are different because here the exact form of
$U(\phi )$ is used.}:
\be{J11>0} J_{11}>0 : \quad
16h>R_{1}>16U_{0}(X)/9>8\Lambda_{eff}\; \ee
and
\be{posit} \Lambda_{eff}>0 : \quad h>R_{1}/16>U_{0}(X)/9>h/3>0\;.
\ee
These inequalities show that for the considered model positive
minimum of the effective potential is possible only in the case of
positive curvature of the internal space $R_1>0$ and in the
presence of the form field $(h>0)$.

To realize which combination of parameters $s, p$ and $q$ ensures
the minimum of the effective potential, one should perform analysis
as in the previous case with $\Lambda_{eff}=0$. However, there is
no need to perform such analysis here because solutions of eq.
\rf{5.1a} coincides with \rf{s1} and all conditions for $X_{p}$ and
$U_0$ are the same as in the previous section. Thus, one obtains
concluding table of the form \rf{XF}. Additionally, it can be
easily seen that expressions of $J_{22}$ in \rf{Dj11} and
\rf{Lj22} exactly coincide with each other if one put $d_1=4$ in
the latter equation\footnote{It follows from the fact that in
\rf{Lj22} $D_0=4$. Although in this equation
the relation \rf{R1} between $h$ and $U_0$ is used, it enters here in the
combination which is proportional to $c$. Thus, this combination
does not contribute if additionally $d_1=4$.}. Hence,
Fig.\ref{Lj22-1} and Fig.\ref{Lj22-2} can be used (for the lines with
$d_1=4$) for analyzing the sign of $J_{22}$. With the help of these
pictures as well as keeping in the mind that $J_{22}(d_1=4, z=z_0)
= 0$, follows that the only combination which ensures the
positive minimum of $U_{eff}$ is: $s=-1\, ,\; p=+1\, ,\; q=+1\,$
and $z\in(z_{0}=1.936,+\infty)$. It is clear that potential
$U_0(X)$ in eqs. \rf{RR}-\rf{posit} is defined by solution of eq.
\rf{5.1a} (i.e. \rf{s1} for $d_1=4$) with this combination of the
parameters. Because $s=-1$ and $z>0$ , the parameters $\mu$ and
$\Lambda_D$ should have the following signs: $\mu<0$ and
$\Lambda_D>0$.

Additionally, it is easy to verify that second solution of eq.
\rf{s1} $X_-$ (with $p=-1$ and $s=-1, \, q=+1$) does not
correspond to the maximum of the effective potential $U_{eff}$.
Indeed, in this case $\partial_{\phi} U_{eff}(X_{-})=0$ but
$\partial_{\varphi} U_{eff}(X_{-})\ne 0$.

\begin{figure}[htbp]
\centerline{\includegraphics[width=4.5in,height=3in]{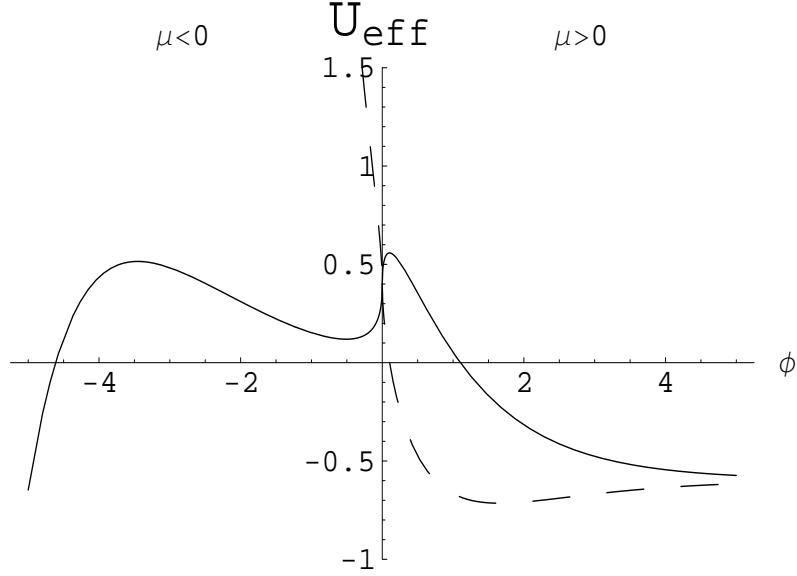}}
\caption{ Profile $\varphi=0$ of the effective potential
$U_{eff}(\varphi ,\phi)$ for parameters
$z=1.99\;(\Lambda_{D}=1.99/2\;,\;|\mu|=1),\;D_{0}=d_{1}=4$ and
$h=U_{0}/6$. The rest of parameters can be found from relations
\rf{RR}. Solid and dashed lines describe branches $q=+1$ and
$q=-1$ respectively. For the solid line, there is only one local
minimum of $U_{eff}$  which is defined by solution $X_{p}$ with
the following parameters: $s=-1,\; p=+1,\; q=+1$. Left ($s=-1,\;
p=-1,\; q=+1$) and right ($s=+1,\; p=+1,\; q=+1$) maxima are not
extrema of $U_{eff}$ because in these points $\partial_{\phi}
U_{eff}=0$ but $\partial_{\varphi} U_{eff}\ne 0$. Analogously, for
right dashed line the solution $X_{p}$ (with $s=+1,\; p=-1,\;
q=-1$) does not correspond to the extremum of $U_{eff}$.
\label{pota}}
\end{figure}

Fig. \ref{pota} demonstrates the typical profile $\varphi =0$ of
the effective potential $U_{eff}(\varphi ,\phi)$ in the case of
positive minimum of $U_{eff}$ considered in the present section.
This picture is in good concordance with the table \rf{XF}.
According to this table, positive extrema of $U_{eff}$ are
possible only for the branch $q=+1$ of the solution \rf{R} (solid
lines in Fig. \ref{pota}). It is obvious that for $z_0<z<2$ one can have 3
such extrema: one for positive $\mu>0 \to s=+1$ and two for
negative $\mu<0 \to s=-1$. These investigations show that in the
left half plane (i.e. for $\mu<0$) the right extremum ($p=+1$ in
eq. \rf{s1}) is the local minimum\footnote{\label{ssil}It can be easily seen
for the branch $q=+1,\; s=-1$ that $U(\phi \to -\infty ) \to
+\infty$ for $z\geq 2$ and $U(\phi \to -\infty ) \to -\infty$ for
$z< 2$. Thus, for $z\geq 2$ this minimum becomes global one.} and
the left maximum ($p=-1$) is not the extremum of $U_{eff}$ because
here $\partial_{\varphi} U_{eff}\ne 0$. Analogously, maximum in
the right half plane $\mu
>0$ (which corresponds to ($p=+1$)-solution \rf{s1}) is not the
extremum of $U_{eff}$. For completeness of picture,
lines corresponding to the branch $q=-1$ (dashed lines in
Fig. \ref{pota}) are also included. The minimum of the right dashed line (for $X_{p}$
with $s=+1,\; p=-1,\; q=-1$) does not describe the extremum of
$U_{eff}$ because again $\partial_{\varphi} U_{eff}\ne 0$.

 \addcontentsline{toc}{subsection}{3.5. Cosmic acceleration and domain walls}
 \setcounter{subsection}{1}
 \medskip\medskip\medskip\medskip
 \textbf{3.5. Cosmic acceleration and domain walls}
  \medskip\medskip\medskip\medskip

Let us consider again the model with $d_1=D_0=4$ in order to
define stages of the accelerating expansion of our Universe. It
was proven that for certain conditions (see \rf{RR}-\rf{posit})
the effective potential $U_{eff}$ has local (for $z_0<z< 2$) or
global (for $z\geq 2$) positive minimum. The position of this
minimum is $\left(\varphi =0,\, \phi = (1/A)\ln
(1+sX^2_{p})\right)$ where $s=-1\, ,\; p=+1\, ,\; q=+1\,$ and
$z\in(z_{0},+\infty)$. Obviously, positive minimum of the
effective potential plays the role of the positive cosmological
constant. Therefore, the Universe undergoes the accelerating
expansion in this position. Thus, one can "kill two birds with one
stone": to achieve the stable compactification of the internal
space and to get the accelerating expansion of our external space.

Let us associate this acceleration with the late-time accelerating
expansion of our Universe. As it follows from eqs. \rf{RR} and
\rf{posit}, positive minimum takes place if the parameters are
positive and the same order of magnitude: $\Lambda_{eff} \sim R_1
\sim U(X) \sim h >0$. On the other hand, in KK models the size of
extra dimensions at present time should be $b_{(0)1} \lesssim
10^{-17}\mbox{cm} \sim 1\mbox{TeV}^{-1}$. In this case $R_1
\gtrsim b_{(0)1}^{-2} \sim 10^{34}\mbox{cm}^{-2}$. Thus, for the
TeV scale of $b_{(0)1} \sim 1$TeV one gets that $\Lambda_{eff} \sim
R_1 \sim U(X) \sim h \sim 1\mbox{TeV}^2$. Moreover, in the case of
natural condition $\Lambda_D\sim \sqrt{|\mu|}$ it follows that the
masses of excitations $m_{\varphi} \sim m_{\phi}\sim 1$TeV. The
above estimates clearly demonstrate the typical problem of the
stable compactification in multidimensional cosmological models
because for the effective cosmological constant one obtains a value
which is in many orders of magnitude greater than observable at
the present time dark energy $\sim 10^{-57}\mbox{cm}^{-2}$. The
necessary small value of the effective cosmological constant can
be achieved only if the parameters $R_1 \, , U(X) \, , h$ are
extremely fine tuned with each other to provide the observed small
value from equation $\Lambda_{eff}(X)=U_{0}(X)/3 -h$. Two
possibilities to avoid this problem can be seen. Firstly, the inclusion of
different form-fields/fluxes may result in a big number of minima
(landscape) \cite{landscape1,landscape4} with sufficient
large probability to find oneself in a dark energy minimum.
Secondly, the restriction $R_1 \sim b_{(0)1}^{-2}
\sim 10^{34}\mbox{cm}^{-2}$ can be avoided if the internal space is Ricci-flat:
$R_1 = 0$. For example, the internal factor-space $M_1$ can be an
orbifold with branes in fixed points (see corresponding discussion
in \cite{Zhuk}).

The WMAP three year data as well as CMB data are consistent with
wide range of possible inflationary models (see e.g.
\cite{WMAP2006}). Therefore, it is of interest to get the stage of
early inflation in the given model. It is well known that it is rather
difficult to construct inflationary models from multidimensional
cosmological models and string theories. The main reason of it
consists in the form of the effective potential which is a
combination of exponential functions (see e.g. eq. \rf{1.7}).
Usually, degrees of these exponents are too large to result in
sufficiently small slow-roll parameters (see e.g. \cite{GZBR}).
Nevertheless, there is a possibility that in the vicinity of
maximum or saddle points the effective potential is flat enough to
produce the topological inflation \cite{topinfl,topinfla,SSTM,R4,R4a1,R4a2,R4a3,R4a4}. Let us
investigate this possibility for the given model.

As stated above, the value $\varphi=0$ corresponds to the internal
space value at the present time. Following this statement, the minimum of the effective potential at this value of
$\varphi$ is found. Obviously, the effective potential can also have
extrema at $\varphi \ne 0$. Let us investigate this possibility
for the model with $d_1=D_0=4$, i.e. for $c=0$. In this case, the
extremum condition of the effective potential reads
\ba{6.1}
\left.\partial_{\varphi}U_{eff}\right|_{\varphi_0,\phi_0}&=&
-\frac{1}{2}R_{1}(a+b)e^{(a+b)\varphi_0}+bU_{0}e^{b\varphi_0}\nn\\
&+& (ad_{1}+b)he^{(ad_{1}+b)\varphi_0}=0
 \ea
and
\begin{equation}\label{6.2}
\left.\partial_{\phi}U_{eff}\right|_{\varphi_0,\phi_0}=\left.
e^{b\varphi_0}\frac{\partial
U}{\partial\phi}\right|_{\phi_0}=0\;;\;\Longrightarrow\;\left.\frac{\partial
U}{\partial\phi}\right|_{\phi_0}=0\; .
\end{equation}
Here, $\varphi_0$ and $\phi_0$ define the extremum position. It
clearly follows from eq. \rf{6.2} that $\phi_0$ is defined by
equation $\partial U/\partial \phi =0$ which does not depend on
$\varphi$. Therefore, extrema of the effective potential may take
place only for $\phi_0$ which correspond to the solutions $X_{p}$
of eq. \rf{5.1a} and different possible extrema should lie on the
sections $X_{p}=const $. So, let us take $X_{+}$ (with $s=-1$ and
$q=+1$) which defines the minimum of $U_{eff}$ in the previous
section. Hence, $U_0$ in eq. \rf{6.1} is the same as for eq.
\rf{RR}.

\begin{figure}[htbp]
\centerline{\includegraphics[width=3.5in,height=3.5in]{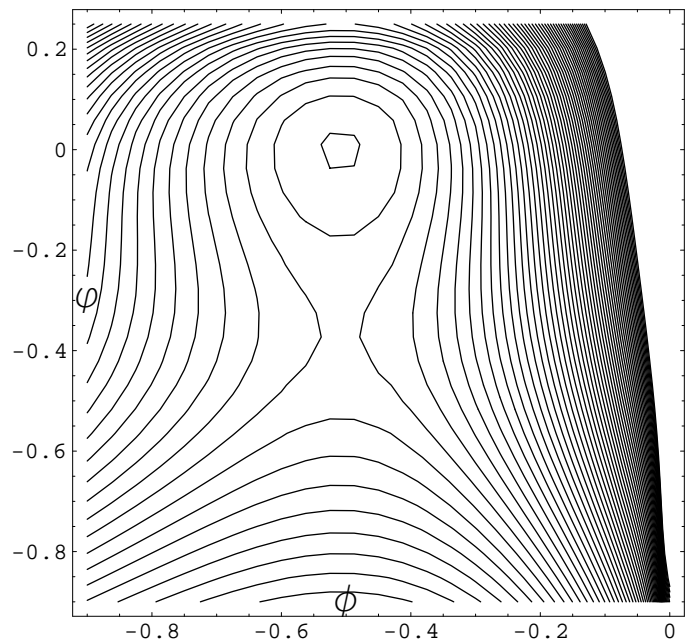}}
 \caption{Contour plot of the effective potential
$U_{eff}(\varphi ,\phi)$ for parameters
$z=1.99\;(\Lambda_{D}=1.99/2\;,\;|\mu|=1),\;D_{0}=d_{1}=4$ and
$h=U_{0}/6$. The rest of parameters follows from relations
\rf{RR}. The branch corresponding to $s=-1,\, q=+1$ is chosen.
This plot clearly shows the minimum and the saddle points of the
effective potential. \label{saddle}}
\end{figure}

Let us define now $\varphi_0$ from eq. \rf{6.1}. With the help of
inequalities \rf{J11>0} and \rf{posit} one can write $h=nU_{0}$
where $n\in (1/9,1/3)$. Taking also into account relations
\rf{RR}, eq. \rf{6.1} can be written as
\be{6.3} y^4 -\left(1+\frac{1}{3n}\right)y +\frac{1}{3n}=0\, , \ee
where the definition is done $y \equiv \exp{(a\varphi_0)} =
\exp{(\varphi_0/\sqrt{3})}$ and put $d_1=D_0=4$. Because $y=1$ is
the solution of eq. \rf{6.3}, the remaining three solutions
satisfy the following cubic equation:
\be{6.4} y^3+y^2+y-\frac{1}{3n}=0\, . \ee
It can be easily verified that the only real solution of this
equation is
\be{6.5} y_{0}=\frac{1}{3}\left(-1-2\nu+\frac{1}{\nu}\right)\; ,
\ee
where
\ba{6.6} \nu &=&\frac{2^{1/3}n}{\left(9n^2 +7n^3 +3\sqrt{9n^4 +14
n^6
+9n^6}\right)^{1/3}}\;;\;\nn\\n&\in&\left(\frac{1}{9},\frac{1}{3}\right)\;
. \ea
Thus $\varphi_{0}(y_0)$ and $\phi_0(X_{+})$ define new extremum of
$U_{eff}$. To clarify the type of this extremum one should check
signs of the second derivatives of the effective potential in this
point. First of all let us remind that in the case $c=0$
mixed second derivative disappears. Concerning second derivative
with respect to $\phi $, one obtains
\be{6.7} J_{22} \equiv \left.\frac{\partial^2 U_{eff}}{\partial
\phi^2}\right|_{\, \varphi_0,\phi_0} =
e^{b\varphi_0}\left.\frac{\partial^2 U}{\partial
\phi^2}\right|_{\, \phi_0} >0 \ee
because in previous section was obtained $\left.\partial^2 U/\partial
\phi^2\right|_{\, \phi_0(X_{+})} >0$. Second derivative with
respect to $\varphi $ reads
\ba{6.8} J_{11} &\equiv &\left.\frac{\partial^2 U_{eff}}{\partial
\varphi^2}\right|_{\, \varphi_0,\phi_0} = y_0^2
U_0\left[-6\left(\frac13 +n\right)y_0 + \frac43 \right.\nn\\&+&
\left.12ny_0^4\right] = 2y_0^2U_0\left[(3n+1)y_0 -
\frac43\right],\ea
where eq. \rf{6.3} is taken into account. Simple analysis shows
that $(3n+1)y_0 < 4/3$ for $n\in (1/9,1/3)$. Keeping in mind that
$U_0>0$ one obtains $J_{11}<0$. Therefore, the extremum is the
saddle surface\footnote{\label{ssil2} Similar analysis performed for the branch
with $s=+1,\, q=-1$ (right dashed line in the Fig. \ref{pota})
shows the existence of the global negative minimum with $\varphi
\ne 0$ along the section $X_{-}=const$.}. Figure \ref{saddle}
demonstrates contour plot of the effective potential in the
vicinity of the local minimum and the saddle point.

Therefore, very interesting  possibility become possible for the
production of an inflating domain wall in the vicinity of the
saddle point. The mechanism for the production of the domain walls
is the following \cite{topinfl,topinfla}. If the scalar field $\varphi$ is
randomly distributed, some part of the Universe will roll down to
$\varphi=0$, while in others parts it will run away to infinity.
Between any two such regions there will appear domain walls. In
Ref. \cite{SSTM}, it was shown for the case of a double-well
potential $V_{dw}(\varphi) = (\lambda/4)(\varphi^2-\xi^2)^2$ that
a domain wall will undergo inflation if the distance $\xi$ between
the minimum and the maximum of $V_{dw}$ exceeds a critical value
$\xi_{cr} =0.33M_{Pl}\, \to \, \kappa_0 \xi_{cr} = 1.65$. In this
case it means that the distance $|\varphi_0|$ between the local
minimum and the saddle point should be greater than $\xi_{cr}$:
$\, |\varphi_0| \geq 1.65$. Unfortunately, for this model
$\varphi_0(n) < 1.65$ if $n\in (1/9,1.3)$. For example, in the
most interesting case $n\to 1/3 $ (where $ \Lambda_{eff} \to 0$
(see eq. \rf{RR})) one obtains $|\varphi_0| \to 1.055$ which is less
than $\xi_{cr}$. Moreover, the domain wall is not thick enough in
comparison with the Hubble radius. The ration of the
characteristic thickness of the wall to the horizon scale is given
by $r_wH \approx \left|U_{eff}/3\partial_{\varphi\varphi}
U_{eff}\right|^{1/2}_{\varphi_0,\phi (X_+)} \to 0.454$ for $n\to
1/3$ which is less than the critical value 0.48 for a double-well
potential. Thus, here there is no a sufficiently large (for
inflation) quasi-homogeneous region of the energy density. And the given
potential is too steep. Obviously, the slaw roll parameter
$\epsilon\approx(1/2)\left(\partial_{\varphi}U_{eff}/U_{eff}\right)^{2}_{\varphi_{0},\phi_{0}}$
is equal to zero in the saddle point. However, another slow roll
parameter
$\eta\approx\left|\partial^{2}_{\varphi\varphi}U_{eff}/U_{eff}\right|
_{\varphi_{0},\phi_{0}}\to -1.617$ for $n\to 1/3$. Therefore, the
domain walls do not inflate in contrast to the case $R^4$ in Ref.
\cite{R4,R4a1,R4a2,R4a3,R4a4}.

\begin{figure}[htbp]
\centerline{\includegraphics[width=4.2in,height=3.2in]{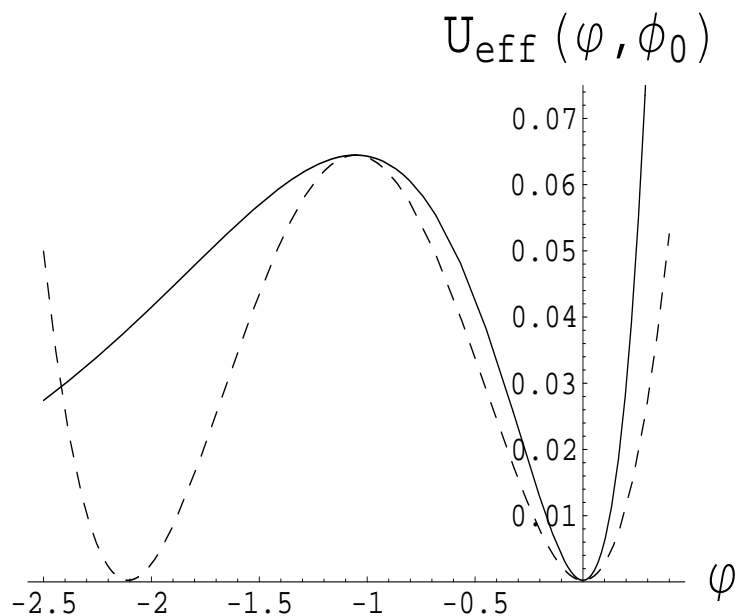}}
 \caption{Comparison of the potential $U_{eff}(\varphi ,\phi_0)$
 with a double well potential for parameters
$z=1.99\;(\Lambda_{D}=1.99/2\;,\;|\mu|=1),\;D_{0}=d_{1}=4$ and
$n=(1-0.001)/3$. \label{Maxhat}}
\end{figure}

In Fig. \ref{Maxhat} the comparison is presented between the potential
(solid line) and a double-well potential (dashed line) in the case
$n=(1-0.001)/3$. It is shown that the given potential is flatter than a
double-well potential around the saddle point. However,
calculations show that it is not enough for inflation.

 \medskip\medskip\medskip\medskip
\textbf{Summary}

It was shown that positive minimum of the effective potential plays
the double role in this model. Firstly, it provides the freezing
stabilization of the internal spaces which enables to avoid the
problem of the fundamental constant variation in multidimensional
models \cite{Zhuk(IJMP),BZ}. Secondly, it ensures the stage
of the cosmic acceleration. However, to get the present-day
accelerating expansion, the parameters of the model should be fine
tuned. Maybe, this problem can be resolved with the help of the idea
of landscape of vacua \cite{landscape1,landscape4}.

It was additionally found that given effective potential has the
saddle point. It results in domain walls which separates regions
with different vacua in the Universe. These domain walls do not
undergo inflation because the effective potential is not flat enough
around the saddle point.

It is worth of noting that minimum in Fig. \ref{pota} (left solid line) is
metastable. In other words, classically it is stable but there is a
possibility for quantum tunnelling both in $\phi$ and in $\varphi$
directions (see Fig. \ref{saddle}). One can avoid this problem in $\phi$
direction in the case of parameters $z\ge 2$ (see footnote \ref{ssil}).
However, tunnelling in $\varphi$ direction (through the saddle) is
still valid because $U_{eff}(\varphi ,\phi_0) \approx e^{b\varphi}
U(\phi_0) \rightarrow 0$ for $\varphi \rightarrow -\infty$ which is
less than any positive $\Lambda_{eff}$. It may result in the
materialization of bubbles of the new phase in the metastable one
(see e.g. \cite{Rubakov}). Thus, late-time acceleration is possible
only if characteristic lifetime of the metastable stage is greater
than the age of the Universe. Careful investigation of this problem
(including gravitational effects) is rather laborious task which
needs a separate consideration. As it was mentioned in footnote \ref{ssil2}, there
is also the global negative minimum for right dashed line in Fig. \ref{pota}
(it corresponds to the point $(\varphi =0.67,\phi =1.66)$ for
parameters taken in Fig. \ref{saddle}). This minimum is stable both in
classical and quantum limits. However, the acceleration is absent
because of its negativity.

Another very interesting feature of the model under consideration
consists in multi-valued form of the effective potential. As it can
be easily seen from eqs. \rf{1.7} and \rf{U}, for each choice of
parameter $\mu$ potential $U(\phi )$ (and consequently $U_{eff}$)
has two branches ($q=\pm 1$) which joint smoothly with each other at
$\phi=0$ (see Fig. \ref{pota}). It gives very interesting possibility
to investigate transitions from one branch to another one by analogy
with catastrophe theory or similar to the phase transitions in
statistical theory. However, as it was mentioned above, in this
particular model the point $\phi =0$ corresponds to the singularity
$\bar R, R \to \pm \infty$. Thus, the analog of the second order
smooth phase transition through the point $\phi =0$ is impossible in
this model. Nevertheless, there is still a possibility for the analog
of the first order transition via quantum jumps from one branch to
another one.

To complete, let us investigate some limiting cases. Firstly,
 the limit $h \to 0$ (for arbitrary $D_0$ and $d_1$)
where the form-fields are absent. From eqs. \rf{U0} - \rf{ppUeff} one
obtains the following system of equations:
\be{c1} R_1 = \frac{2b}{a+b} U_0(X)\, , \quad \left.
U_{eff}\right|_{extr} = \frac{a}{a+b}U_0(X) \ee
and
\ba{c2} J_{11} &=& -ab\, U_0(X)\, , \quad J_{21} = 0\, ,\;
\nn\\J_{22} &=& \left[B(A-B)-\frac{sAB}{2X^2}(1+sX^2)\right]U_0(X)\,
.\qquad \ea
Since for minimum should hold true the condition $J_{11}>0$, one
arrives at the conclusion: $R_1,\, U_0, \, U_{eff} <0$. Consequently,
the minimum of the effective potential as well as the effective
cosmological constant is negative and accelerating expansion is
absent in this limit. Therefore, the presence of the form-fields is
the necessary condition for the acceleration of the Universe in the
position of the freezing stabilization of the internal spaces.
Additionally, it can be easily seen that the extremum position
equation takes the same form as \rf{5.1a}. Simple analysis show that
minimum takes place for the branch: $s=+1$ (i.e. $\mu>0$), $p=-1, \,
q=-1$ and $z\in (-\infty ,+\infty )$. If additionally is imposed that $z
\to 0$ (i.e. $\Lambda_D \to 0 $ and $\mu$ is fixed) then the results of Ref. \cite{GZBR} are reproduced.

 \newpage
 \addcontentsline{toc}{section}{CHAPTER 4. INFLATION IN MULTIDIMENSIONAL \\ COSMOLOGICAL MODELS}
 \section*{\center{CHAPTER 4\\ $\;$ \\
INFLATION IN MULTIDIMENSIONAL COSMOLOGICAL MODELS}}
 \setcounter{section}{4}
 \setcounter{equation}{0}

\medskip\medskip
In this chapter, a multidimensional cosmological models are considered with linear, nonlinear
quadratic $\bar{R}^2$ and quartic $\bar{R}^4$ actions with a monopole form field as a matter source; and pure gravitational model
with  $\bar{R}^2+\bar{R}^4$ nonlinearities. The inflation is investigated
in these models. It is shown that $\bar{R}^2$ and $\bar{R}^4$ models
can have up to 10 and 22 e-foldings, respectively. These values are
not sufficient to solve the homogeneity and isotropy problem but big
enough to explain the recent CMB data. Additionally, $\bar{R}^4$ model can
provide conditions for eternal topological inflation. However, the
main drawback of the given inflationary models consists in a value
of spectral index $n_s$ which is less than observable now
$n_s\approx 1$. For example, in the case of $\bar{R}^4$ model one finds $n_s
\approx 0.61$.

For the model   $\bar{R}^2+\bar{R}^4$ , the effective scalar degree of freedom $\phi$ (scalaron) has a multi-valued potential $U(\phi)$ consisting of a number of branches. These branches are fitted with each other in the branching and monotonic points.
In the case of four-dimensional space-time, it is shown that the monotonic points are penetrable for scalaron while in the vicinity of the branching points scalaron has the bouncing behavior and cannot cross these points. Moreover, there are branching points
where scalaron bounces  an infinite number of times with decreasing amplitude and the Universe asymptotically approaches the de Sitter stage. Such accelerating behavior is called bouncing inflation. For this accelerating expansion there is no need for original potential $U(\phi)$ to have a minimum or to check the slow-roll conditions. A necessary condition for such inflation is the existence of the branching points. This is a new type of inflation. It is shown that bouncing inflation takes place both in the Einstein and Brans-Dicke frames.


 \addcontentsline{toc}{subsection}{4.1. Linear model}
 \setcounter{subsection}{1}
 \medskip\medskip\medskip\medskip
 \textbf{4.1. Linear model}
  \medskip\medskip\medskip\medskip

To start with, let us define the topology of the given models and consider
a factorizable $D$-dimensional metric

\be{2.1} g^{(D)}=  g^{(0)}(x) + L_{Pl}^2 e^{2\beta^1 (x)}g^{(1)}\, ,
\ee
which is defined on a warped product manifold $M = M_0\times M_1$.
$M_0$ describes external $D_0$-dimensional space-time (usually $D_0=4$) and $M_1$ corresponds to
$d_1$-dimensional internal space which is a flat
orbifold\footnote{For example, $S^1/Z_2$ and $T^2/Z_2$ which
represent circle and square folded onto themselves due to $Z_2$
symmetry.} with branes in fixed points. Scale factor of the internal
space depends on coordinates $x$ of the external space-time: $a_1(x)
= L_{Pl}e^{\beta^1(x)}$, where $L_{Pl}$ is the Planck length.

First, let us consider the linear model $f(R)=R$ with $D$-dimensional
action of the form
\be{2.2} S =\frac{1}{2\kappa_D^2}\int_M
d^Dx\sqrt{|g^{(D)}|}\left\{R[g^{(D)}]-2\Lambda_D\right\} +S_m+S_b \,
. \ee
$\Lambda_D$ is a bare cosmological constant\footnote{Such
cosmological constant can originate from $D$-dimensional form field
which is proportional to the $D$-dimensional world-volume:
$F^{MN\ldots Q}=(C/\sqrt{|g^{(D)}|})\epsilon^{MN\ldots Q}$. In this
case the equations of motion gives $C=const$ and $F^2$ term in
action is reduced to $(1/D!)F_{MN\ldots Q}F^{MN\ldots Q}= -C^2$.}.
In the spirit of Universal Extra Dimension models \cite{UED,UEDa1,UEDa2,UEDa3,UEDa5}, the
Standard Model fields are not localized on the branes but can move
in the bulk. The compactification of the extra dimensions on
orbifolds has a number of very interesting and useful properties,
e.g. breaking (super)symmetry and obtaining chiral fermions in four
dimensions (see e.g. paper by H.-C. Cheng at al in \cite{UED,UEDa1,UEDa2,UEDa3,UEDa5}). The
latter property gives a possibility to avoid famous no-go theorem of
KK models (see e.g. \cite{no-go,no-goa}). Additional arguments in favor of
UED models are listed in \cite{Kundu}.

Following a generalized Freund-Rubin ansatz \cite{FR} to achieve a
spontaneous compactification $M \rightarrow M = M_0\times M_1$, one
endows the extra dimensions with real-valued solitonic form field
$F^{(1)}$ with an action:
\be{2.3} S_m=-\frac12 \int_M
d^Dx\sqrt{|g^{(D)}|}\frac{1}{d_1!}\left(F^{(1)}\right)^2\, , \ee
This form field is nested in $d_1$-dimensional factor space $M_1$,
i.e. $F^{(1)}$ is proportional to the world-volume of the internal
space. In this case $(1/d_1!)\left(F^{(1)}\right)^2=\bar
f_1^2/a_1^{2d_1}$, where $\bar f_1$ is a constant of integration
\cite{GMZ2}.

Branes in fixed points contribute in action functional \rf{2.2} in
the form \cite{Zhuk}:
\be{2.4} S_b=\sum_{\phantom{x}^{fixed}_{points}}\left. \int_{M_0}
d^4x \sqrt{ |g^{(0)}(x)}|\; L_b
\right|_{\phantom{x}^{fixed}_{point}}\, , \ee
where $g^{(0)}(x)$ is induced metric (which for the given geometry
\rf{2.1} coincides with the metric of the external space-time  in
the Brans-Dicke frame) and $L_b$ is the matter Lagrangian on the
brane. In what follows, that the case where branes are only
characterized by their tensions $L_{b(k)} = -\tau_{(k)}\, ,\,
k=1,2,\ldots ,m$ is considered, where $m$ is the number of branes.

Let $\beta_0^1$ be the internal space scale factor at the present
time and $\bar\beta^{1} = \beta^1 - \beta^1_0$ describes
fluctuations around this value. Then after dimensional reduction of
the action \rf{2.1} and conformal transformation to the Einstein
frame $ g^{(0)}_{\mu \nu } = \left( e^{d_1\bar
\beta^1}\right)^{-2/(D_0-2)} \tilde g^{(0)}_{\mu \nu}$, one arrives at
effective $D_0$-dimensional action of the form
\ba{2.5} S_{eff} &=&
\frac{1}{2\kappa^2_0}\int_{M_0}d^{D_0}x\sqrt{|\tilde
g^{(0)}|}\left\{R[\tilde g^{(0)}]\right.\nn\\&-&\left. \tilde
g^{(0)\mu\nu}\partial_{\mu}\varphi\partial_{\nu}\varphi-2U_{eff}(\varphi
)\right\}\, , \ea
where scalar field $\varphi$ is defined by the fluctuations of the
internal space scale factor:
\be{2.6} \varphi \equiv  \sqrt{\frac{d_1(D-2)}{D_0-2}}\; \bar
\beta^1\, \ee
and
$G := \kappa^2_0/8\pi := \kappa^2_D/(8\pi V_{d_1})$ ($V_{d_1}$ is
the internal space volume at the present time) denotes the
$D_0$-dimensional gravitational constant. The effective potential
$U_{eff}(\varphi )$ reads (hereafter $D_0=4$):
\ba{2.7} U_{eff}(\varphi ) &=& e^{-\, \sqrt{\frac{2d_1}{d_1+2}}\;
\varphi}\left[\Lambda_D + f_1^2e^{-2\, \sqrt{\frac{2d_1}{d_1+2}}\;
\varphi}\right.\nn\\&-&\left.\lambda e^{-\,
\sqrt{\frac{2d_1}{d_1+2}}\; \varphi} \right]\, , \ea
where $f_1^2 \equiv \kappa^2_D\bar f_1^2/a^{2d_1}_{(0)1}$ and
$\lambda \equiv -\kappa^2_0\sum^m_{k=1}\tau_{(k)} $.

Now, this potential should be investigated from the point of the
external space inflation and the internal space stabilization.
First, it is clear that internal
space is stabilized if $U_{eff}(\varphi )$ has a minimum with
respect to $\varphi$. The position of minimum should correspond to
the present day value $\varphi =0$. Additionally, it needs to be demanded that
the value of the effective potential in the minimum position is
equal to the present day dark energy value $U_{eff}(\varphi =0) \sim
\Lambda_{DE} \sim 10^{-57}\mbox{cm}^{-2}$. However, it results in
very flat minimum of the effective potential which in fact
destabilizes the internal space \cite{Zhuk}. To avoid this problem,
the case of zero minimum $U_{eff}(\varphi =0)=0$ shall be considered.

\begin{figure*}[htbp]
\centerline{\includegraphics[width=5in,height=3in]{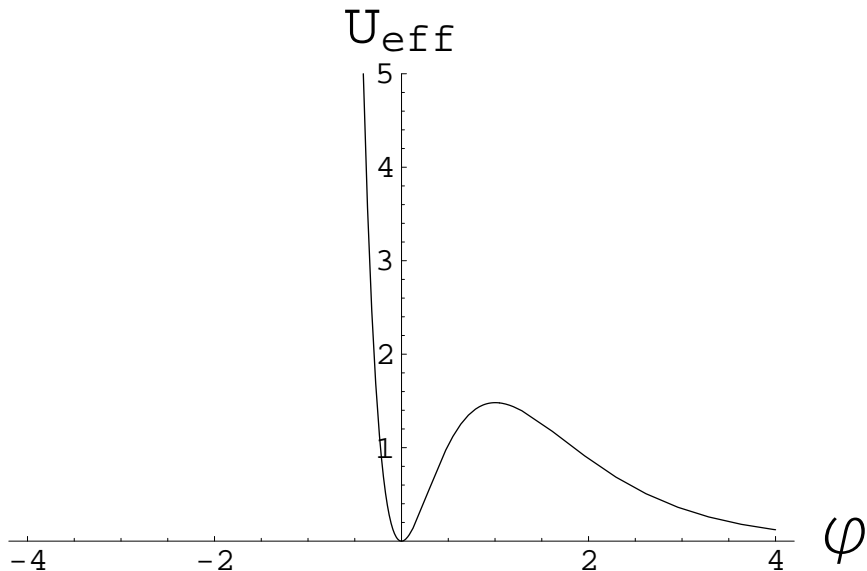}}
\caption {The form of the effective potential \rf{2.7} in the case
$d_1=3$ and  $\Lambda_D = f_1^2 = \lambda /2 =10$.\label{potf}}
\end{figure*}

The extremum condition $\left.dU_{eff}/d\varphi \right|_{\varphi =0}
= 0$ and zero minimum condition $U_{eff}(\varphi =0) =0$ result in a
system of equations for parameters $\Lambda_D, f_1^2$ and $\lambda$
which has the following solution:
\be{2.8} \Lambda_D = f_1^2 = \lambda /2\, . \ee
For the mass of scalar field excitations (gravexcitons/radions) one
obtains: $m^2 = \left.d^2U_{eff}/d \varphi^2\right|_{\varphi =0} =
(4d_1/(d_1+2))\Lambda_D$. In Fig. \ref{potf} the effective
potential \rf{2.7} is presented in the case $d_1=3$ and $\Lambda_D = 10$. It is
worth of noting that usually scalar fields in the present paper are
dimensionless\footnote{To restore dimension of scalar fields one
should multiply their dimensionless values by $M_{Pl}/\sqrt{8\pi}$.
} and $U_{eff}, \Lambda_D, f_1^2, \lambda$ are measured in
$M_{Pl}^2$ units.

Let us turn now to the problem of the external space inflation. As
far as the external space corresponds to our Universe, the
metric $\tilde g^{(0)}$ is taken in the spatially flat
Friedmann-Robertson-Walker form with scale factor $a(t)$. Scalar
field $\varphi $ depends also only on the synchronous/cosmic time
$t$ (in the Einstein frame).

It can be easily seen that for $\varphi
>>0$ (more precisely, for $\varphi
> \varphi_{max} = \sqrt{(d_1+2)/2d_1}\ln 3$) the potential
\rf{2.7} behaves as
\be{2.9} U_{eff}(\varphi ) \approx \Lambda_D e^{-\sqrt{q}\, \varphi}
\, , \ee
with
\be{2.10} q:=\frac{2d_1}{d_1+2}\, . \ee
It is well known (see e.g. \cite{RatraPeebles,EM,Ratra1,Stornaiolo})
that for such exponential potential scale factor has the following
asymptotic form:
\be{2.11} a(t) \sim t^{2/q}\, . \ee
Thus, the Universe undergoes the power-law inflation if $q<2$.
Precisely this condition holds for eq. \rf{2.10} if $d_1 \ge 1$.

It can be easily verified that $\varphi
> \varphi_{max}$ is the only region of the
effective potential where inflation takes place. Indeed, in the
region $\varphi <0$ the leading exponents are too large, i.e. the
potential is too steep. The local maximum of the effective potential
$\left.U_{eff}\right|_{max} = (4/27)\Lambda_D$ at $\varphi_{max} =
\sqrt{(d_1+2)/2d_1}\ln 3$ is also too steep for inflation because
the slow-roll parameter $\eta_{max}= \left.
\frac{1}{U_{eff}}\frac{d^2U_{eff}}{d\varphi^2}\right|_{max} =
-\frac{3d_1}{d_1+2}\; \; \Rightarrow \; 1\leq |\eta_{max}|<3$ and
does not satisfy the inflation condition $|\eta |<1$. Topological
inflation is also absent here because the distance between global
minimum and local maximum  $\varphi_{max} = \sqrt{(d_1+2)/2d_1}\ln 3
\leq 1.35$ is less than critical value $\varphi_{cr} \geq 1.65$ (see
\cite{Ellis,SSTM,ZhukSaidov}). It is worth of noting that
$\eta_{max}$ and $\varphi_{max}$ depend only on the number of
dimensions $d_1$ of the internal space and do not depend on the
hight of the local maximum (which is proportional to $\Lambda_D$).

Therefore, there are two distinctive regions in this model. In the
first region, at the left of the maximum in the vicinity of the
minimum, scalar field undergoes the damped oscillations. These
oscillations have the form of massive scalar fields in our Universe
(in \cite{GZ1} these excitations were called gravitational excitons
and later (see e.g. \cite{ADMR}) these geometrical moduli
oscillations were also named radions). Their life-time with respect
to the decay $\varphi \to 2\gamma $ into radiation is
\cite{GSZ,Cho,Cho2} $\tau \sim (M_{Pl}/m)^3T_{Pl}$. For example, in given case one
obtains $\tau \sim 10\, \mbox{s} , 10^{-2}\, \mbox{s}$ for $m \sim
10\, \mbox{TeV} , 10^2\, \mbox{TeV}$ correspondingly, where $m^2 = (4d_1/(d_1+2))\Lambda_D$.
Therefore, this is the
graceful exit region. Here, the internal space scale factor, after
decay its oscillations into radiation, is stabilized at the present
day value and the effective potential vanishes due to zero minimum.
In second region, at the right of the maximum of the potential, our
Universe undergoes the power-low inflation. However, it is
impossible to transit from the region of inflation to the graceful
exit region because given inflationary solution satisfies the
following condition $\dot \varphi >0$. There is also serious
additional problem connected with obtained inflationary solution.
The point is that for the exponential potential of the form
\rf{2.9}, the spectral index reads
\cite{RatraPeebles,Ratra1}\footnote{With respect to conformal time,
solution \rf{2.11} reads $a(\eta) \sim \eta^{1+\beta}$ where $\beta
=-(4-q)/(2-q)$. It was shown in \cite{MSch} that for such
inflationary solution (with $q<2$) the spectral index of density
perturbation is given by $n_s=2\beta +5$ resulting again in
\rf{2.12}.}:
\be{2.12} n_s=\frac{2-3q}{2-q}\, . \ee
In the case \rf{2.10}, it results in $n_s = 1-d_1$. Obviously, for
$d_1\ge 1$ this value is very far from observable data $n_s\approx
1$. Therefore, it is necessary to generalize given linear model.

 \addcontentsline{toc}{subsection}{4.2. Quadratic model}
 \setcounter{subsection}{1}
 \medskip\medskip\medskip\medskip
 \textbf{4.2. Quadratic model}
  \medskip\medskip\medskip\medskip

One of possible ways for generalizing of the
effective potential making it more complicated and having more reach
structure is an introduction of an additional minimal scalar field $\phi$.
It is possible to do "by hand"$\:$,  inserting  minimal scalar field $\phi$ with
a potential $U(\phi )$ in linear action \rf{2.2}\footnote{If such
scalar field is the only matter field in these models, it is known
(see e.g. \cite{GZ2,GMZ1}) that the effective potential can has only
negative minimum. i.e. the models are asymptotical AdS. To uplift
this minimum to nonnegative values, it is necessary to add
form-fields \cite{GMZ2}.}. Then, effective potential takes the form
\ba{3.1} U_{eff}(\varphi ,\phi ) &=& e^{-\,
\sqrt{\frac{2d_1}{d_1+2}}\; \varphi}\left[U(\phi ) + f_1^2e^{-2\,
\sqrt{\frac{2d_1}{d_1+2}}\; \varphi}\right. \nn\\ &-&\left.\lambda
e^{-\, \sqrt{\frac{2d_1}{d_1+2}}\; \varphi} \right]\, , \quad \ea
where $\Lambda_D =0$ in \rf{2.2}.

However, it is well known that scalar field $\phi$ can naturally
originate from the nonlinearity of higher-dimensional models where
the Hilbert-Einstein linear lagrangian $\bar{R}$ is replaced by nonlinear
one $f(\bar{R})$. These nonlinear theories are equivalent to the linear
ones with a minimal scalar field (which represents additional degree
of freedom of the original nonlinear theory). It is not difficult to
verify (see e.g. \cite{GMZ1,GMZ2}) that nonlinear model
\ba{3.2} S &=&\frac{1}{2\kappa^2_D}\int_M d^Dx\sqrt{|\ov
g^{(D)}|}f(\ov R) \nn\\&-&\frac12 \int_M
d^Dx\sqrt{|g^{(D)}|}\frac{1}{d_1!}\left(F^{(1)}\right)^2 \nn\\
&-&\sum_{k=1}^m \int_{M_0} d^4x \sqrt{ |g^{(0)}(x)}|\; \tau_{(k)}
\ea
is equivalent to a linear one with conformally related metric
\be{3.3} g_{ab}^{(D)} = e^{2A\phi /(D-2)}\ov g_{ab}^{(D)} \ee
plus minimal scalar field $\phi =\ln [df/d\ov R\, ]/A$ with a
potential
\be{3.4} U(\phi )=\frac12 e^{-B\phi }\left[\ov R(\phi )e^{A\phi
}-f(\ov R(\phi) )\right]\, , \ee
where
\ba{qwer2} A&=&\sqrt{\frac{(D-2)}{(D-1)}}=\sqrt{\frac{(d_1+2)}{d_1+3}}\;; \nn\\
B&=&\frac{D}{\sqrt{(D-2)(D-1)}}=A\frac{(d_1+4)}{(d_1+2)}\;.
\ea
After dimensional
reduction of this linear model, one  obtains an effective
$D_0$-dimensional action of the form
\ba{3.5} S_{eff} &=&
\frac{1}{2\kappa^2_0}\int_{M_0}d^{D_0}x\sqrt{|\tilde
g^{(0)}|}\left[R[\tilde g^{(0)}]- \tilde
g^{(0)\mu\nu}\partial_{\mu}\varphi\partial_{\nu}\varphi\right.
\nn\\&-&\left.\tilde
g^{(0)\mu\nu}\partial_{\mu}\phi\partial_{\nu}\phi -2U_{eff}(\varphi
,\phi )\right]\, , \ea
with effective potential exactly of the form \rf{3.1}. It is worth
to note that it is supposed that matter fields are coupled to the metric
$g^{(D)}$ of the linear theory (see also analogous approach in
\cite{DSS}). Because in all considered below models both fields
$\varphi$ and $\phi$ are stabilized in the minimum of the effective
potential, such convention results in a simple
redefinition/rescaling of the matter fields and effective
four-dimensional fundamental constants. After such stabilization,
the Einstein and Brans-Dicke frames are equivalent each other
(metrics $g^{(0)}$ and $\tilde g^{(0)}$ coincide with each other),
and linear $g^{(D)}$ and nonlinear $\ov g^{(D)}$ metrics in \rf{3.3}
are related via constant prefactor (models became asymptotically
linear)\footnote{However, small quantum fluctuations around the
minimum of the effective potential distinguish these metrics.}.

Let us consider first the quadratic theory
\be{3.6} f(\bar R) = \bar R +\xi \bar R^2 -2\Lambda_D\, . \ee
For this model the scalar field potential \rf{3.4} reads:
\be{3.7} U(\phi )=\frac12
e^{-B\phi}\left[\frac{1}{4\xi}\left(e^{A\phi}
-1\right)^2+2\Lambda_D\right]\, . \ee

It was proven \cite{GZ2} that the internal space is stabilized if
the effective potential \rf{3.1} has a minimum with respect to both
fields $\varphi$ and $\phi$. It can be easily seen from the form of
$U_{eff}(\varphi ,\phi )$ that minimum $\phi_0$ of the potential
$U(\phi )$ coincides with the minimum of $U_{eff}(\varphi ,\phi ) \;
: \left.dU/d\phi \right|_{\phi_0} =0 \to
\left.\partial_{\phi}U_{eff}\right|_{\phi_0}=0$. For minimum
$U(\phi_0 )$ one obtains \cite{GMZ1}:
\be{3.8} U(\phi_0) = \frac{1}{8\xi
}x_0^{\frac{-D}{D-2}}\left[(x_0-1)^2 + 8\xi\Lambda_D\right]\, , \ee
where the denotation $x_0 := \exp
(A\phi_0)$ is applied, taking into account that  $ \exp
(A\phi_0)=\left(A-B+\sqrt{A^2+(2A-B)8sB\,\xi\Lambda_D}\,
\right)$ $/\left(2A-B\right)$.
It is the global minimum and the only extremum of $U(\phi )$.
Nonnegative minimum of the effective potential $U_{eff}$ takes place
for positive $\xi ,\Lambda_D >0$. If $\xi ,\Lambda_D>0$, the
potential $U(\phi)$ has asymptotic behavior $U(\phi )\to +\infty $
for $\phi \to \pm \infty$.

The relations \rf{2.8}, where one should make the substitution
$\Lambda_D \to U(\phi_0)$, are the necessary and sufficient
conditions of the zero minimum of the effective potential
$U_{eff}(\varphi ,\phi)$ at the point $(\varphi =0, \phi =\phi_0)$.
Thus, if parameters of the quadratic models satisfy the conditions
$U(\phi_0)=f_1^2=\lambda/2$, one arrives at zero global minimum:
$U_{eff}(0, \phi_0)=0$.

\begin{figure*}[htbp]

\centerline{\includegraphics[width=5.4in,height=3.5in]{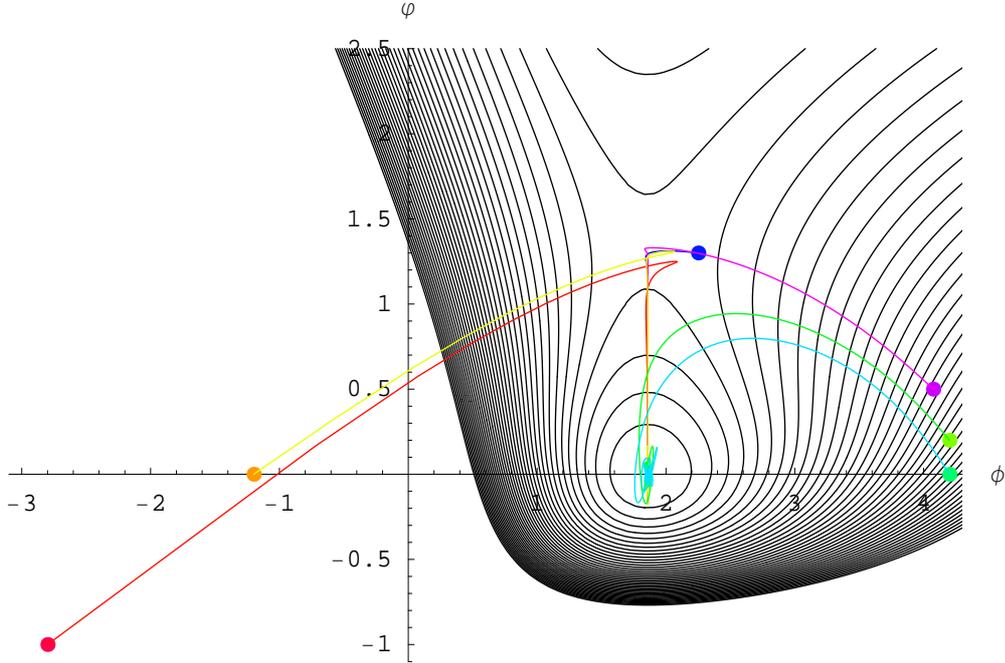}}

\caption {Contour plot of the effective potential $U_{eff}(\varphi
,\phi )$ \rf{3.1} with potential $U(\phi )$ of the form \rf{3.7} for
parameters $d_1=1\, , \xi\Lambda_D  =1 $ and relations $U(\phi_0) =
f_1^2 = \lambda /2$ . This plot clearly shows the global minimum and
the saddle. The colored lines describe trajectories for scalar
fields starting at different initial conditions. \label{effpotr2}}
\end{figure*}

It is clear that profile $\phi =\phi_0$ of the effective potential
$U_{eff}$ has a local maximum in the region $\varphi >0$ because
$U_{eff}(\varphi ,\phi =\phi_0 ) \to 0$ if $\varphi \to +\infty$.
Such profile has the form shown in Fig. \ref{potf}. Thus, the effective
potential $U_{eff}$ has a saddle point $(\varphi =\varphi_{max},\phi
=\phi_0 )$ where $\varphi_{max}=\sqrt{(d_1+2)/2d_1}\ln 3$. At this
point $\left.U_{eff}\right|_{max} = (4/27)U(\phi_0)$. The Figure
\ref{effpotr2} demonstrates the typical contour plot of the
effective potential \rf{3.1} with the potential $U(\phi )$ of the
form \rf{3.7} in the vicinity of the global minimum and the saddle
point.

Let us discuss now a possibility of the external space inflation in
this model. It can be easily realized that for all models of the
form \rf{3.1} in the case of local zero minimum at $(\varphi
=0,\phi_0)$, the effective potential will also have a saddle point
at $(\varphi = \varphi_{max}, \, \phi_0)$ with
$\varphi_{max}=\sqrt{(d_1+2)/2d_1}\ln 3 < \varphi_{cr}=1.65$ and the
slow-roll parameter $|\eta_{\varphi}|$ in this point cannot be less
than 1: $|\eta_{\varphi}| =3d_1/(d_1+2)\geq 1$. Therefore, such
saddles are too steep (in the section $\phi =\phi_0$) for the
slow-roll and topological inflations. However, as one shall see below, a short period of de-Sitter-like
inflation is possible if one start not precisely at the saddle point
but first move in the vicinity of the saddle along the line $\varphi
\approx \varphi_{max}$ with subsequent turn into zero minimum along
the line $\phi \approx \phi_0$. Similar situation happens for
trajectories from different regions of the effective potential which
can reach this saddle and spend here a some time (moving along the
line $\varphi \approx \varphi_{max}$).

Let us consider now regions where the following conditions take
place:
\be{3.9} U(\phi ) \gg f_1^2e^{-2\, \sqrt{\frac{2d_1}{d_1+2}}\;
\varphi}\; ,\;\: \lambda e^{-\, \sqrt{\frac{2d_1}{d_1+2}}\;
\varphi}. \ee
For the potential \rf{3.7} these regions exist both for negative and
positive $\phi$. In the case of positive $\phi$ with $\exp (A\phi)
\gg \, \mbox{max}\; \left\{1\, , (8\xi\Lambda_D)^{1/2}\right\}$ one
obtains
\be{3.10} U_{eff} \approx \frac{1}{8\xi}e^{-\sqrt{q}\;
\varphi}e^{\sqrt{q_1}\; \phi}\, , \ee
where $q$ is defined by Eq. \rf{2.10},  $q_1:=(2A-B)^2 =
d_1^2/[(d_1+2)(d_1+3)]$ and $q>q_1$. For potential \rf{3.10} the
slow-roll parameters are\footnote{\label{slowroll}In the case of $n$
scalar fields $\varphi_i\, (i=1,\ldots ,n)$ with a flat
($\sigma-$model) target space, the slow-roll parameters for the
spatially flat Friedmann Universe read (see e.g. \cite{GZ2,GMZ1}):
$\epsilon \equiv \frac{2}{H^2}\sum_{i=1}^n\left(\partial_i
H\right)^2 \approx \frac{1}{2}|\partial U|^2 / U^2 \, ; \quad \eta_i
\equiv -\ddot \varphi_i / (H\dot\varphi_i)= 2\partial^2_{ii} H /H
\approx -\epsilon + \sum_{j=1}^n
\partial^2_{ij}U \partial_j U / (U
\partial_i U)$, where $\partial_i := \partial /\partial \varphi_i$ and $|\partial
U|^2=\sum_{i=1}^n\left(\partial_i U\right)^2$. In some papers (see
e.g. \cite{multi-inflation,multi-inflationa1,multi-inflationa2}) it was introduced a "cumulative"
parameter $\eta \equiv  - \sum_{i=1}^n\ddot \varphi_i\dot\varphi_i
/( H|\dot \varphi|^2) \approx
-\epsilon+\sum_{i,j=1}^n(\partial_{ij}^2 U)(\partial_i U)(\partial_j
U)/(U |\partial U |^2)$ , where $|\dot \varphi|^2=\sum_{i=1}^n
\dot\varphi_i^2$. One can easily find that for the potential
\rf{3.10} parameter $\eta$ coincides exactly with parameters
$\eta_1$ and $\eta_2$.}:
\be{3.11} \epsilon \approx \eta_1\approx \eta_2 \approx \frac{q}{2}
+\frac{q_1}{2}\, \ee
and satisfy the slow-roll conditions $\epsilon, \eta_1, \eta_2 <1$.
As far as it is known, there are no analytic solutions for such
two-scalar-field potential. Anyway, from the form of the potential
\rf{3.10} and condition $q>q_1$ one can get an estimation $a \approx
t^{s}$ with $s\gtrsim 2/q$ (e.g. $2/q = 3,2,5/3$ for $d_1=1,2,3$,
respectively). Thus, in these regions it is possible to achieve a period of
power-law inflation. In spite of a rude character of these
estimates, one shall see below that external space scale factors
undergo power-law inflation for trajectories passing through these
regions.

Now, let us investigate dynamical behavior of scalar fields and the
external space scale factor in more detail. There are no analytic
solutions for considered model. So, let us use numerical calculations with
applying of a Mathematica package proposed in \cite{KP},
adjusting it to given models and notations (see appendix 3).

\begin{figure*}[htbp]

\centerline{\includegraphics[width=3.5in,height=2.5in]{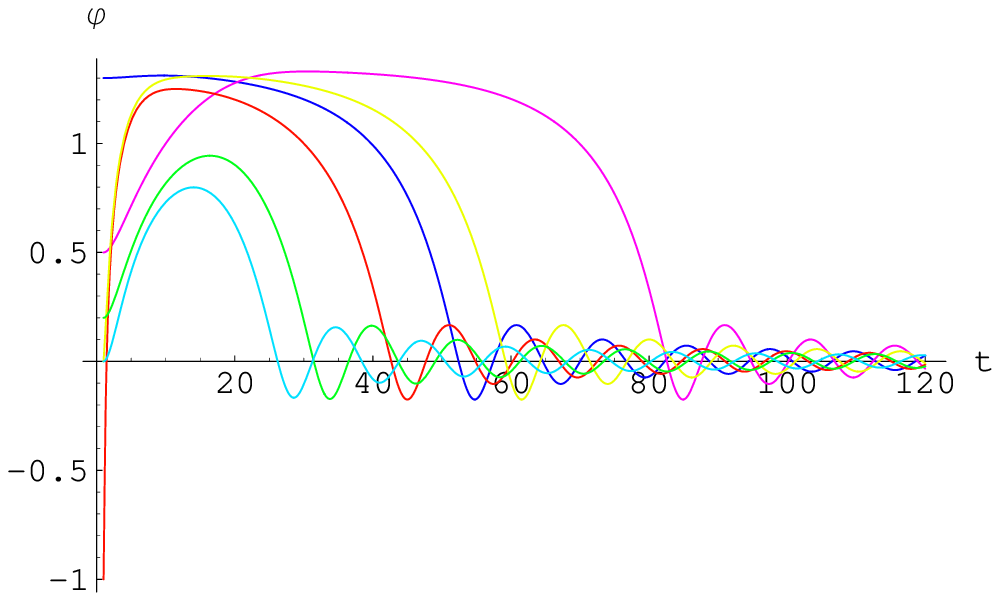}
\includegraphics[width=3.0in,height=2.5in]{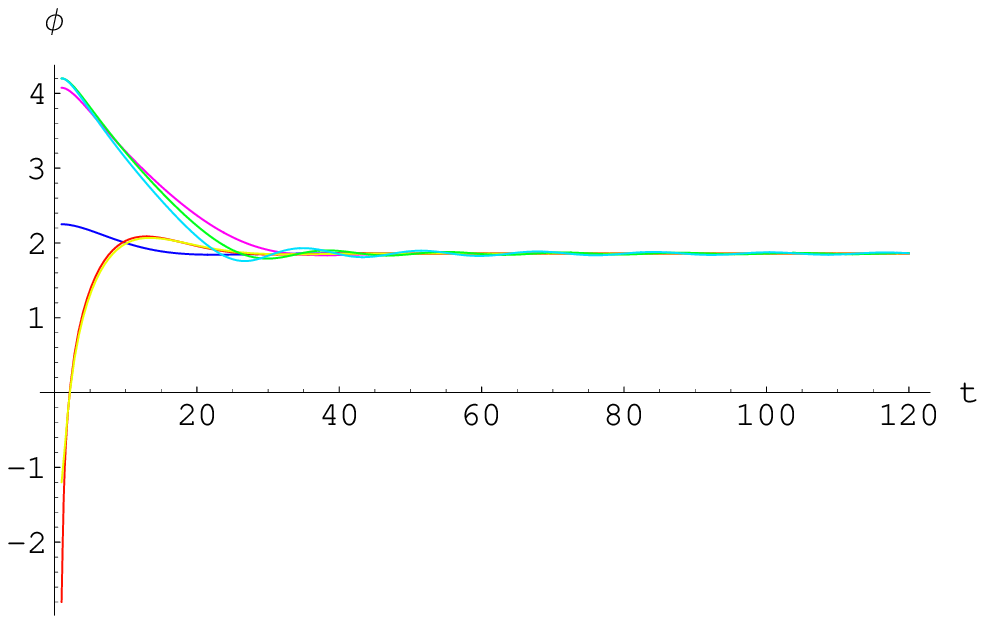}}
\caption {Dynamical behavior of scalar fields $\varphi$ (left panel)
and $\phi$ (right panel) with corresponding initial values denoted
by the colored dots in Fig. \ref{effpotr2}. \label{fields2}}
\end{figure*}

The colored lines on the contour plot of the effective potential in
Fig. \ref{effpotr2} describe trajectories for scalar fields
$\varphi$ and $\phi$ with different initial values (the colored
dots). The time evolution of these scalar fields\footnote{Let us remind
that $\varphi$ describes fluctuations of the internal space scale
factor and $\phi$ reflects the additional degree of freedom of the
original nonlinear theory.} is drawn in Fig. \ref{fields2}. Here,
the time $t$ is measured in the Planck times and classical evolution
starts at $t=1$. For given initial conditions, scalar fields
approach the global minimum of the effective potential along spiral
trajectories.

\begin{figure*}[htbp]
\centerline{\includegraphics[width=3.0in,height=2.5in]{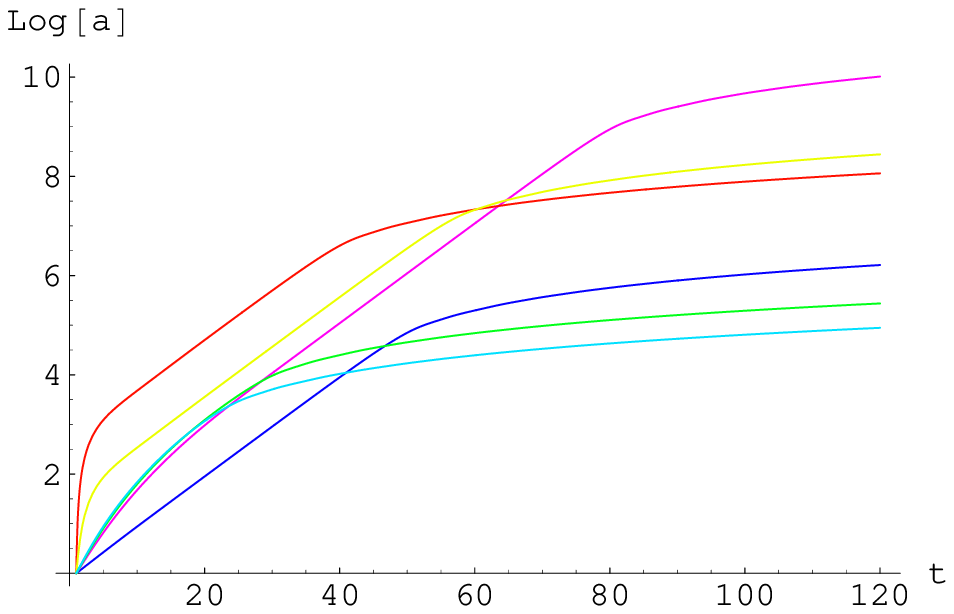}
\includegraphics[width=3.0in,height=2.5in]{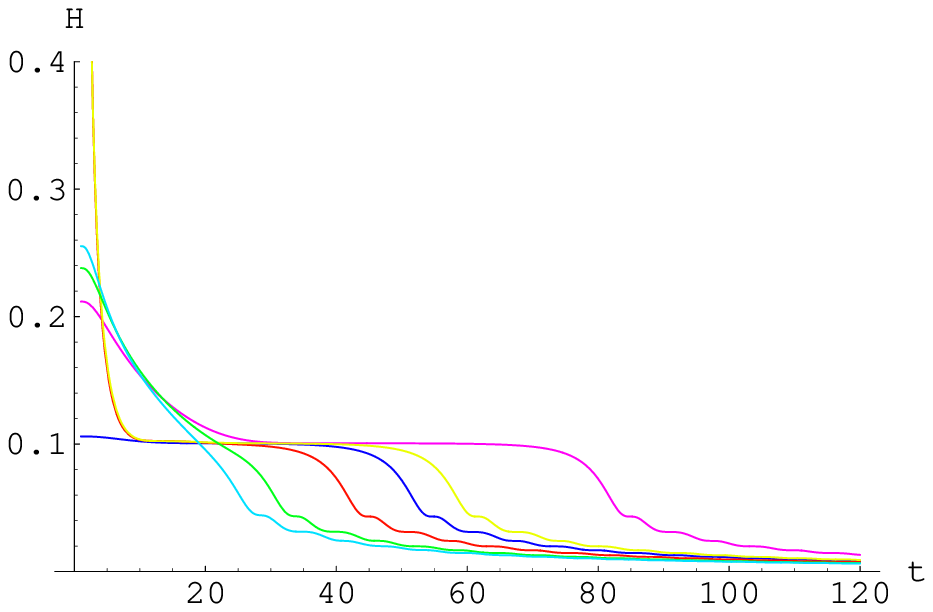}}
\caption {The number of e-folds (left panel) and the Hubble
parameter (right panel) for the corresponding trajectories.
\label{H2}\label{efolds2}}
\end{figure*}

\begin{figure*}[htbp]

\centerline{\includegraphics[width=3.0in,height=2.5in]{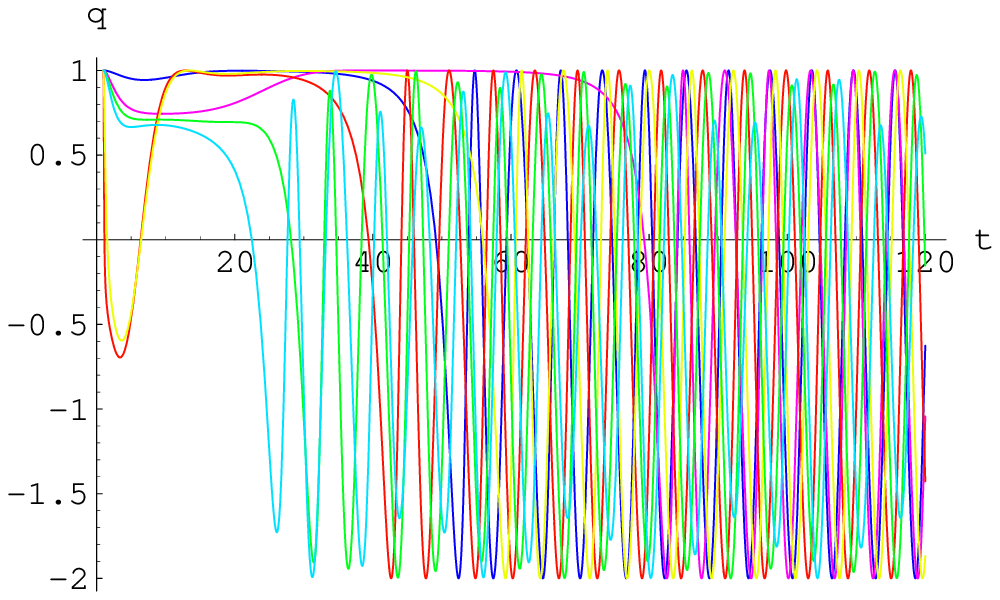}
\includegraphics[width=3.0in,height=2.5in]{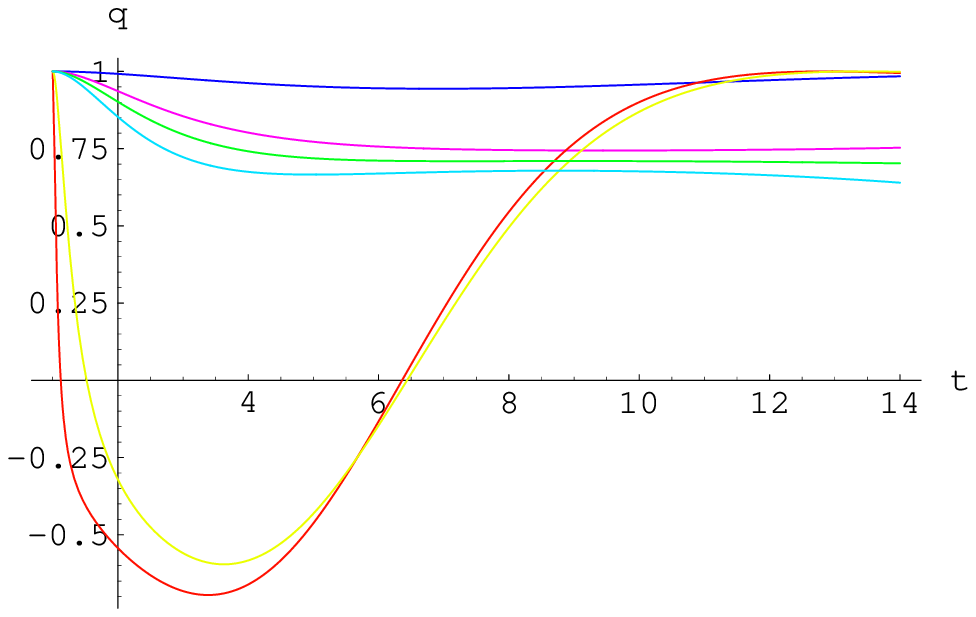}}
\caption {The parameter of acceleration (left panel) and its
magnification for early times (right panel). There are two different
form of acceleration with $q\approx 1$ (De Sitter-like inflation)
and $q\approx 0.75$ (power-law inflation with $s\approx 4$)
accordingly. The averaging of $q$ over a few periods of oscillations
results in $\bar q =-0.5$ which corresponds to the matter dominated
decelerating Universe. \label{q2}}
\end{figure*}

In Figure \ref{H2} the evolution of the logarithms of the
scale factor $a(t)$ (left panel) and the evolution of the Hubble
parameter $H(t)$ (right panel) are presented and in Fig. \ref{q2} the evolution of
the parameter of acceleration $q(t)$.

Because for initial condition the value $a(t=1) = 1$ (in the
Planck units) is used, then $\log a(t)$ gives the number of e-folds: $\log
a(t) =N(t)$.
The Figure \ref{efolds2} shows that for considered trajectories one
can reach the maximum of e-folds of the order of 10. Clearly, 10
e-folds is not sufficient to solve the horizon and flatness problems
but it can be useful to explain a part of the modern CMB data. For
example, the Universe inflates by $\triangle N \approx 4 $ during
the period that wavelengths corresponding to the CMB multipoles
$2\leq l \leq 100$ cross the Hubble radius \cite{CMB,CMBa}. However, to
have the inflation which is long enough for all modes which
contribute to the CMB to leave the horizon, it is usually supposed
that $\triangle N \ge 15$ \cite{WMAP5}.

The Figure \ref{H2} for the evolution of the Hubble parameter (right
panel) demonstrates that the red, yellow, dark blue and pink lines
have a plateau $H \approx const$. It means that the scale factor
$a(t)$ has a stage of the De Sitter expansion on these plateaus.
Clearly, it happens because these lines reach the vicinity of the
effective potential saddle point and spend there some time.

The Fig. \ref{q2} for the acceleration parameter defined in \rf{6}
confirms also the above conclusions. According to Eq. \rf{6b}, $q=1$
for the De Sitter-like behavior. Indeed, all these 4 lines have
stages $q\approx 1$ for the same time intervals when $H$ has a
plateau. Additionally, the magnification of this picture at early
times (the right panel of the Figure \ref{q2}) shows that pink,
green and blue lines have also a period of time when $q$ is
approximately constant less than one: $q \approx 0.75$. In
accordance with Eq. \rf{6b}, it means that during this time the
scale factor $a(t)$ undergoes the power-law inflation $a(t) \propto
t^s$ with $s\approx 4$. This result confirms rude estimates made
above for the trajectories which go through the regions where the
effective potential has the form \rf{3.10}. After stages of the
inflation, the acceleration parameter starts to oscillate. Averaging
$q$ over a few periods of oscillations, one obtains $\bar q = -0.5$.
Therefore, the scale factor behaves as for the matter dominated
Universe: $a(t) \propto t^{2/3}$. Clearly, it corresponds to the
times when the trajectories reach the vicinity of the effective
potential global minimum and start to oscillate there. It is worth
of noting, that there is no need to plot dynamical behavior for the
equation of state parameter $\omega (t)$ because it is linearly
connected with $q$ (see Eq. \rf{6a}) and its behavior can be easily
understood from the pictures for $q(t)$.

As it is shown above for considered quadratic model, the maximal
number of e-folds is near 10. Can this number be increased? To answer
this question, let us consider a new model with a higher degree of
nonlinearity, i.e. the nonlinear quartic model.


 \addcontentsline{toc}{subsection}{4.3. Quartic model}
 \setcounter{subsection}{1}
 \medskip\medskip\medskip\medskip\medskip\medskip\medskip\medskip
 \textbf{4.3. Quartic model}
  \medskip\medskip\medskip\medskip

In this section let us consider the nonlinear quartic model
\be{4.1} f(\bar{R})=\bar{R}+\gamma\bar{R}^{4}-2\Lambda_{D}\, . \ee
For this model the scalar field potential \rf{3.4} reads
\cite{GZBR}:
\be{4.2} U(\phi )= \frac12 e^{-B\phi }\left[\frac34 (4\gamma
)^{-1/3}(e^{A\phi }-1)^{4/3} +2\Lambda_D\right]\, . \ee
Here, the scalar curvature $\bar{R}$ and scalar field $\phi$ are
connected as follows: $e^{A\phi} \equiv f'= 1+4\gamma\bar{R}^{3}
\quad \Leftrightarrow \quad
\bar{R}=\left[(e^{A\phi}-1)/4\gamma\right]^{1/3}\; .$

Let us look for a solution which has a nonnegative minimum of the
effective potential $U_{eff}(\varphi ,\phi )$ \rf{3.1} where
potential $U(\phi)$ is given by Eq. \rf{4.2}. If $\phi_0$
corresponds to this minimum, then, as it was mentioned above (see also
\cite{Zhuk}), $U(\phi_0),\lambda $ and $f_1^2$ should be positive.
To get zero minimum of the effective potential, these positive
values should satisfy the relation of the form of \rf{2.8}:
$U(\phi_0) = f_1^2 = \lambda /2$.

Additionally, it is important to note that positiveness of
$U(\phi_0)$ results in positive expression for $\bar R (\phi_0)>0$
\cite{GZBR}.

Eq. \rf{4.2} shows that potential $U(\phi)$ has the following
asymptotes for positive $\gamma$ and $\Lambda_D$\footnote{Negative
values of $\Lambda_D$ and $\gamma$ may lead either to negative
minima, resulting in asymptotically AdS Universe, or to infinitely
large negative values of $U_{eff}$ \cite{GZBR}. In the present consideration
let us to avoid both of these possibilities. Therefore, one shall
consider the case of $\Lambda_D ,\gamma >0$. See also footnote
\ref{gamma}.} : $\phi \to -\infty \Longrightarrow U(\phi ) \approx
\frac12e^{-B\phi}\left[\frac34(4\gamma)^{-1/3}+2\Lambda_D\right] \to
+\infty$ and $\phi\to +\infty \Longrightarrow U(\phi) \approx
\frac38(4\gamma)^{-1/3}e^{(-B+4A/3)\phi}\to +0$. For the latter
asymptote $-B+4A/3=
(D-8)/3\sqrt{(D-2)(D-1)}<0$ is to be taken into account for $D<8$. Obviously, the total number
of dimensions $D=8$ plays the critical role in quartic nonlinear
theories (see \cite{GZBR,BR,ZhukSaidov2}) and investigations for
$D<8, D=8$ and $D>8$ should be performed separately. Now, let us consider the case $D<8$ (i.e.
$d_1=1,2,3$).

It is worth of noting that for considered signs of parameters, the
effective potential $U_{eff}(\varphi ,\phi)$ \rf{3.1} acquires
negative values when $\phi \to +\infty$ (and $U(\phi) \to 0$). For
example, if $U(\phi_0) = f_1^2 = \lambda /2$ (the case of zero
minimum of the effective potential), the effective potential
$U_{eff}(\varphi , \phi \to \infty) <0$ for $0<e^{-b\varphi}<2$ and
the lowest negative asymptotic value $\left.U_{eff}\right|_{min} \to
-(16/27)\lambda$ takes place along the line $e^{-b\varphi}=4/3$.
Therefore, zero minimum of $U_{eff}$ is local\footnote{It is not
difficult to show that the thin shell approximation is valid for
considered model and a tunnelling probability from the zero local
minimum to this negative $U_{eff}$ region is negligible.}.

As it was mentioned above, extremum positions $\phi_i$ of the potential
$U(\phi )$ coincide with extremum positions of $U_{eff}(\varphi
,\phi ) \; : \left.dU/d\phi \right|_{\phi_i} =0 \to
\left.\partial_{\phi}U_{eff}\right|_{\phi_i}=0$. The condition of
extremum for the potential $U(\phi)$ reads:
\be{4.3} \frac{d U}{d \phi}=0\;\Longrightarrow \;
\bar{R}^{4}-\frac{(2+d_{1})}{\gamma(4-d_{1})}\bar{R}+2\Lambda_{D}\frac{(4+d_{1})}{\gamma(4-d_{1})}=0
\, . \ee
For positive $\gamma$ and $\Lambda_D$ this equation has two real
roots:
\begin{equation}\label{4.4}
\bar{R}_{0(1)}=\frac{\Lambda_{D}}{2}\left(-\sqrt{\frac{2(2+d_{1})}{(4-d_{1})k\sqrt{M}}-M}+\sqrt{M}\right)\;
,
\end{equation}
\begin{equation}\label{4.5}
\bar{R}_{0(2)}=\frac{\Lambda_{D}}{2}\left(\sqrt{\frac{2(2+d_{1})}{(4-d_{1})k\sqrt{M}}-M}+\sqrt{M}\right)\;
,
\end{equation}
where a dimensionless parameter $k$ is introduces as
\be{4.6} k:=\gamma\Lambda_D^3\, , \ee
which is positive for positive $\gamma$ and $\Lambda_D$, and
quantities $M ,\; \omega$ read
\ba{4.7}
M&\equiv&-2^{10/3}\frac{(4+d_{1})}{\omega^{1/3}}-\frac{1}{3\cdot2^{1/3}k}\frac{\omega^{1/3}}{(4-d_{1})}\;,\\
\label{4.8} \omega&\equiv&
k\left[-27(4-d_{1})(2+d_{1})^{2}\right.\nn\\&+&\left.\sqrt{27^{2}(4-d_{1})^{2}(2+d_{1})^{4}-4\cdot24^{3}k(16-d_{1}^2)^{3}}\;
\right].\nn\\
\ea
It can be easily seen that for $k>0$ follows $\omega <0$ and $M\ge 0
$. To have real $\omega $, parameter $k$ should satisfy the
following condition
\be{4.9}
k\leq\frac{27^{2}(4-d_{1})^{2}(2+d_{1})^{4}}{4\cdot24^{3}(16-d_{1}^2)^{3}}\equiv
k_0\, . \ee
It is not difficult to verify that roots $\bar R_{0(1,2)}$ are real
and positive if $0<k\le k_0$ and they degenerate for $k \to k_0\, :
\quad \bar{R}_{0(1,2)}\to (\Lambda_D/2)\sqrt{M}$. In this limit the
minimum and maximum of $U(\phi )$ merge into an inflection point.
Now, let us define which of these roots corresponds to minimum of
$U(\phi)$ and which to local maximum. The minimum condition
\be{4.10} \left.\frac{d^2U(\phi)}{d\phi^2}\right|_{\phi_0} >0\;
\Longrightarrow \;
\gamma\left[(d_{1}+2)-4\gamma\bar{R}_{0}^{3}(4-d_{1})\right]>0 \ee
results in the following inequality\footnote{As it was already
mentioned above, the condition  $U(\phi_0) >0$ leads to the
inequality $\bar R (\phi_0)>0$ \cite{GZBR}. Taking into account the
condition $d_1<4$, it is obvious that inequality $
(d_{1}+2)+4|\gamma|\bar{R}_{0}^{3}(4-d_{1})<0$ for $\gamma <0$
cannot be realized. This is an additional argument in favor of
positive sign of $\gamma$.\label{gamma}}:
\be{4.11} \gamma >0\; : \quad
(d_{1}+2)-4\gamma\bar{R}_{0}^{3}(4-d_{1})>0\, . \ee
Thus,  the root $\bar R_{0}$ which corresponds to the minimum of
$U(\phi)$ should satisfy the following condition:
\be{4.12} 0< \bar
R_{0}<\left(\frac{d_{1}+2}{4\gamma(4-d_{1})}\right)^{1/3}\, . \ee
Numerical analysis shows that $\bar{R}_{0(1)}$ satisfies these
conditions and corresponds to the minimum. For $\bar{R}_{0(2)}$ one
obtains that $\bar{R}_{0(2)}
> \left(\frac{d_{1}+2}{4\gamma(4-d_{1})}\right)^{1/3}$ and
corresponds to the local maximum of $U(\phi)$. In what follows one
shall use the notations:
\ba{4.13} \phi_{min}&=&\frac{1}{A}\ln\left[1+4\gamma
\bar{R}_{0(1)}^{3}\right]\, ,\\
\label{4.14} \phi_{max}&=&\frac{1}{A}\ln\left[1+4\gamma
\bar{R}_{0(2)}^{3}\right] \ea
and $U(\phi_{min} )\equiv U_{min}\, ,U(\phi_{max} )\equiv U_{max} $.
Let us note that $\phi_{min}\, ,\phi_{max}$ and the ratio
$U_{max}/U_{min}$ depend on the combination $k$ \rf{4.6} rather than
on $\gamma$ and $\Lambda_D$ taken separately.

Obviously, because potential $U(\phi )$ has two extrema at
$\phi_{min}$ and $\phi_{max}$, the effective potential
$U_{eff}(\varphi ,\phi)$ may have points of extrema only on the
lines $\phi =\phi_{min}$ and $\phi=\phi_{max}$ where $\partial
U_{eff}/\partial\phi |_{\phi_{min},\phi_{max}}=0$. To find these
extrema of $U_{eff}$, it is necessary to consider the second
extremum condition $\partial U_{eff}/\partial \varphi =0$ on each
line separately:
\be{4.15} \frac{\partial U_{eff}}{\partial \varphi} = 0
\Longrightarrow \left\{\begin{array}{cc}
-U_{min}-3f_1^{2}\chi^{2}_{1}+2\lambda\chi_{1}=0 \;, \\
\\
-U_{max}-3f_1^{2}\chi^{2}_{2}+2\lambda\chi_{2}=0 \;,  \\
\end{array}\right.
\ee
where $\chi_{1}\equiv \exp \left(-\sqrt{2d_1/(d_1+2)}\,
\varphi_{1}\right)$ $>0$ and $\chi_{2} \equiv\exp
\left(-\sqrt{2d_1/(d_1+2)}\, \varphi_{2}\right)$ $>0$; $\varphi_1$ and
$\varphi_2$ denote positions of extrema on the lines $\phi
=\phi_{min}$ and $\phi=\phi_{max}$, respectively. These equations
have solutions
\ba{4.16}
\chi_{1(\pm)}=\alpha&\pm& \sqrt{\alpha^{2}-\beta}\,,\nn\\
\alpha&\geq&\sqrt{\beta}\equiv\alpha_{1}\,;\\
\label{4.17} \chi_{2(\pm)}=\alpha&\pm&
\sqrt{\alpha^{2}-\beta\frac{U_{max}}{U_{min}}}\,,\nn\\
\alpha&\geq&\sqrt{\beta\frac{U_{max}}{U_{min}}}\equiv\alpha_{2}>\alpha_1\,;
\ea
where
$\alpha\equiv\lambda/(3f_1^{2})$ and $\beta\equiv
U_{min}/(3f_1^{2})$. These equations show that there are 5 different
possibilities which are listed in the Table \ref{ttable}.
\begin{table*}
\centering \caption{The number of extrema of the effective potential
$U_{eff}$ depending on the relation between parameters.
}\label{ttable}

\vspace{0.1cm}
\begin{tabular}{|c|c|c|c|c|}
  \hline
$0<\alpha<\alpha_1 $&$ \alpha=\alpha_1$ & $\alpha_1<\alpha<\alpha_2$
&$ \alpha=\alpha_2 $&$ \alpha>\alpha_2 $
 \\
\hline
\begin{tabular}{c}
\textbf{no}\\ \textbf{extrema}
  \\
  \\
\end{tabular} &
\begin{tabular}{c}
\textbf{one} \\ \textbf{extremum}
\\
  (point of
  \\
inflection
 on \\the line
 \\ $\phi=\phi_{min}$)
\end{tabular}  &
\begin{tabular}{c}
\textbf{two }\\ \textbf{extrema}
\\
 (one minimum
 \\and one saddle
 \\ on the line
 \\ $\phi=\phi_{min}$ )
\end{tabular}  &
\begin{tabular}{c}
\textbf{three} \\ \textbf{extrema}
\\
(minimum and
\\
saddle on the line
\\$\phi=\phi_{min}$,
\\
inflection on
\\the
line
\\   $\phi=\phi_{max}$)
\end{tabular}  &
\begin{tabular}{c}
 \textbf{four} \\ \textbf{extrema}
 \\
 (minimum and
\\saddle on
\\the
line
\\ $\phi=\phi_{min}$
\\
maximum and
\\
saddle on
\\the
line
\\ $\phi=\phi_{max}$)
\end{tabular}
\\
\hline\end{tabular}
\end{table*}

To clarify which of solutions \rf{4.16} and \rf{4.17} correspond to
minima of the effective potential (with respect to $\varphi $) one
should consider the minimum condition
\be{4.18}
\left.\frac{\partial^{2}U_{eff}}{\partial^{2}\varphi}\right|_{min}>0\quad
\Longrightarrow \quad U_{extr}+\chi^{2}9f_1^{2}-4\lambda\chi>0\, ,
\ee
where $U_{extr}$ is either $U_{min}$ or $U_{max}$ and $\chi$ denotes
either $\chi_1$ or $\chi_2$. Taking into account relations
\rf{4.15}, one obtains
\be{4.19} \chi^2 3f_1^{2} -\chi \lambda>0 \quad \Longrightarrow\quad
\chi>\frac{\lambda}{3f_1^2}=\alpha .
\end{equation}
Thus, roots $\chi_{1,2(+)}$ define the positions of local minima of
the effective potential with respect to the variable $\varphi$ and
$\chi_{1,2(-)}$ correspond to local maxima (in the direction of
$\varphi$).

Now, let us fix the minimum $\chi_{1(+)}$ at the point $\varphi =0 $. It
means that in this local minimum the internal space scale factor is
stabilized at the present day value. In this case
\be{4.20} \left.\chi_{1(+)}\right|_{\varphi=0}=1=\alpha +
\sqrt{\alpha^{2}-\beta}\quad \Longrightarrow\quad
\alpha=\frac{1+\beta}{2}\, . \ee
Obviously, one can do it only if\footnote{Particular value $\alpha
=1$ corresponds to the case $\alpha=\alpha_1=1$ where the only
extremum is the inflection point with $ \chi_{1(-)}=\chi_{1(+)} =
\alpha =1$. Here, $\lambda = U_{min}=3f_1^2$ and
$U_{eff}(\varphi=0, \phi = \phi_{min})= -\lambda+U_{min}+f_1^2
>0$.} $\alpha <1 \Rightarrow \beta \in
[0,1)$. For $\chi_{1(-)}$ follows:  $\chi_{1(-)}=\beta$.

Additionally, the local minimum of the effective potential at the
point $(\varphi=0, \phi = \phi_{min})$ should play the role of the
nonnegative four-dimensional effective cosmological constant. Thus,
the following conditions appear:
\ba{4.21} \Lambda_{eff}&\equiv& U_{eff}(\varphi=0, \phi =
\phi_{min})\nn\\&=&
-\lambda+U_{min}+f_1^2\geq0\nn\\&\Longrightarrow&
-\alpha+\beta+\frac{1}{3}\geq0\, . \ea
From the latter inequality and equation \rf{4.20} takes place
$\beta\in\left[\frac{1}{3},1\right)$. It can be easily seen that
$\beta =1/3 $ (and, correspondingly, $\alpha =2/3$) results in
$\Lambda_{eff}=0$ and one obtains the mentioned above relations:
$U_{min} = f_1^2 = \lambda /2$. In general, it is possible to demand
that $\Lambda_{eff}$ coincides with the present day dark energy
value $10^{-57}\mbox{cm}^{-2}$. However, it leads to very flat local
minimum which means the decompactification of the internal space
\cite{Zhuk}. In what follows, the case of
zero $\Lambda_{eff}$ is to be mainly considered although all obtained results are trivially
generalized to $\Lambda_{eff} = 10^{-57}\mbox{cm}^{-2}$.

Summarizing results, in the most interesting case of $\alpha
>\alpha_2$ the effective potential has four extrema: local minimum
at $\left(\left.\varphi\right|_{\chi_{1(+)}}=0,\phi_{min}\right)$,
local maximum at $\left(\varphi|_{\chi_{2(-)}},\phi_{max}\right)$
and two saddle-points at
$\left(\varphi|_{\chi_{1(-)}},\phi_{min}\right)$, and
$\left(\left.\varphi\right|_{\chi_{2(+)}},\phi_{max}\right)$ (see
Fig. \ref{effpotr4}).

\begin{figure}[htbp]
\centerline{ \includegraphics[width=3in,height=3in]{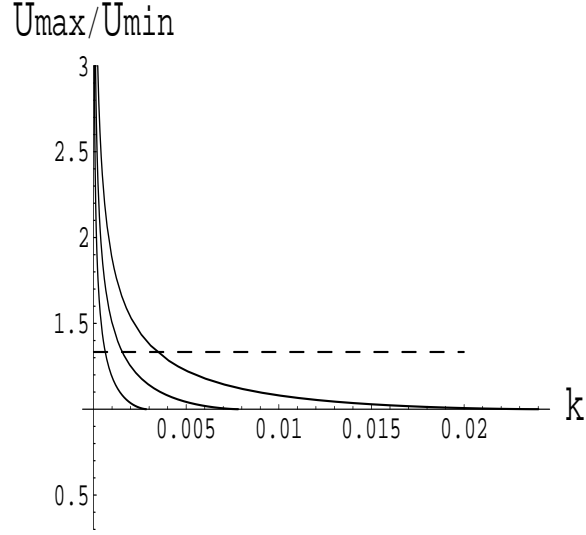}}
\caption{The form of $U_{max}/U_{min}$ as a function of
$k\in(0,k_0]$ for $d_{1}=1,2,3$ from left to right, respectively.
Dashed line corresponds to
 $U_{max}/U_{min}=4/3$. \label{u}}
\end{figure}

\begin{figure*}[htbp]
\centerline{\includegraphics[width=5in,height=3.5in]{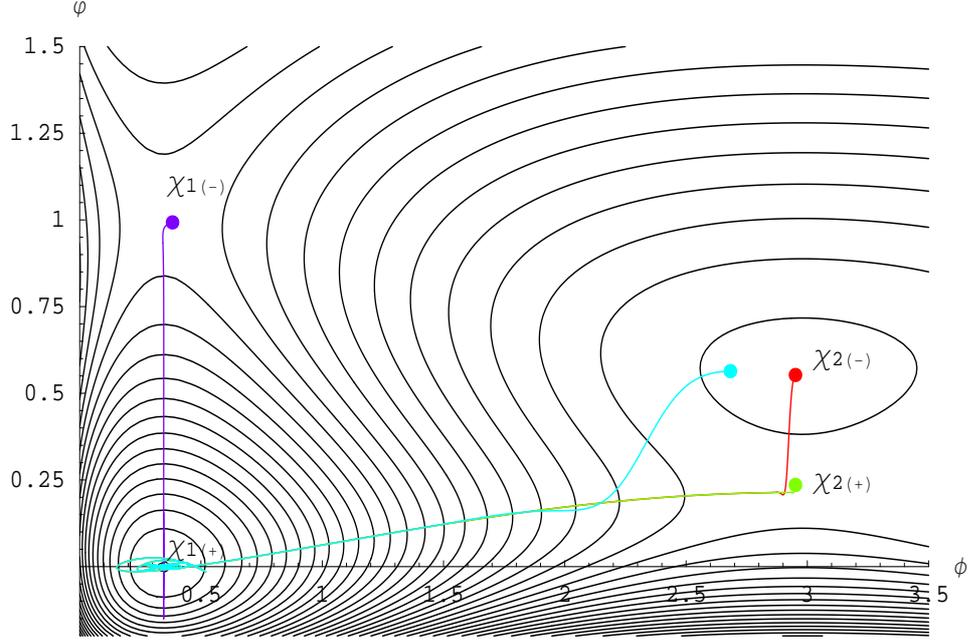}}
\caption {Contour plot of the effective potential $U_{eff}(\varphi
,\phi )$ \rf{3.1} with potential $U(\phi )$ of the form \rf{4.2} for
parameters $\beta = 1/3$, $d_1=3$ and $k=0.004$. This plot shows the
local zero minimum, local maximum and two saddles. The colored lines
describe trajectories for scalar fields starting at different
initial conditions. \label{effpotr4}}
\end{figure*}

Let us pay particular attention to the case of zero local minimum, when
$U_{eff}(\left.\varphi\right|_{\chi_{1(+)}} =0,\phi_{min})=0$ and where
$\beta = 1/3\Longrightarrow \alpha = (1+\beta)/2 =2/3$. To satisfy
the four-extremum condition $\alpha > \alpha_2$, next condition has to be satisfied
\be{4.22} \frac{U_{max}}{U_{min}} < \frac43\; . \ee
The fraction $U_{max}/U_{min}$ is the function of $k$ and depends
parametrically only on the internal space dimension $d_1$.
Inequality \rf{4.22} provides the lower bound on $k$ and numerical
analysis (see Fig. \ref{u})
gives $\tilde{k}(d_{1}=1)\approx0.000625\;;\;\tilde{k}(d_{1}=2)
\approx0.00207\;;\;\tilde{k}(d_{1}=3)\approx0.0035\, .$ Therefore,
effective potentials with zero local minimum will have four extrema
if $k\in (\tilde{k},k_0)$ (where $k_0$ is defined by Eq. \rf{4.9}).
The limit $k \to \tilde k$ results in merging $\chi_{2(-)}
\leftrightarrow \chi_{2(+)}$ and the limit $k\to k_0$ results in
merging $\chi_{1(-)} \leftrightarrow \chi_{2(-)}$ and $\chi_{1(+)}
\leftrightarrow \chi_{2(+)}$. Such merging results in transformation
of corresponding extrema into inflection points. For example, from
Fig. \ref{u} follows that $U_{max}/U_{min} \to 1$ for $k \to k_0$.

The typical contour plot of the effective potential with four
extrema in the case of zero local minimum is drawn in Fig.
\ref{effpotr4}. Here, for $d_1=3$ it is taken $k=0.004 \in
(\tilde{k},k_0)$ which gives $\alpha_2 \approx 0.655$. Thus, $\alpha
= 2/3 \approx 0.666> \alpha_2$.

Let us investigate now a possibility of inflation for considered
potential. First of all, taking into account the comments in
previous section (see Eq. \rf{3.9}), it is clear
that topological inflation in the saddle point $\chi_{1(-)}$ as well
as the slow rolling from there in the direction of the local minimum
$\chi_{1(+)}$ are absent. It is not difficult to verified that the
generalized power-low inflation discussed in the case of the
nonlinear quadratic model is also absent here. Indeed, from Eqs.
\rf{3.1} and \rf{4.2} follows that nonlinear potential $U(\phi)$ can
play the leading role in the region $\phi \to -\infty$ (because
$U(\phi)\to 0$ for $\phi \to +\infty$). In this region $U_{eff}
\propto \exp{(-\sqrt{q}\varphi)}\exp{(-\sqrt{q_2}\phi)}$ where
$q=2d_1/(d_1+2)$ and $q_2=B^2=(d_1+4)^2/[(d_1+2)(d_1+3)]$. For these
values of $q$ and $q_2$ the slow-roll conditions are not satisfied:
$\epsilon \approx \eta_1\approx \eta_2 \approx {q}/{2} +{q_2}/{2}
>1$. However, there are two promising regions where the stage of
inflation with subsequent stable compactification of the internal
space may take place. These are the local maximum $\chi_{2(-)}$ and
the saddle $\chi_{2(+)}$ (see Fig. \ref{effpotr4}). Let us estimate
the slow roll parameters for these regions.

\begin{figure*}[htbp]
\centerline{ \includegraphics[width=3in,height=2.5in]{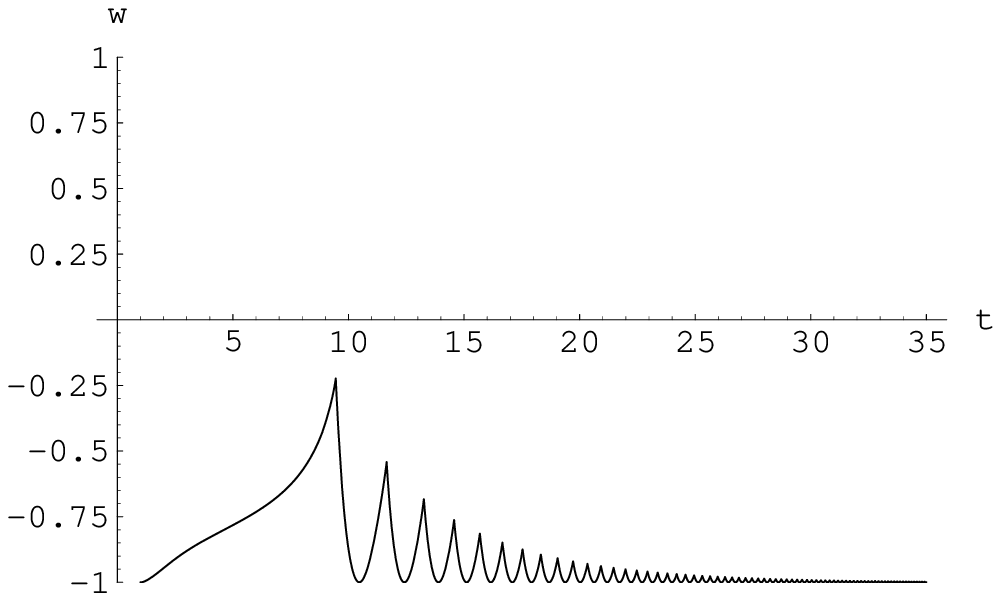}
\includegraphics[width=3in,height=2.5in]{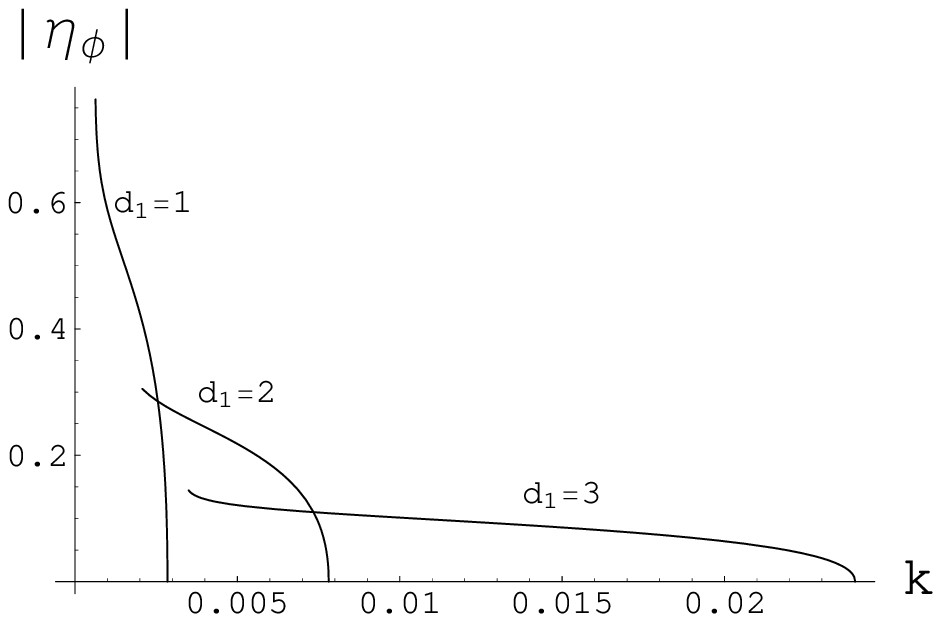}}
\caption{Graphs of $|\eta_{\varphi}|$ (left panel) and
$|\eta_{\phi}|$ (right panel) as functions of $k\in (\tilde{k},k_0)$
for local maximum $\chi_{2(-)}$ and parameters $\beta=1/3$ and
$d_{1}=1,2,3$.\label{w1}}
\end{figure*}

First the local maximum $\chi_{2(-)}$. It is obvious
that the parameter $\epsilon$ is equal to zero here. Additionally,
from the form of the effective potential \rf{3.1} it is clear that
the mixed second derivatives are also absent in extremum points.
Thus, the slow roll parameters $\eta_1$ and $\eta_2$, defined in the
footnote \rf{slowroll}, coincide exactly with $\eta_{\varphi}$ and
$\eta_{\phi}$. In Fig. \ref{w1} a typical form of these
parameters is presented as functions of $k\in (\tilde k, k_0)$ in the case
$\beta=1/3$ and $d_1=1,2,3$.
These plots show that, for considered parameters, the slow roll
inflation in this region is possible for $d_1=1,3$.

\begin{figure*}[htbp]
\centerline{ \includegraphics[width=3in,height=2.5in]{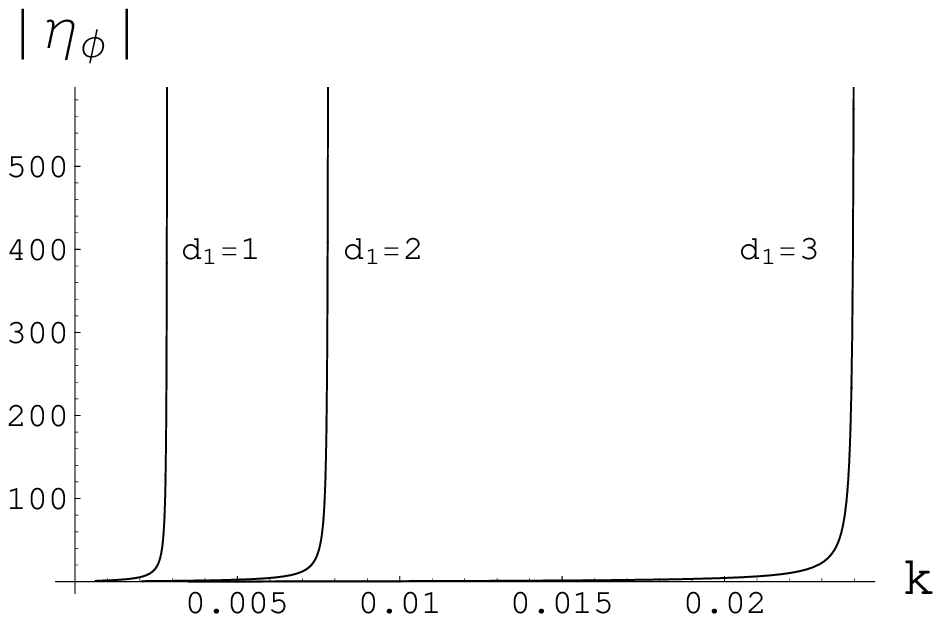}
\includegraphics[width=3in,height=2.5in]{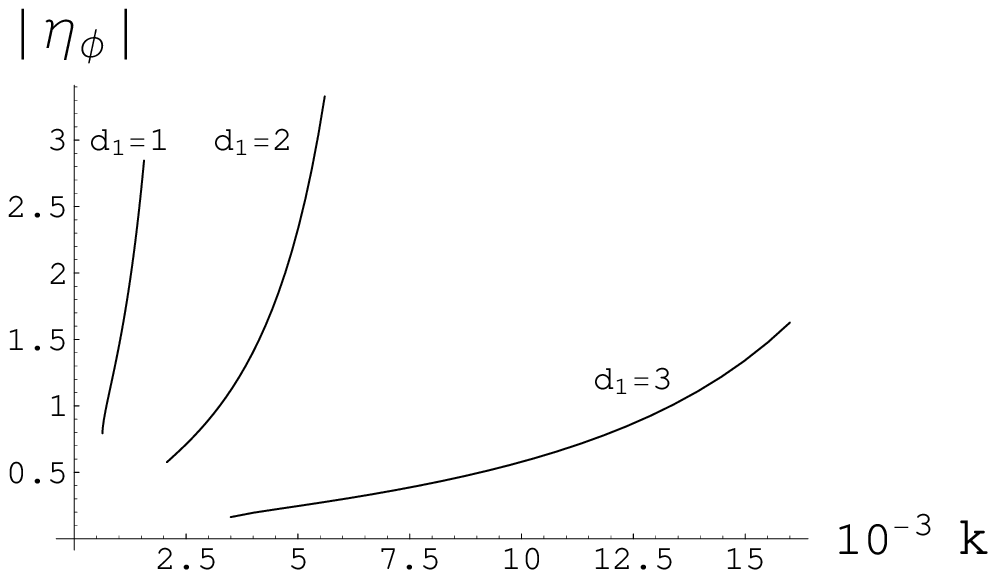}} \caption{Graphs of
$|\eta_{\phi}|$ as functions of $k$ for saddle point $\chi_{2(+)}$
and parameters $\beta=1/3$ and $d_{1}=1,2,3$. Left panel
demonstrates the whole region of variable $k\in (\tilde{k},k_0)$ and
right panel shows detailed behavior for small $k$.\label{f1a1}}
\end{figure*}

The vicinity of the saddle point $\chi_{2(+)}$ is another promising
region. Obviously, if one starts from this point, a test particle will
roll mainly along direction of $\phi$. That is why it makes sense to
draw only $|\eta_{\phi}|$. In Fig. \ref{f1a1}  typical form of
$|\eta_{\phi}|$ is plotted in the case $\beta=1/3$ and $d_1=1,2,3$. Left panel
represents general behavior for the whole range of $k\in
(\tilde{k},k_0)$ and right panel shows detailed behavior in the most
interesting region of small $k$. It shows that $d_1=3$ is the most
promising case in this region.

\begin{figure*}[htbp]

\centerline{\includegraphics[width=3.5in,height=2.5in]{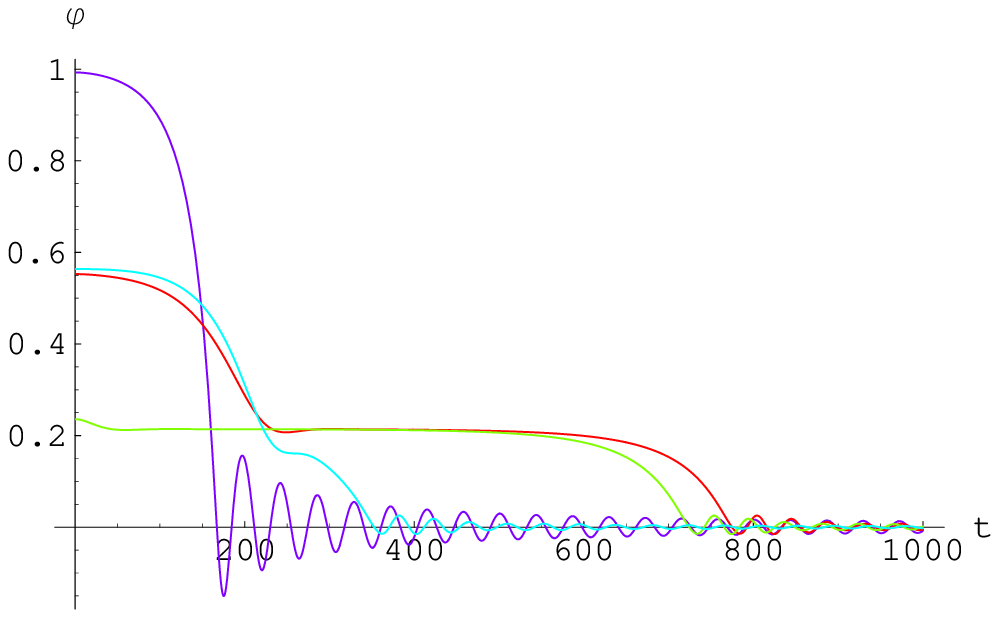}
\includegraphics[width=3.0in,height=2.5in]{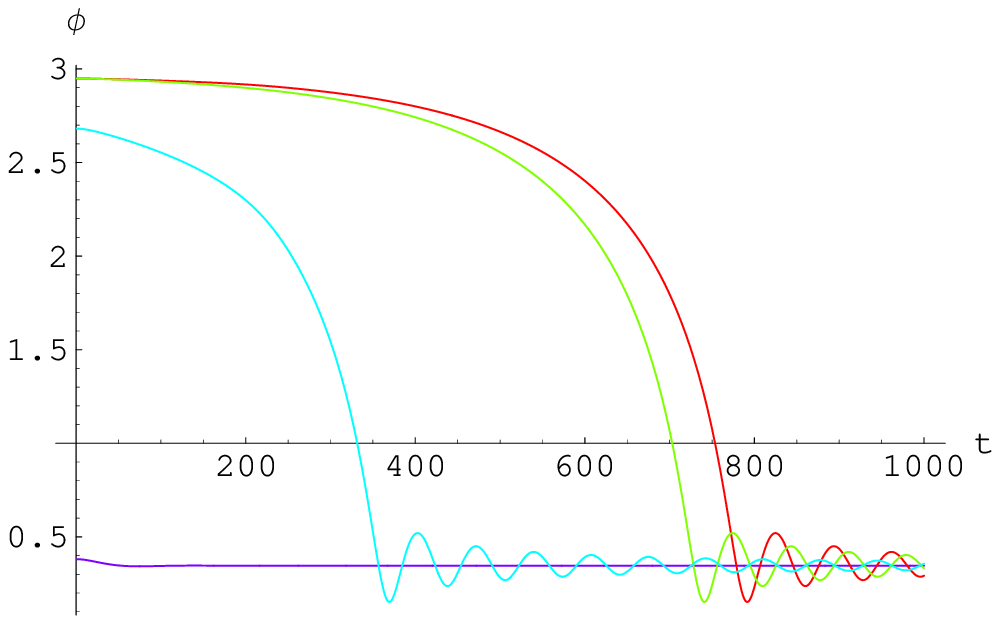}}
\caption {Dynamical behavior of scalar fields $\varphi$ (left panel)
and $\phi$ (right panel) with corresponding initial values denoted
by the colored dots in Fig. \ref{effpotr4}. \label{fields4}}
\end{figure*}

Now, let us investigate numerically the dynamical behavior of scalar
fields and the external space scale factor for trajectories which
start from  the regions $\chi_{1(-)}, \chi_{2(-)}$ and
$\chi_{2(+)}$. All numerical calculations perform for $\beta=1/3,
d_1=3$ and $k=0.004$. The colored lines on the contour plot of the
effective potential in Fig. \ref{effpotr4} describe trajectories for
scalar fields $\varphi$ and $\phi$ with different initial values
(the colored dots) in the vicinity of these extrema points. The time
evolution of these scalar fields is drawn in Fig. \ref{fields4}. For
given initial conditions, scalar fields approach the local minimum
$\chi_{1(+)}$ of the effective potential along the spiral
trajectories.

\begin{figure*}[htbp]
\centerline{\includegraphics[width=3.0in,height=2.5in]{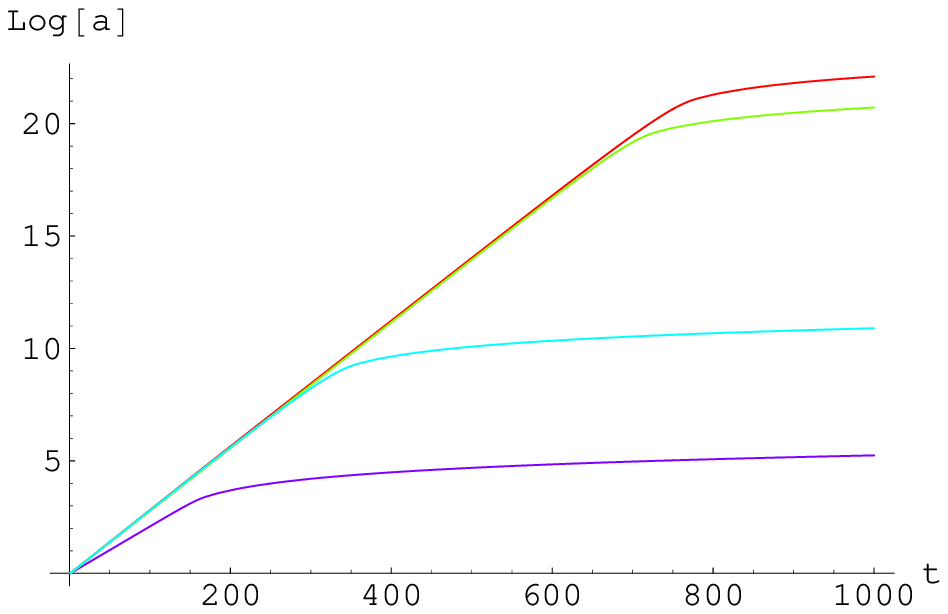}
\includegraphics[width=3.0in,height=2.5in]{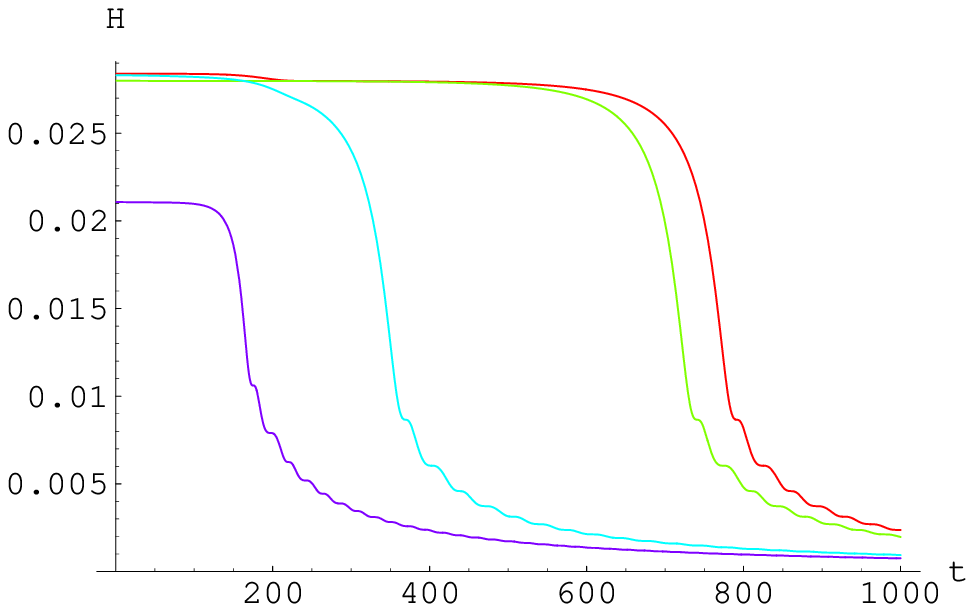}}
\caption {The number of e-folds (left panel) and the Hubble
parameter (right panel) for the corresponding trajectories.
\label{H4}\label{efolds4}}
\end{figure*}

\begin{figure*}[htbp]

\centerline{\includegraphics[width=3.0in,height=2.5in]{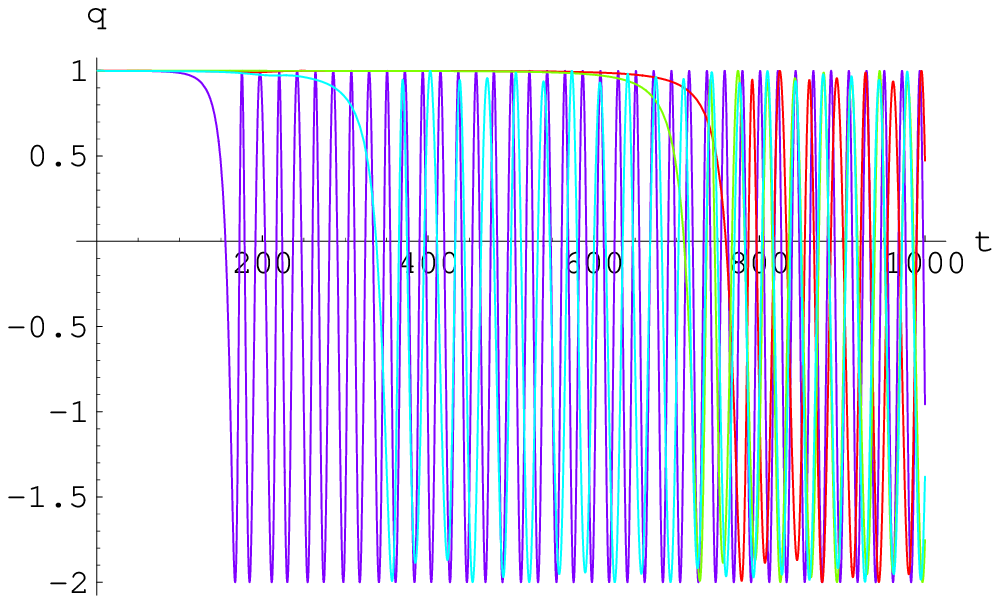}
\includegraphics[width=3.0in,height=2.5in]{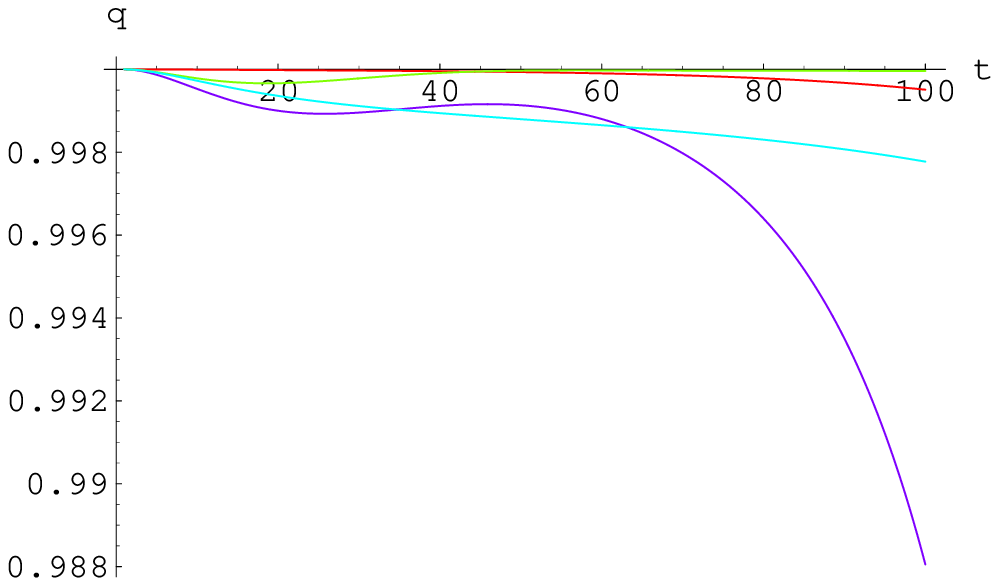}}
\caption {The parameter of acceleration (left panel) and its
magnification for early times (right panel). \label{q4}}
\end{figure*}

In Figure \ref{H4} the evolution of the logarithm of the
scale factor $a(t)$ (left panel) is plotted, which gives directly the number of
e-folds and the evolution of the Hubble parameter $H(t)$ (right
panel) and in Fig. \ref{q4} the evolution of the parameter of
acceleration $q(t)$.

The Figure \ref{efolds4} shows that for considered trajectories
the maximum of e-folds that can be reached is of the order of 22 which is long
enough for all modes which contribute to the CMB to leave the
horizon.

The Figure \ref{H4} for the evolution of the Hubble parameter (right
panel) demonstrates that all lines have plateaus $H \approx const$.
However, the red, yellow and blue lines which pass in the vicinity
of the saddle $\chi_{2(+)}$ have  bigger value of the Hubble
parameter with respect to the dark blue line which starts from the
$\chi_{1(-)}$ region. Therefore, the scale factor $a(t)$ has stages
of the De Sitter-like expansion corresponding to these plateaus
which last approximately from 100 (dark blue line) up to 800 (red
line) Planck times.

The Fig. \ref{q4} for the acceleration parameter confirms also the
above conclusions. All 4 lines have stages $q\approx 1$ for the same
time intervals when $H$ has plateaus. After stages of inflation, the
acceleration parameter starts to oscillate. Averaging $q$ over a few
periods of oscillations, one obtains $\bar q = -0.5$. Therefore, the
scale factor behaves as for the matter dominated Universe: $a(t)
\propto t^{2/3}$. Clearly, it corresponds to the times when the
trajectories reach the vicinity of the effective potential local
minimum $\chi_{1(+)}$ and start to oscillate there.

Let us investigate now a possibility of the topological inflation
\cite{Linde,Vilenkin} if scalar fields $\varphi, \phi$ stay in the
vicinity of the saddle point $\chi_{2(+)}$. As it was mentioned before,
topological inflation in the case of the double-well
potential takes place if the distance between a minimum and local
maximum bigger than $\Delta\phi_{cr}=1.65$. In this case domain wall
is thick enough in comparison with the Hubble radius. The critical
ratio of the characteristic thickness of the wall to the horizon
scale in local maximum is $r_wH\approx |U/3\partial U_{\phi
\phi}|^{1/2} \approx 0.48$ \cite{SSTM} and for topological inflation
it is necessary to exceed this critical value. Therefore, the saddle $\chi_{2(+)}$ should be checked
from the point of these criteria.

\begin{figure*}[htbp]
\centerline{
\includegraphics[width=3in,height=2.5in]{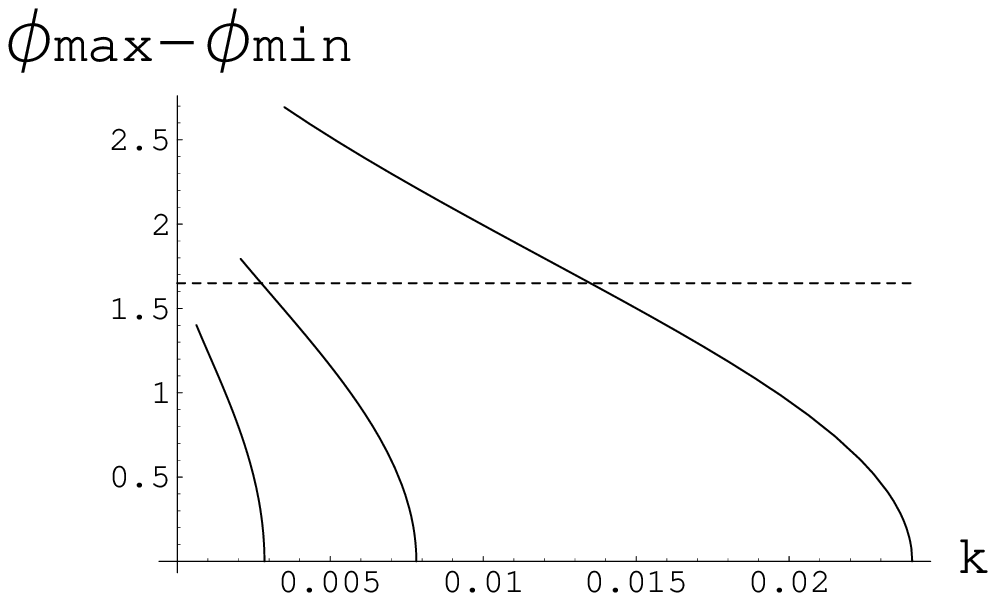}
\includegraphics[width=3in,height=2.5in]{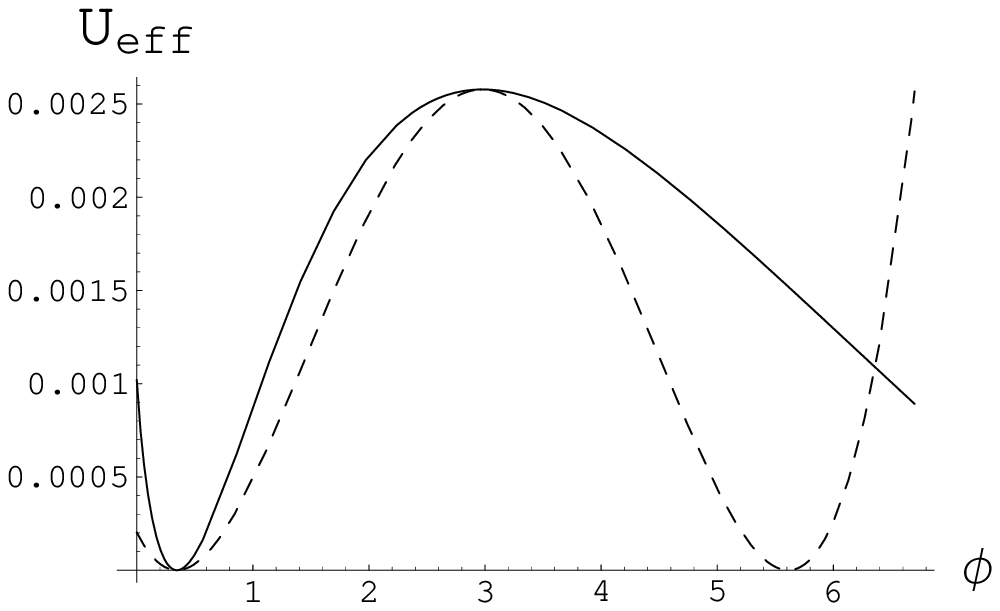}}
 \caption{Left panel demonstrates the difference  $\phi_{max}-\phi_{min}$ (for the profile $\varphi = \varphi|_{\chi_{2(+)}}$)
 as a functions of $k\in (\tilde{k},k_0)$ for parameters
 $\beta=1/3$, and $d_{1}=1,2,3$ (from left to right respectively).
 Dashed line corresponds to $\phi_{max}-\phi_{min}=1.65$. Right panel shows the comparison of the potential $U_{eff}(\varphi|_{\chi_{2(+)}},\phi)$ with a double-well potential for parameters $\beta=1/3, k=0.004$ and $d_1=3$.\label{n}}
\end{figure*}

In Fig. \ref{n} (left panel) the difference  $\Delta \phi
=\phi_{max}-\phi_{min}$ is drawn for the profile $\varphi =
\varphi|_{\chi_{2(+)}}$
 as a functions of $k\in (\tilde{k},k_0)$ in the case
 $\beta=1/3$ for dimensions $d_{1}=1,2,3$. This picture shows that this difference can exceed the critical value if the number of the internal dimensions is $d_1=2$ and $d_1=3$. Right panel of Fig. \ref{n} confirms this conclusion. Here the case $\beta =1/3, \quad k=0.004$ and $d_1=3$ is considered. For chosen values of the parameters, $\Delta \phi = 2.63$ which is considerably bigger than the critical value 1.65 and the ratio of the thickness of the wall to the horizon scale is 1.30 which again bigger than the critical value 0.48. Therefore, topological inflation can happen for considered model. Moreover, due to quantum fluctuations of scalar fields, inflating domain wall will have fractal structure: it will contain many other inflating domain walls and each of these domain walls again will contain new inflating domain walls and so on \cite{Linde}. Thus, from this point, such topological inflation is the eternal one.

To conclude this section, let us draw the attention to one
interesting feature of the given model. From above consideration
follows that in the case of zero minimum of the effective potential
the positions of extrema are fully determined by the parameters $k$
and $d_1$, and for fixed $k$ and $d_1$ do not depend on the choice
of $\Lambda_D$. The same takes place for the slow roll parameters.
On the other hand, if $k$ and $d_1$ are kept, the hight of the
effective potential is defined by $\Lambda_D$ (see Appendix 2).
Therefore, the hight of extrema can be changed with the help of
$\Lambda_D$ but preserve the conditions of inflation for given $k$
and $d_1$.

However, the dynamical characteristics of the model (drawn in
Figures \ref{fields4} - \ref{q4}) depend on variations of
$\Lambda_D$ by the self-similar manner. It means that the change of
hight of the effective potential via transformation $\Lambda_D \to
c\Lambda_D$ ($c$ is a constant) with fixed $k$ and $d_1$ results in
rescaling of Figures \ref{fields4} - \ref{q4} in $1/\sqrt{c}$ times
along the time axis.

 \addcontentsline{toc}{subsection}{4.4. Bouncing inflation in $R^2+R^4$ model}
 \setcounter{subsection}{1}
 \medskip\medskip\medskip\medskip\medskip\medskip\medskip\medskip
 \textbf{4.4. Bouncing inflation in $R^2+R^4$ model}

 \addcontentsline{toc}{subsection}{4.4.1. The fitting procedure}
 \setcounter{subsection}{1}
 \medskip\medskip\medskip\medskip
 \textbf{4.4.1. The fitting procedure}
  \medskip\medskip\medskip\medskip

Here let us continue the investigation of the quadratic-quartic model started in section 2.4
for an arbitrary number of dimensions
$D$, but in the most particular examples $D=4$ will be used (unless stated otherwise).
First of all, let us define the relation between the
scalar curvature $\bar{R}$ and the scalaron field $\phi$.

\begin{figure}
  \center
    \includegraphics[width=3.6in,height=3in]{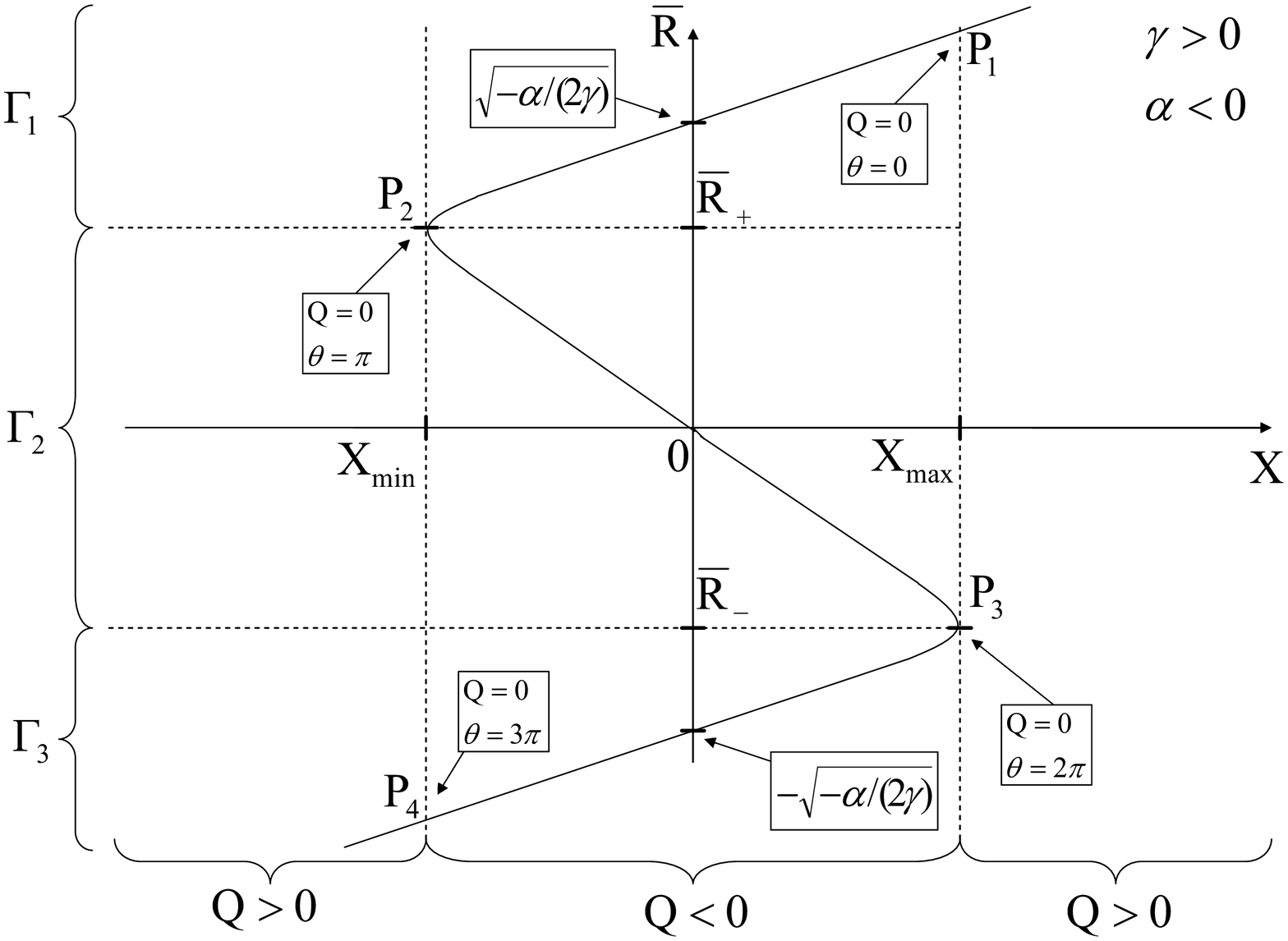}\\
\caption {The schematic drawing of the  real solution branches
and the matching points $P_{1,2,3,4}$ This figure shows that points $P_{2,3}$ (correspondingly, $\theta=\pi,2\pi$) and points
$P_{1,4}$ (correspondingly, $\theta=0,3\pi$) have different nature. So,  $P_{2,3}$ and  $P_{1,4}$ one shall call branching points and
monotonic points (in the sense that function $\bar R$ is monotonic in the vicinity of these points), respectively.\label{zigzag}}
\end{figure}

 Starting by considering a concrete example, for
definiteness, let us assume $\gamma>0,\alpha<0$. The pairwise
fitting of the various solution branches should be performed at
points where $Q=0$ and different branches of the three-solution
sector are fitted with each other or to the branches of the
one-solution sector. The points with $Q=0$ correspond to the
$X-$values $X_{min}$ and $X_{max}$. Explicitly from
\rf{s15a} follows
\be{f1} X(\bar R_\pm)=\frac 43\alpha\bar R_\pm=\pm\frac
43\alpha\sqrt{-\frac{\alpha}{6\gamma}} \ee
 and for the concrete
configuration $\gamma>0,\alpha<0$
\ba{f2} X_{max}&=&X(\bar R_-)=-\frac
43\alpha\sqrt{-\frac{\alpha}{6\gamma}}\ge 0, \; \nn\\X_{min}&=&X(\bar
R_+)=\frac 43\alpha\sqrt{-\frac{\alpha}{6\gamma}}\le 0\,.\ea
 Next,
from the defining equation \rf{s11} for the angle
$\vartheta$ that at $Q=0$ holds
\be{f3} \cos (\vartheta)=\frac{r}{|b|}=\frac{r}{|r|} \ee
so that
\ba{f4}  X_{max}\ge 0 &\Longrightarrow& r>0 \;
\Longrightarrow \; \cos(\vartheta)=1 \; \nn\\&\Longrightarrow&
\vartheta=2\pi m,\ m\in \ZZ\,,\nn\\
X_{min}\le 0 &\Longrightarrow& r<0 \;
\Longrightarrow \;\cos(\vartheta)=-1 \;\nn\\&\Longrightarrow&
 \vartheta=\pi+2\pi n,\ n\in \ZZ\,. \ea
   Now, the fitting of
the various solution branches can be performed as follows (see Fig. \ref{zigzag}).
Let us
start with the branch $\Gamma_1:=(R_+\le \bar R\le
+\infty,X_{min}<X<+\infty)$. Moving in on this branch from
$X\approx +\infty$ one is working in the one-solution sector $Q>0$
with
\be{f5} \bar
R(\Gamma_1;Q)=\left[r+Q^{1/2}\right]^{1/3}+\left[r-Q^{1/2}\right]^{1/3}\ee
 until $Q=0$ is hitten at $X=X_{max}$. At this point
$P_1:=(\Gamma_1,X=X_{max})\in \Gamma_1$ one has to perform the
first fitting. Due to $r>0$ the following may be chosen
\be{f6} \bar R(\Gamma_1;Q=0)=2r^{1/3}=2|b|^{1/3}\ee
 so that as
simplest parameter choice in \rf{s13} comes
\be{f7} P_1=(\Gamma_1,X=X_{max},Q=0)\ \mapsto \ \vartheta=0,\
k=0\,. \ee
 Hence, the parametrization for $(\Gamma_1,Q<0)$ will be
given as
\be{f8} \bar R(\Gamma_1,Q<0)=2|b|^{1/3}\cos (\theta/3). \ee
 For
later convenience, the $\vartheta$ from the
equations \rf{s12}, \rf{s13} is replaced by
$\theta$. The reason will become clear from the subsequent
discussion. In should be noted that on this $\Gamma_1-$segment one may set
$\vartheta=\theta$. Let us further move on $\Gamma_1$ until its
end at $X_{min}$, where again $Q=0$. Because there was no other
point with $Q=0$ on this path, the smoothly changing $\theta$ can
at this local minimum $P_2=\Gamma_1\cap \Gamma_2=(X=X_{min},\bar
R=\bar R_+)$ only take one of the values $\theta=\pm \pi$. For
definiteness $\theta (P_2)=\pi$ is chosen. Hence, it holds
\ba{f9} \bar R(P_2)&=&2|b|^{1/3}\cos
(\pi/3)=|b|^{1/3}\nn\\&=&\sqrt{-q}=\sqrt{-\alpha/(6\gamma)}=\bar R_+\qquad\ea
as it should hold. For convenience, let us parameterize the
movement on the cubic curve by simply further increasing $\theta$.
This gives for moving on $\Gamma_2=(\bar R_+\ge \bar R\ge \bar
R_-,X_{min}\le X \le X_{max})$ from the local minimum at $P_2$ to
the local maximum at $P_3=\Gamma_2\cap \Gamma_3=(X=X_{max},\bar
R=\bar R_-)$ a further increase of $\theta$ by $\pi$ up to $\theta
(P_3)=2\pi$. Accordingly, the complete consistency condition looks as
\ba{f10} \bar R(P_3)&=&2|b|^{1/3}\cos
(2\pi/3)=-|b|^{1/3}\nn\\&=&-\sqrt{-q}=-\sqrt{-\alpha/(6\gamma)}=\bar
R_-\,. \ea
By further increasing $\theta$ up to $\theta=3\pi$
 the point $P_4=(X=X_{min},Q=0)\in \Gamma_3$  can be reached with \be{f11}
\bar R(P_4)=2|b|^{1/3}\cos (3\pi/3)=-2|b|^{1/3}=-2|r|^{1/3}.
\ee
Because of $r<0$ it can be smoothly fitted it to the one-solution branch
\be{f12} \bar
R(\Gamma_4,Q)=\left[r+Q^{1/2}\right]^{1/3}+\left[r-Q^{1/2}\right]^{1/3}
\ee
 by setting trivially
 \be{f13} \bar
R(P_4)=2(-|r|)^{1/3}=-2|r|^{1/3}.
\ee
  Summarizing, a very simple and transparent branch fitting picture is achieved, where all
the movement in the three-solution sector can be parameterized by
choosing the effective angle as $\theta \in [0,3\pi]$. Finally, this picture have to be fitted in terms of smoothly varying
$\theta\in[0,3\pi]$ with the three-solutions $\bar R_k$ from
\rf{s13}. For this purpose let us note that the single value
$\vartheta\in [0,\pi]$ in \rf{s13} is a projection of the smoothly
varying $\theta\in[0,3\pi]$. Fixing an arbitrary $\vartheta$ one
easily finds the following correspondences
\be{f14} \theta(\Gamma_1,\vartheta)=\vartheta,\;
\theta(\Gamma_2,\vartheta)=2\pi-\vartheta,\;
\theta(\Gamma_3,\vartheta)=2\pi+\vartheta
\ee
 and hence
\ba{f15}
\bar R[\theta(\Gamma_1,\vartheta)]&=&2|b|^{1/3}\cos\left(\frac \vartheta 3\right)=
\bar{R}_{(k=0)}=\overline{\mathcal{R}}_{3}\;,\nn\\
\bar
R[\theta(\Gamma_2,\vartheta)]&=&2|b|^{1/3}\cos\left(\frac{2\pi-
\vartheta} 3\right)\nn\\&=&
2|b|^{1/3}\cos\left(\frac{\vartheta-2\pi} 3\right)=\bar R_{(k=-1)}=\overline{\mathcal{R}}_{2},\nn\\
\bar
R[\theta(\Gamma_3,\vartheta)]&=&2|b|^{1/3}\cos\left(\frac{2\pi+\vartheta}3\right)\nn\\&=&\bar
R_{(k=1)}=\overline{\mathcal{R}}_{1}\,. \ea
Analogically, the rules for fitting procedure can be obtained in the
case $\gamma<0,\alpha>0$.
So, all the fitting mechanism is clear now and
can be used in further considerations.

 \addcontentsline{toc}{subsection}{4.4.2. Dynamics of the Universe and scalaron in $R^2+R^4$ model}
 \setcounter{subsection}{1}
 \medskip\medskip\medskip\medskip
 \textbf{4.4.2. Dynamics of the Universe and scalaron in $R^2+R^4$ model}
  \medskip\medskip\medskip\medskip

\begin{figure}
  \center
    \includegraphics[width=4in,height=2.5in]{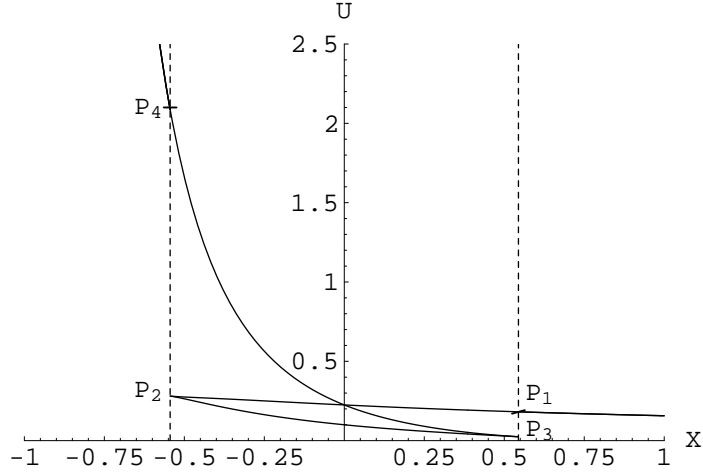}\\
\caption {The form of the potential \rf{1.8} as a multi-valued function of $X=e^{A\phi}-1$ in the case
$D=4$, $\Lambda_4=0.1$, $\gamma =1$ and $\alpha =-1$. Points $P_{1,2,3,4}$ are defined in Fig. \ref{zigzag}.\label{U(X)}}
\end{figure}

\begin{figure}
  \center
    \includegraphics[width=4in,height=3in]{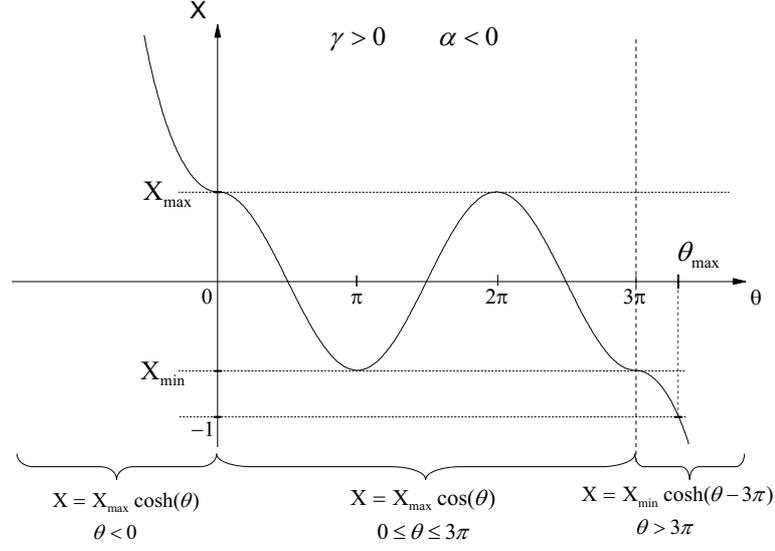}\\
\caption {The schematic drawing of Eq. \rf{5.3} in the case $X_{min}>-1$. Here, $X_{max} =\sqrt{(1/\gamma )\left(2|\alpha |/3\right)^3}\, ,\;
X_{min}=-\sqrt{(1/\gamma )\left(2|\alpha |/3\right)^3}$ and $\theta_{max}$ is defined by Eq. \rf{5.6}.\label{X(theta)}}
\end{figure}

To study the dynamics of the Universe in given model, let us assume that the four-dimensional metric $g$ in \rf{1.9} is
spatially flat Friedmann-Robertson-Walker one:
\be{5.1a1}
g=-dt\otimes dt + a^2(t)d\vec{x}\otimes d\vec{x}\, .
\ee
Thus, scalar curvatures
$R$ and $\bar R$ and scalaron $\phi$ are functions of time. Therefore, Eq. \rf{1.11ab} for homogeneous field $\phi$ reads
\be{5.2}
\ddot{\phi}+3H\dot{\phi}+\frac{d U}{d\phi}=0\, ,
\ee
where the Hubble parameter $H=\dot{a}/a$ and the dotes denote the differentiation with respect to time $t$. Potential $U$ is defined by
Eq. \rf{1.8a1}. Because $U$ depends on $\bar R$ which is a multi-valued function of $\phi$ (or, equivalently, of $X$), the potential $U$ is
also a multi-valued function of $X$ (see Fig. \ref{U(X)})\footnote{In spite of the divergency of $d\bar R/d X$ in the branching points $P_{2,3}$, the derivatives $dU/dX$ are finite in these points.  Moreover, $\bar R$ and $X$ have the same values in branching points for different branches. Therefore, the branches arrive at the branching points with the same values of $dU/dX$ and Fig. \ref{U(X)} clearly shows it.
}.
However, previous analysis shows that this problem can be avoided by making $X$ and $\bar R$ single-valued
functions of a new field $\theta$ (one would remind that the particular case of $\alpha \gamma <0$ is considered when $\alpha <0,\, \gamma >0$):
\be{5.3}
X(\theta) =
\left\{\begin{array}{ccc}
\sqrt{(1/\gamma )\left(2|\alpha |/3\right)^3}\; \cosh (\theta ) \qquad\,,\theta <0&;&\; \\
\sqrt{(1/\gamma )\left(2|\alpha |/3\right)^3}\; \cos (\theta ) \,, 0\leq\theta \leq 3\pi&;&\;\\
-\sqrt{(1/\gamma )\left(2|\alpha |/3\right)^3}\; \cosh (\theta -3\pi ) \,, \theta > 3\pi &;&\;  \\
\end{array}\right.
\ee
and
\be{5.4}
\bar R(\theta) =
\left\{\begin{array}{ccc}
2\sqrt{|\alpha|/(6\gamma)}\; \cosh (\theta /3 ) \,,\quad\qquad \theta <0&;&\; \\
2\sqrt{|\alpha|/(6\gamma)}\; \cos (\theta /3) \,, \quad 0\leq\theta \leq 3\pi&;&\;\\
-2\sqrt{|\alpha|/(6\gamma)}\; \cosh [(\theta /3) -\pi ] \,,\quad  \theta > 3\pi &.&\;  \\
\end{array}\right.
\ee
The function $X = X(\theta)$ is schematically given in Fig. \ref{X(theta)}.
It is necessary to keep in mind that the case under consideration is
$f'>0 \rightarrow X>-1$. If condition $X_{min} >-1$ is imposed (in opposite case the graphic $X(\theta)$ will be cut into two disconnected parts) then
the parameters $\alpha$ and $\gamma$ should satisfy the inequality:
\be{5.5}
X_{min}=-\sqrt{(1/\gamma )\left(2|\alpha |/3\right)^3} >-1 \; \Rightarrow \;|\alpha| \leq \frac32 \gamma^{1/3}\, .
\ee
The maximal value of $\theta $ (which is greater than $3\pi$ in the case $X_{min} >-1$) is defined from the transcendental equation
\ba{5.6}
&&\frac12\left[\left(c+\sqrt{c^2-1}\, \right)^{1/3}+ \left(c-\sqrt{c^2-1}\, \right)^{1/3}\right]=\nn\\ &&= \cosh [(\theta_{max} /3) -\pi ]\, ,
\ea
where $c := \left(\sqrt{(1/\gamma )\left(2|\alpha |/3\right)^3}\right)^{-1} $. The limit $X\to -1$ corresponds to the limit $\theta \to \theta_{max}$.
With the help of Eq. \rf{5.3} and formula
\be{5.7}
\frac{d \phi}{d \theta} = \frac{1}{A(X+1)}\frac{d X}{d \theta}
\ee
the following useful expressions can be also obtained:
\be{5.8}
\left. \frac{d \phi}{d \theta}\right|_{\theta =0,\pi,2\pi,3\pi}  = \left.\frac{d X}{d \theta}\right|_{\theta =0,\pi,2\pi,3\pi} =0\, .
\ee

It can be easily verified that field $\theta$ satisfies the equation:
\be{5.9}
\ddot \theta +3 H\dot \theta + \Gamma (\theta){\dot \theta}^2 + G (\theta )\frac{d U}{d \theta} =0\, .
\ee
Here, $G (\theta )\equiv G^{11} = (G_{11})^{-1}=(d\phi /d\theta)^{-2}$ is the one dimensional metric on the moduli space with the
corresponding Christoffel symbol $\Gamma (\theta) \equiv \Gamma^{1}_{11}=(1/2)G^{11}(G_{11})_{,\, \theta} = (d^2\phi /d\theta^2)/(d\phi /d\theta)$.

 \addcontentsline{toc}{subsection}{4.4.3. Properties of the potential $U(\theta )$}
 \setcounter{subsection}{1}
 \medskip\medskip\medskip\medskip
 \textbf{4.4.3. Properties of the potential $U(\theta )$}
  \medskip\medskip\medskip\medskip

As it was mentioned above, in current case the potential \rf{1.8a1} as a function of $\theta $ is a single-valued one.
Now, let us investigate analytically some general properties of $U(\theta )$. In this section, $D$ is an arbitrary number of dimensions and signs of
$\alpha$ and $\gamma$ are not fixed if it is not specified particularly.

First, let us consider the extrema of the potential $U(\theta )$. To find the extremum points the equation have to be solved
\be{5.10}
\frac{d U}{d\theta} = \frac{dU}{d\phi}\frac{d\phi}{d\theta}=\frac{dU}{d\phi}\frac{1}{A(X+1)}\frac{dX}{d\theta}=0\, .
\ee
Therefore, the extrema correspond either to the solutions of the equation \rf{1.11} $dU/d\phi =0$ for finite $dX/d\theta$ ($X>-1$) or
to the solutions of the equation $dX/d\theta=0$ ($X >-1$) for finite $dU/d\phi$.  The form of the potential $U$ (see Eq. \rf{1.8a1})
shows that this potential and its derivative $dU/d\phi$ is finite for $X>-1$. Thus, as it follows from Eq. \rf{5.8}, the potential $U(\theta )$ has extrema at the matching points $\theta =0,\pi,2\pi,3\pi$.
Additional extremum points are real solutions of the equation \rf{extR}. For the given model \rf{4.1a} this
equation reads:
\ba{5.11}
\bar R^{4}\gamma\left(\frac{D}{2}-4\right)+\bar
R^{2}\alpha\left(\frac{D}{2}-2\right)+\bar
R\left(\frac{D}{2}-1\right)-D\Lambda_{D} =0\;.
\ea
The form of this equation shows that there are two particular cases: $D=8$ and $D=4$. The $D=8$ case was considered previously. Let us consider now the
case $D=4$:
\be{5.12}
 \bar R^4 -\frac{1}{2\gamma}\bar R +\frac{2\Lambda_4}{\gamma}=0\, .
\ee
It is worth of noting that parameter $\alpha$ disappeared from this equation. Thus $\alpha$ has no an influence on a number
of additional extremum points.
To solve this quartic equation, one should consider an auxiliary cubic equation
\be{5.13}
u^3-\frac{8\Lambda_4}{\gamma}u-\frac{1}{4\gamma^2}=0\, .
\ee
The analysis of this equation can be performed in similar manner as it was done it for the cubic equation \rf{s1a}.
Let us introduce the notations:
\ba{5.14}
&&\bar q := -\frac{8}{3}\frac{\Lambda_4}{\gamma}\; ,\bar r := \frac{1}{8}\frac{1}{\gamma^2}\, ,
\nn\\&&\bar Q:= \bar r^2 +\bar q^3 = \frac{1}{\gamma^4}\left(\frac{1}{8^2}-\gamma\left(\frac{8\Lambda_4}{3}\right)^3\right).
\ea
It make sense to consider two separate cases.

1. $\sign \gamma = -\sign \Lambda_4 \quad \Rightarrow \quad \bar Q>0\, .$

In this case only one real solution of Eq. \rf{5.13} is needed:
\be{5.15}
u_1 = \left[\bar r +\bar Q^{1/2}\right]^{1/3} + \left[\bar r -\bar Q^{1/2}\right]^{1/3} >0\, .
\ee
Then, solutions of the quartic \rf{5.12} are the real roots of two quadratic equations
\be{5.16}
\bar R^2 \pm \sqrt{u_1}\bar R +\frac12\left(u_1\pm\epsilon\sqrt{u_1^2+3q} \right)=0\, ,\quad \epsilon = \sign \gamma\, .
\ee
Simple analysis shows that for any sign of $\gamma$ two real solutions can be obtained:
\be{5.17}
\begin{array}{ccc}
\gamma <0  \Rightarrow  \bar R^{(+)}_{1,2}=-\frac{1}{2}u_1^{1/2}\pm \sqrt{-\frac{1}{4}u_1+\frac{1}{2}(u_1^2+3q)^{1/2}}\,,\\
\phantom{\int}\\
\gamma >0 \Rightarrow
\bar R^{(-)}_{1,2}=\frac{1}{2}u_1^{1/2}\pm \sqrt{-\frac{1}{4}u_1+\frac{1}{2}(u_1^2+3q)^{1/2}}\, .\\
\end{array}
\ee

2. $\sign \gamma = \sign \Lambda_4 \quad \Rightarrow \quad \bar Q\gtrless0\, .$

It is not difficult to show that in this case the real solutions of the form of \rf{5.17}
(where the evident substitution is done $u_1^2+3q \rightarrow u_1^2-3|q|$)
takes place if
\be{5.18}
\bar Q >0 \quad \Rightarrow \quad |\gamma |^{1/3} < \frac{3}{32|\Lambda_4|}\, .
\ee

Now, let us investigate zeros of the potential $U$. For $f'\neq 0 \Rightarrow X\neq -1$, the condition of zeros of the
potential \rf{1.8-1} is:
\be{5.19}
\bar R f'-f =0 \quad \Rightarrow \quad 3\gamma \bar R^4 +\alpha \bar R^2 +2\Lambda_D=0\, .
\ee
Therefore, zeros are defined by equation:
\be{5.20}
\bar R^2 = -\frac{\alpha}{6\gamma} \pm \left[\left(\frac{\alpha}{6\gamma}\right)^2-\frac{2\Lambda_D}{3\gamma}\right]^{1/2}\, .
\ee
Obviously, the necessary conditions for zeros are:
\be{5.21}
 \begin{array}{ccc}
\gamma > 0 \quad &\Rightarrow& \quad \Lambda_D \leq \alpha^2/(24\gamma)\, , \\
&\phantom{\int}&\\
\gamma < 0 \quad &\Rightarrow& \quad \Lambda_D \geq -\alpha^2/(24|\gamma |)\, .\\
\end{array}
\ee
Additionally, the positiveness of the r.h.s. of the equation \rf{5.20} should be checked.

Let us consider now asymptotical behavior of the potential $U(\theta)$.  Here, the limits $\theta \to \theta_{max}$ and $\theta \to -\infty$. In the former case it appears:
\be{5.22}
\theta \to \theta_{max} \quad \Rightarrow \quad U(\theta ) \to -\sign (f(\theta_{max}))\times \infty\, .
\ee
In the latter case:
\ba{5.23}
&&\theta \to -\infty \Rightarrow \nn\\
&&U(\theta )\sim \exp\left(\frac{8-D}{D-2}\; \theta \right)\to
\left\{\begin{array}{ccc}
+\infty  \,,\;\:\qquad D>8 \;; \\
\const >0 \,, D = 8 \;; \\
+0  \,,\quad 2<D<8 \;; \\
\end{array}\right.\nn\\
\ea
where Eqs. \rf{5.3} and \rf{5.4} have been used. This result coincides with conclusions of appendix 1 in \cite{ZhukSaidov2}.

\begin{figure}
  \center
    \includegraphics[width=3.4in,height=2.1in]{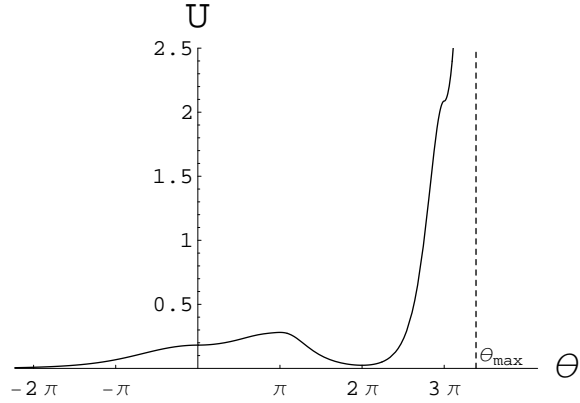}\\
\caption {The form of the potential \rf{1.8a1} as a function of $\theta$ in the case
$D=4$, $\Lambda_4=0.1$, $\gamma =1$ and $\alpha =-1$. For these values of the parameters, all extrema
correspond to the matching points $\theta = 0, \pi, 2\pi, 3\pi$. In the branching points $\theta =\pi,2\pi$ the potential has
local maximum and local non-zero minimum, respectively, and the monotonic points $\theta=0,3\pi$ are the inflection ones. Potential tends asymptotically to $+\infty$
when $\theta$ goes to $\theta_{max}$ and to zero when $\theta \rightarrow -\infty$.\label{pot}}
\end{figure}

To illustrate the described above properties, let us draw the potential $U(\theta )$ in Fig. \ref{pot} for the following parameters:
$D=4$, $\Lambda_4=0.1$, $\gamma =1$ and $\alpha =-1$.
These parameters contradict to the inequalities \rf{5.18} and \rf{5.21}. Therefore, $\theta = 0, \pi, 2\pi, 3\pi$ are
the only extremum points of the potential $U(\theta )$ and zeros are absent. These parameters are also satisfy the condition \rf{5.5}.
The absence of zeros means that all minima of the potential $U(\theta)$ are positive.

\begin{figure*}[htbp]
\center{\includegraphics [width = 3in, height = 2.1in ]{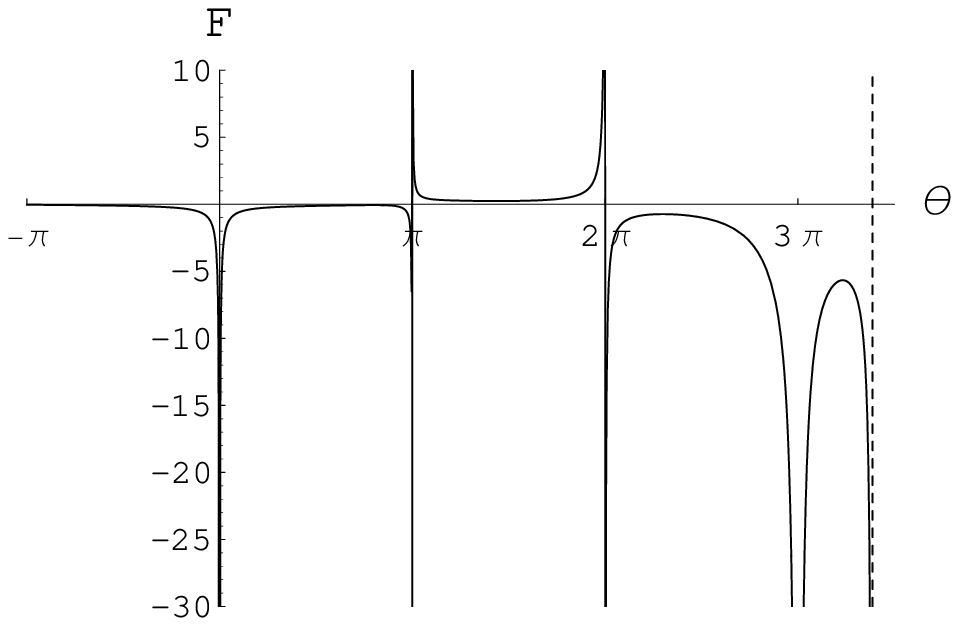}
\includegraphics [width = 3in, height = 2.1in ]{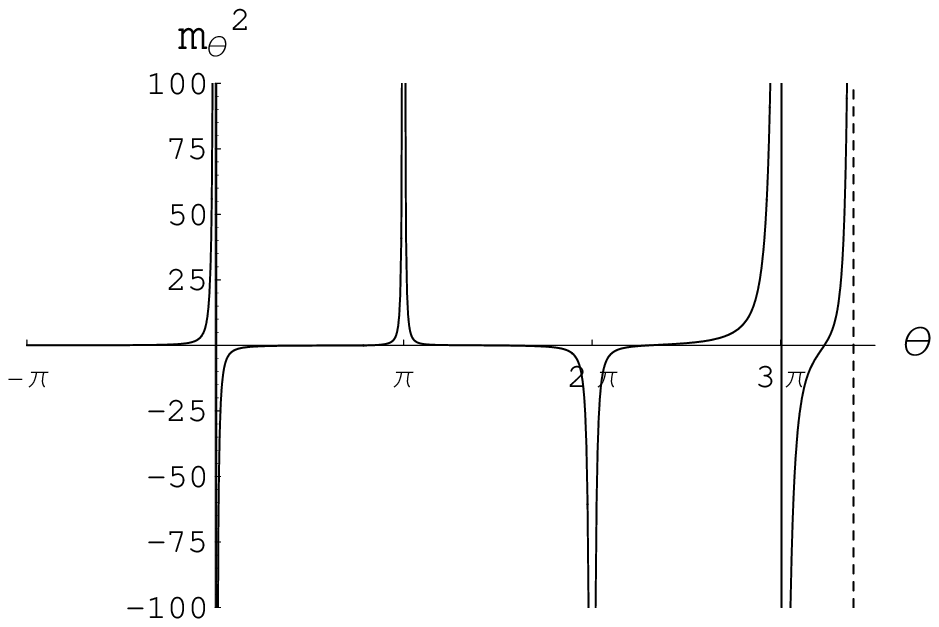}}
\caption{The effective force \rf{5.24} (left panel) and the mass squared \rf{5.25} (right panel) for the potential
$U(\theta )$ drawn in Fig. \ref{pot}. These pictures clearly show singular behavior of $F$ and $m^2_{\theta}$
in the matching points $\theta = 0, \pi, 2\pi, 3\pi$.
\label{force-mass}}
\end{figure*}

For the subsequent investigations, it is useful also to consider an effective force and mass squared of the field $\theta $. As it follows from Eq. \rf{5.9}, the effective force:
\be{5.24}
F= -G(\theta) \frac{dU}{d\theta}\, .
\ee
Varying Eq. \rf{5.9} with respect to field $\theta $ one obtains dynamical equation for small fluctuation $\delta \theta$ where
mass squared reads:
\be{5.25}
m^2_{\theta} = G(\theta )\frac{d^2 U}{d\theta^2}+\frac{d G(\theta)}{d\theta}\frac{dU}{d\theta }\, .
\ee
In Fig. \ref{force-mass} the effective force and the mass squared are presented as functions of $\theta$ for the potential drawn in Fig. \ref{pot}.
These figures
indicate that field $\theta $ may have very nontrivial behavior. This non-triviality  follows from two reasons. First, the field $\theta $ has non-canonical
kinetic term which result in appearing of non-flat moduli space metric $G(\theta )$ and derivative of $G(\theta )$ in Eq. \rf{5.9}. Second, the function $G(\theta ) $ has singular
behavior at the matching points $\theta = 0, \pi, 2\pi, 3\pi$.
Thus, the given intuition does not work
when one wants to predict dynamical behavior of fields with equations of the form of \rf{5.9} with potential drawn in Fig. \ref{pot}, especially when field approaches the matching points. It is necessary to solve equations analytically or to investigate them numerically.  Such analysis for the given model is performed in the next section where
the concentration is focused on the case,  where all extrema correspond to the matching points $\theta = 0, \pi, 2\pi, 3\pi$.

 \addcontentsline{toc}{subsection}{4.4.4. Dynamical behavior of the Universe and field $\theta$}
 \setcounter{subsection}{1}
 \medskip\medskip\medskip\medskip
 \textbf{4.4.4. Dynamical behavior of the Universe and field $\theta$}
  \medskip\medskip\medskip\medskip

Let us investigate dynamical behavior of scalar field $\theta$ and the
scale factor $a$ in more detail. There are no analytic
solutions for considered model. So, numerical calculations are used.
To do it, a Mathematica package proposed in \cite{KP} is applied and
adjusted to the given models and notations in appendix 3 of the paper \cite{SZPRD2009}. According to these notations, all dimensional quantities in the graphics
are given in the Planck units. Additionally, one should remember that metric on the moduli space is
not flat and defined in Eq. \rf{5.9}. For example, the canonical momenta and the kinetic energy read:
\ba{5.26}
P_{\theta} &=& \frac{a^3}{\kappa_4^2}G_{11}\dot \theta = \frac{a^3}{\kappa_4^2}\left(\frac{d\phi}{d\theta}\right)^2\dot\theta\, ,\\
\nn\\E_{kin} &=& \frac{1}{2\kappa_4^2} G_{11}\dot \theta^2=\frac{\kappa_4^2}{2a^6}G^{11}P_{\theta}^2 = \frac{1}{2\kappa_4^2} \left(\frac{d\phi}{d\theta}\right)^2\dot\theta^2\,,\nn
\ea
where $8\pi G \equiv \kappa_4^2$ and $G$ is four-dimensional Newton constant. To understand the dynamics of the Universe, one shall also draw the
Hubble parameter:
\be{5.27}
3 \left(\frac{\dot a}{a}\right)^2 \equiv 3 H^2 =
\frac12 G_{11} \dot \theta^2 +U(\theta)\,
\ee
and the acceleration parameter:
\be{5.28}
q \equiv \frac{\ddot a}{H^2 a} = \frac{1}{6H^2} \left(-4 \times
\frac12 G_{11} \dot \theta^2 + 2U(\theta)\right)\, .
\ee

Fig. \ref{force-mass} shows that the effective force changes its sign and the mass squared preserves the sign when $\theta$ crosses the branching points
$\pi, 2\pi$ and vise verse, the effective force preserves the sign and the mass squared changes the sign when $\theta$ crosses the monotonic points
$0, 3\pi$. Therefore, it make sense to consider these cases separately.

 \addcontentsline{toc}{subsection}{4.4.5. Branching points
$\theta =\pi, 2\pi$}
 \setcounter{subsection}{1}
 \medskip\medskip\medskip\medskip
 \textbf{4.4.5. Branching points
$\theta =\pi, 2\pi$}
  \medskip\medskip\medskip\medskip

\begin{figure*}[htbp]
\centerline{\includegraphics[width=3.0in,height=1.9in]{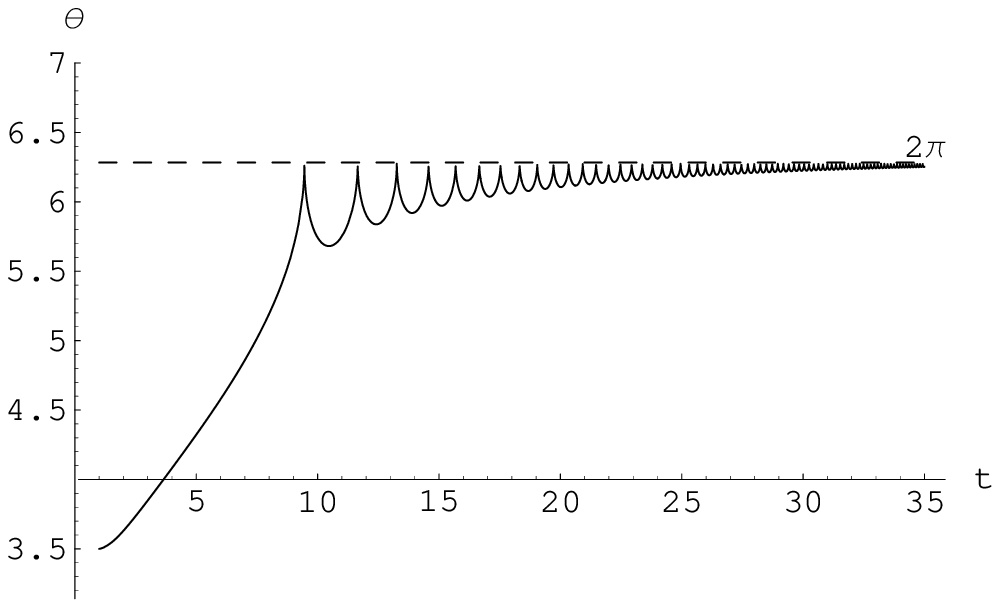}
\includegraphics[width=3.0in,height=1.9in]{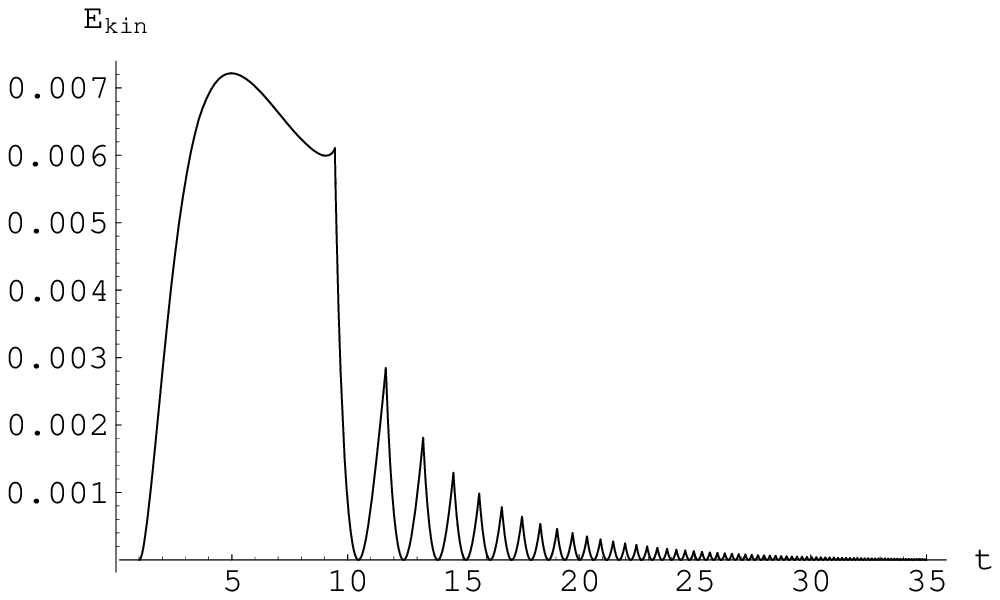}}
\caption {Dynamical behavior of scalar field $\theta (t)$ (left panel) and its
kinetic energy $E_{kin}(t)$ (right panel) in the vicinity of the branching point $\theta =2\pi$. \label{theta}}
\end{figure*}

First, the dynamical behavior of the Universe and scalaron is considered in the vicinity of the branching point $\theta =2\pi$ which is the local minimum
of the potential in Fig. \ref{pot}. The time evolution of scalaron field $\theta$ and its kinetic energy $E_{kin}$ are drawn in Fig. \ref{theta}.
Here and in all pictures below, the same parameters are used as in Fig. \ref{pot}. The time $t$ is measured in the Planck times and classical evolution starts at $t=1$.
For the initial value of $\theta$ it is taken that $\theta_{initial}=3.5$.
In Fig. \ref{H,a} the evolution of the logarithms of the
scale factor $a(t)$ (left panel) and the evolution of the Hubble
parameter $H(t)$ (right panel) is plotted.
\begin{figure*}[htbp]
\centerline{\includegraphics[width=2.1in,height=1.9in]{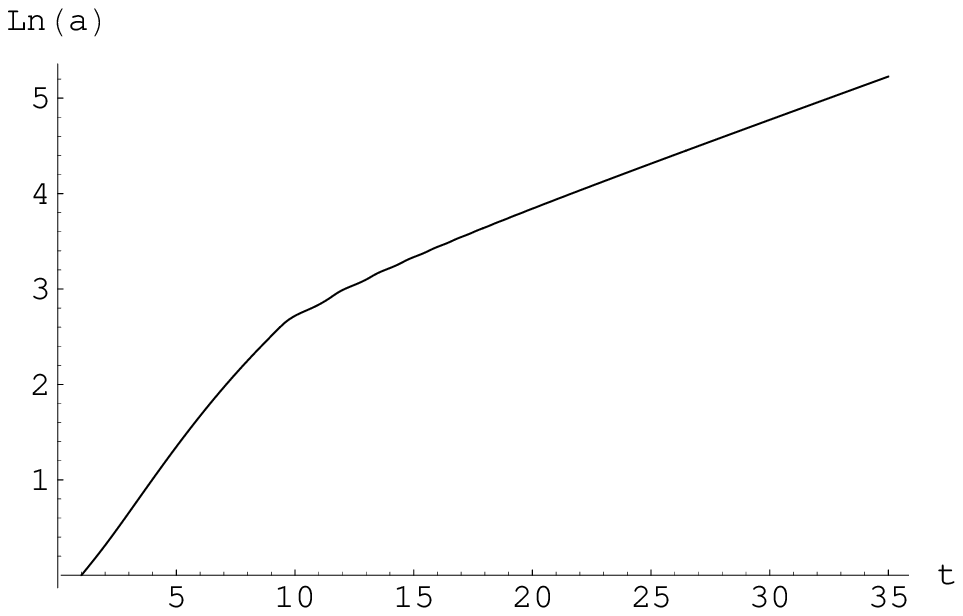}
\includegraphics[width=2.1in,height=1.9in]{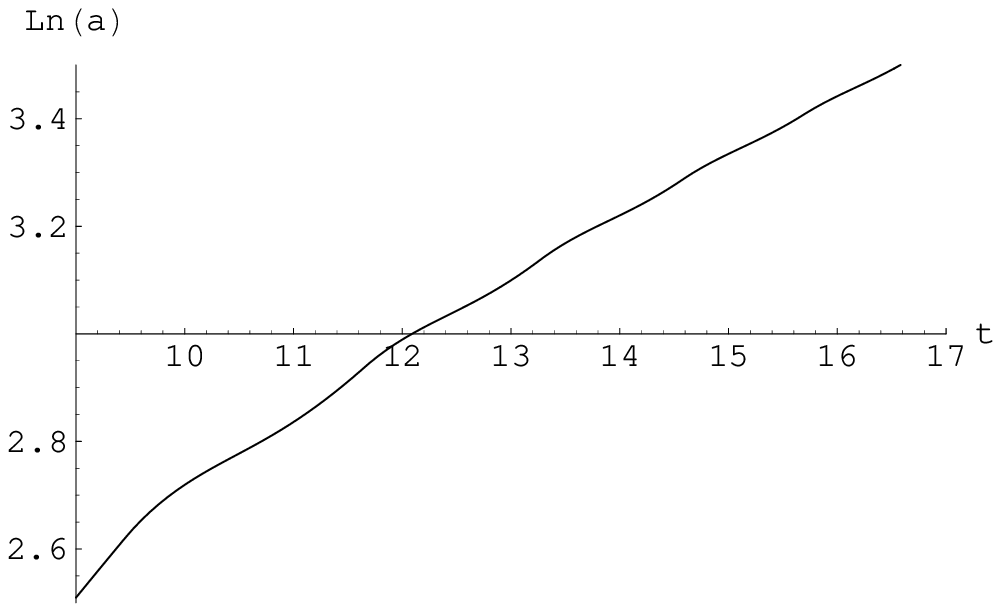}
\includegraphics[width=2.1in,height=1.9in]{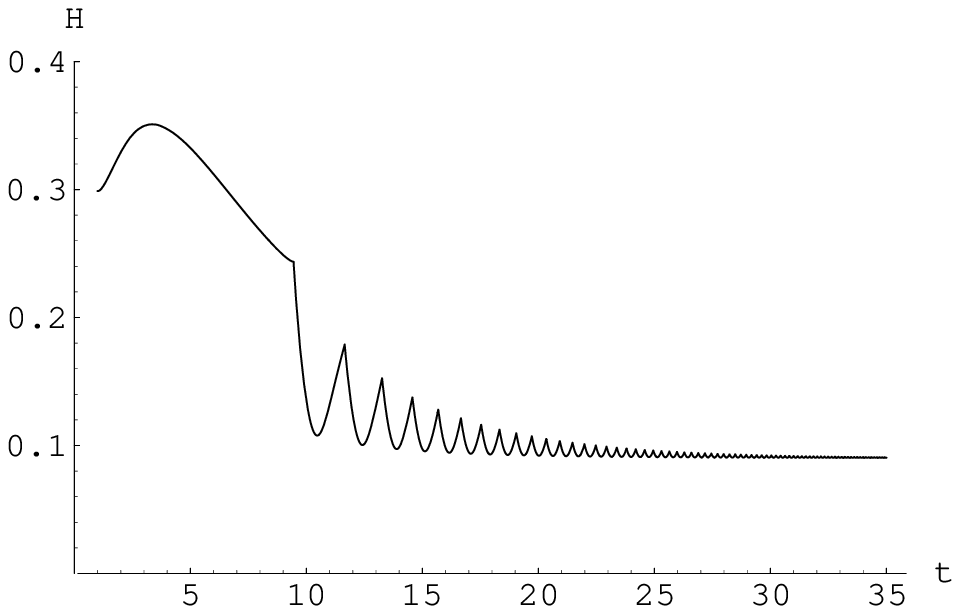}}
\caption {The time evolution of the logarithms of the scale factor $a(t)$ (left panel) and the Hubble
parameter $H(t)$ (right panel) for the trajectory drawn in Fig. \ref{theta}. A slightly visible oscillations of $\ln (a)$  (caused by bounces) can be seen  by magnification of this picture (middle panel).
\label{H,a}}
\end{figure*}
and in Fig. \ref{q} the evolution
of the parameter of acceleration $q(t)$ (left panel) and the equation of state parameter $\omega (t)= [2q(t)+1]/3 $ (right panel).
\begin{figure*}[htbp]
\centerline{\includegraphics[width=3.0in,height=2.0in]{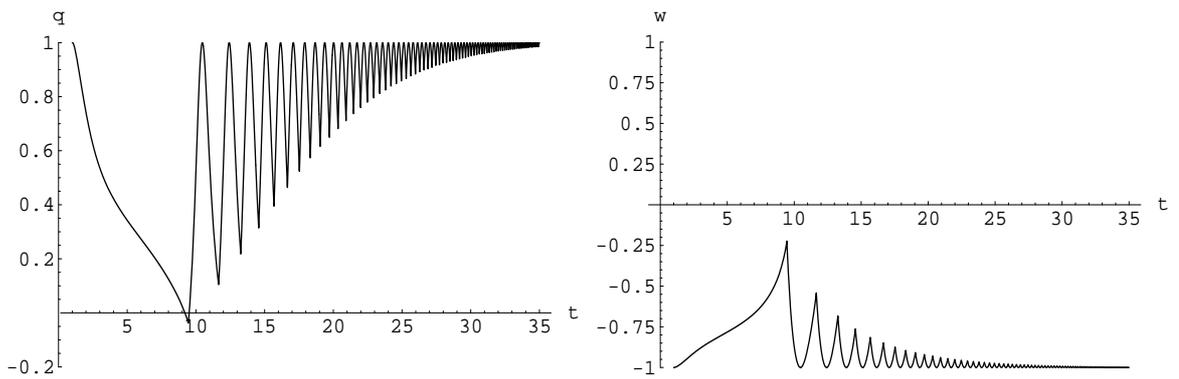}
\includegraphics[width=3.0in,height=2.0in]{w.eps}}
\caption {The parameter of acceleration $q(t)$ (left panel) and the equation of state parameter $\omega (t)$ (right panel) for the scale factor in Fig. \ref{H,a}.
\label{q}}
\end{figure*}

Fig. \ref{theta} demonstrates that scalar field $\theta$
bounces  an infinite number of times with decreasing amplitude
in the vicinity of the branching point
$\theta =2\pi$. $\theta$ cannot cross this point.  From Figs. \ref{H,a} and \ref{q} it can be seen that the Universe asymptotically approaches the de Sitter
stage: $H\rightarrow \const, \quad q \rightarrow +1, \quad$ and $\omega \rightarrow -1$. Such accelerating behavior can be called {\em bouncing inflation}.

Concerning the dynamical behavior in the vicinity of the branching point $\theta =\pi$, the analysis (similar performed above) shows that the scalaron field
$\theta $ cannot cross this local maximum regardless of the magnitude of initial velocity in the direction of $\theta=\pi$. It bounces back from this point.

 \addcontentsline{toc}{subsection}{4.4.6. Monotonic points
$\theta =0, 3\pi$}
 \setcounter{subsection}{1}
 \medskip\medskip\medskip\medskip
 \textbf{4.4.6. Monotonic points
$\theta =0, 3\pi$}
  \medskip\medskip\medskip\medskip

\begin{figure*}[htbp]
\centerline{\includegraphics[width=2.1in,height=2.0in]{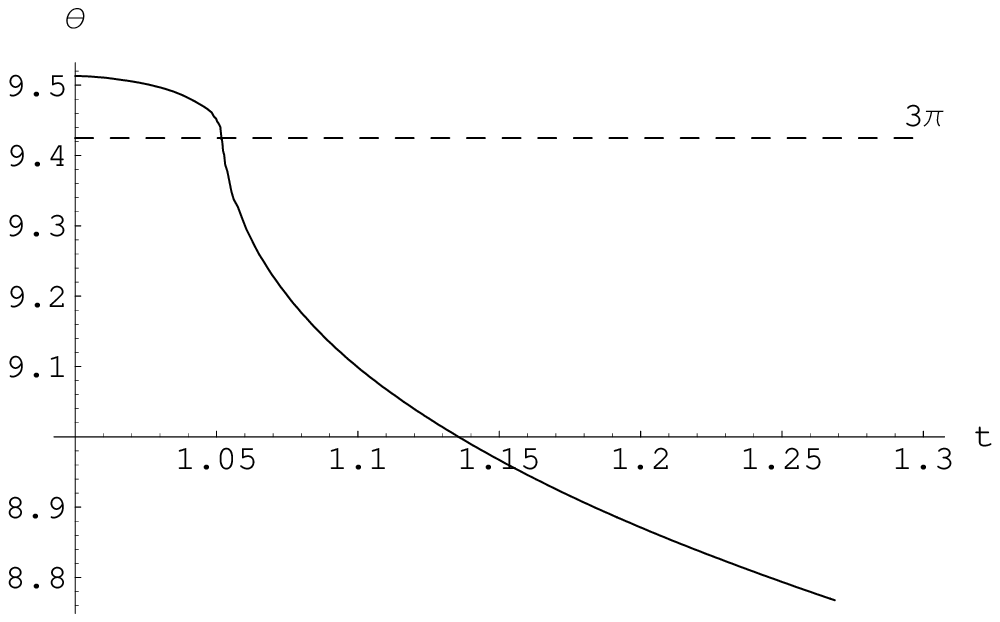}
\includegraphics[width=2.1in,height=2.0in]{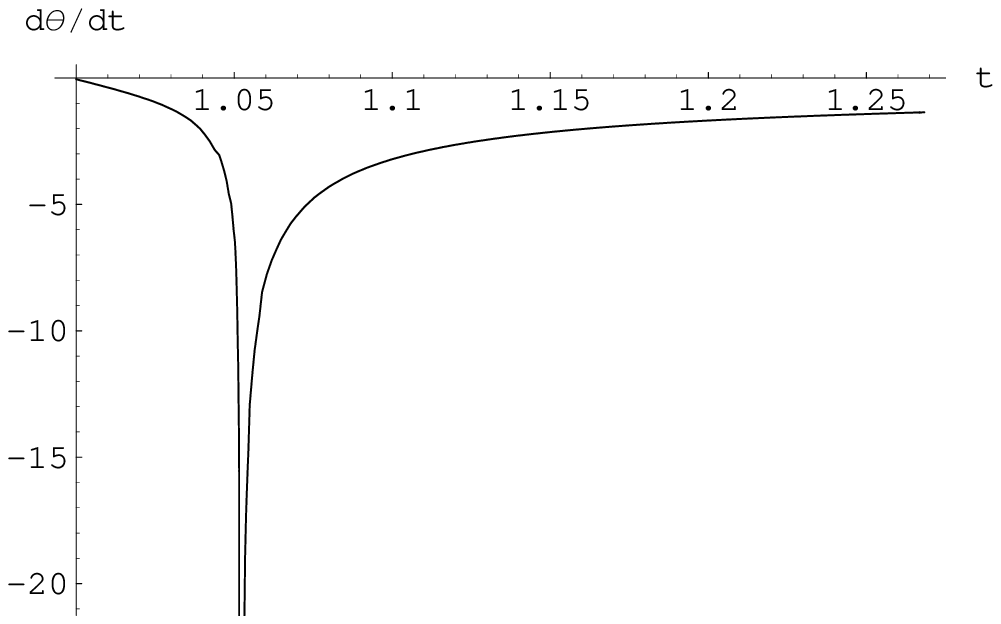}
\includegraphics[width=2.1in,height=2.0in]{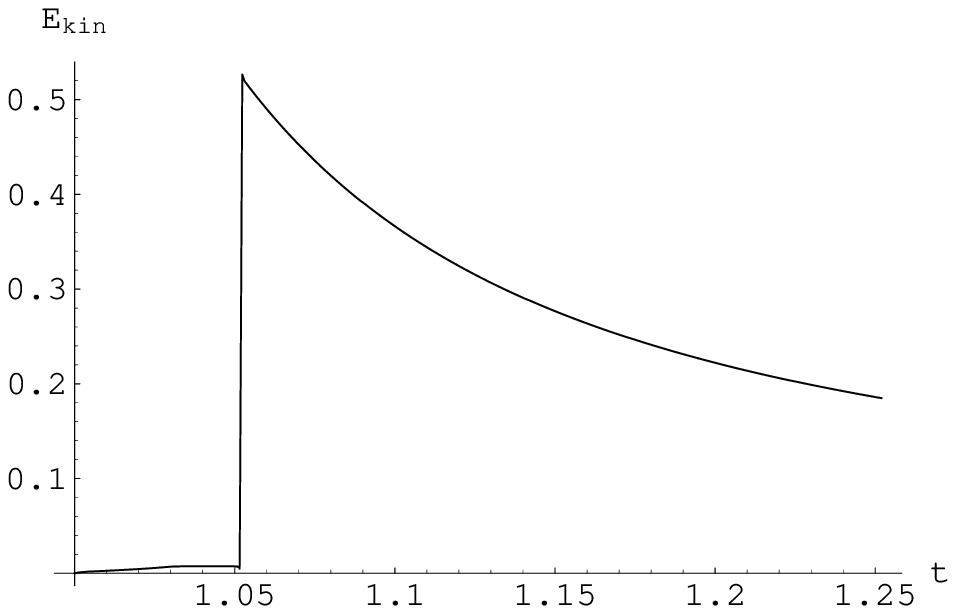}}
\caption {Dynamical behavior of scalar field $\theta (t)$ (left panel) and its time derivative $d\theta /dt$ (middle panel)
and kinetic energy $E_{kin}(t)$ (right panel) for the case of crossing of the inflection point $\theta =3\pi$.
\label{theta 3P}}
\end{figure*}

Now, let us investigate the dynamical behavior of the model in the vicinity of the monotonic points $\theta=0,3\pi$ which are the points
of inflection of the potential in Fig. \ref{pot}. Figs. \ref{pot} and \ref{force-mass} show that for both of these points
the model has the similar dynamical behavior. Therefore, for definiteness, let us consider the point $\theta=3\pi$.
To investigate numerically the crossing of the monotonic point $3\pi$, it is necessary to take very small value of a step $\Delta t$. It can be achieved
if very large value of the maximum number of steps is chosen. Thus, for the given value of the maximum number of steps, the closer to $3\pi$ the initial value $\theta_{initial}$ is taken the smaller step $\Delta t$ is obtained. For current calculation let us choose $\theta_{initial}=9.513$.
Fig. \ref{theta 3P} demonstrates that scalar field $\theta $ slowly crosses the monotonic point $3\pi$ with nearly zero kinetic energy\footnote{The derivative $d\theta/dt$ goes to $-\infty$ when $\theta \to 3\pi$ (with different speed on  different sides of $3\pi$) but $d\phi/d\theta =0$ at $3\pi$ and kinetic energy is finite (see Eq. \rf{5.26}).\label{theta 3Pi}}. Then, just after the crossing,
the kinetic energy has its maximum value and starts to decrees gradually when $\theta$ moves to the direction $2\pi$.
\begin{figure*}[htbp]
\centerline{\includegraphics[width=3.0in,height=2.0in]{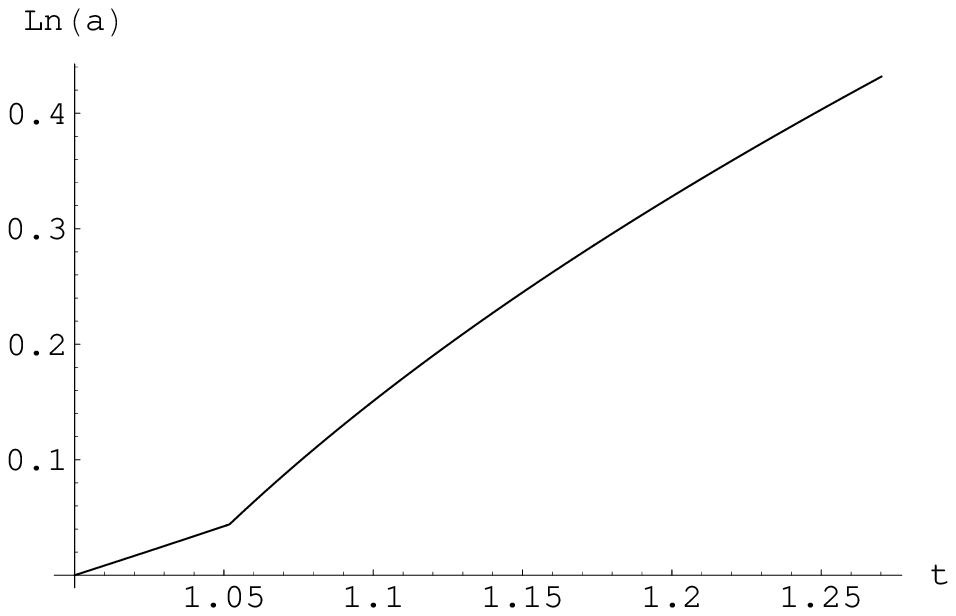}
\includegraphics[width=3.0in,height=2.0in]{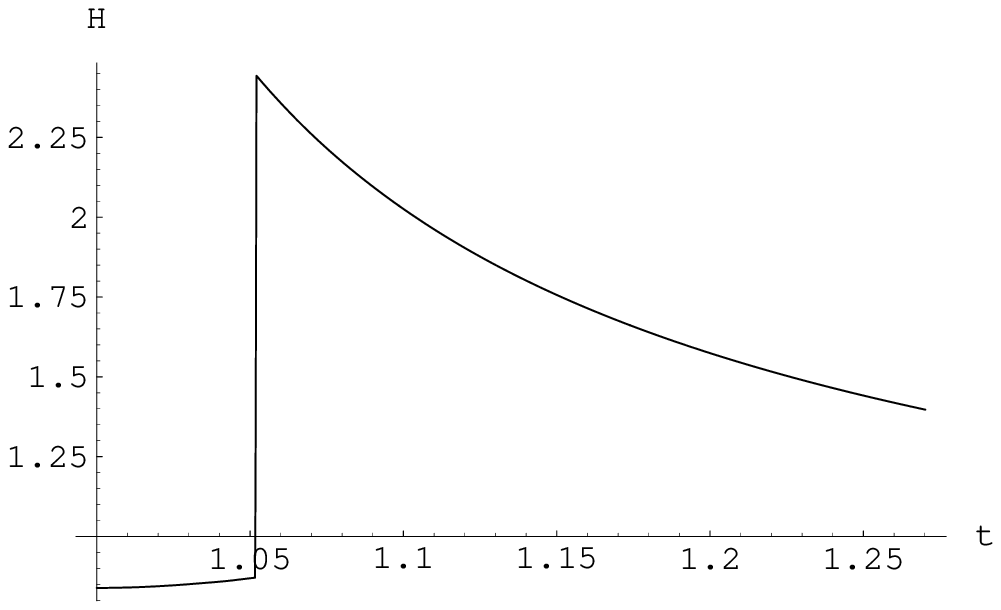}}
\caption {The time evolution of the logarithms of the scale factor $a(t)$ (left panel) and the Hubble
parameter $H(t)$ (right panel) for the trajectory drawn in Fig. \ref{theta 3P}.
\label{H,a3P}}
\end{figure*}
\begin{figure*}[htbp]
\centerline{\includegraphics[width=3.0in,height=2.0in]{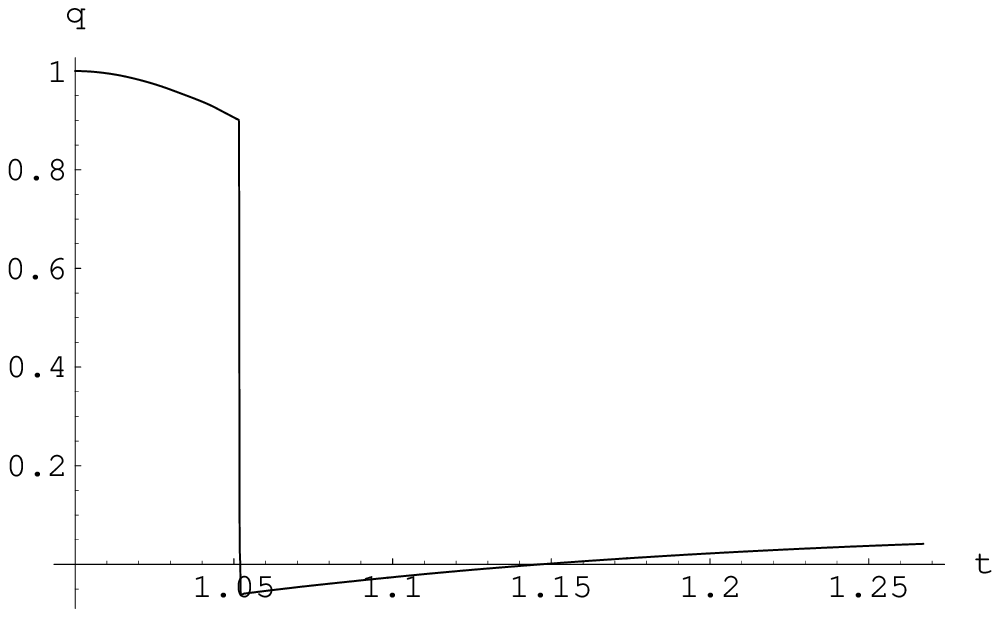}
\includegraphics[width=3.0in,height=2.0in]{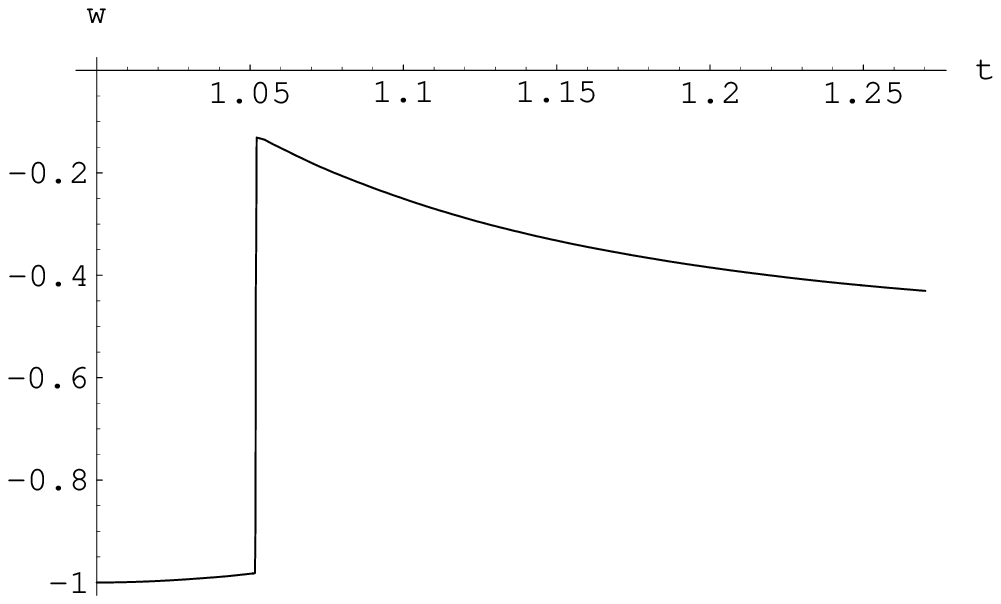}}
\caption {The parameter of acceleration $q(t)$ (left panel) and the equation of state parameter $\omega (t)$ (right panel) for the scale factor in Fig. \ref{H,a3P}.
\label{q3P}}
\end{figure*}
\begin{figure*}[!]
\centerline{\includegraphics[width=3.0in,height=2.0in]{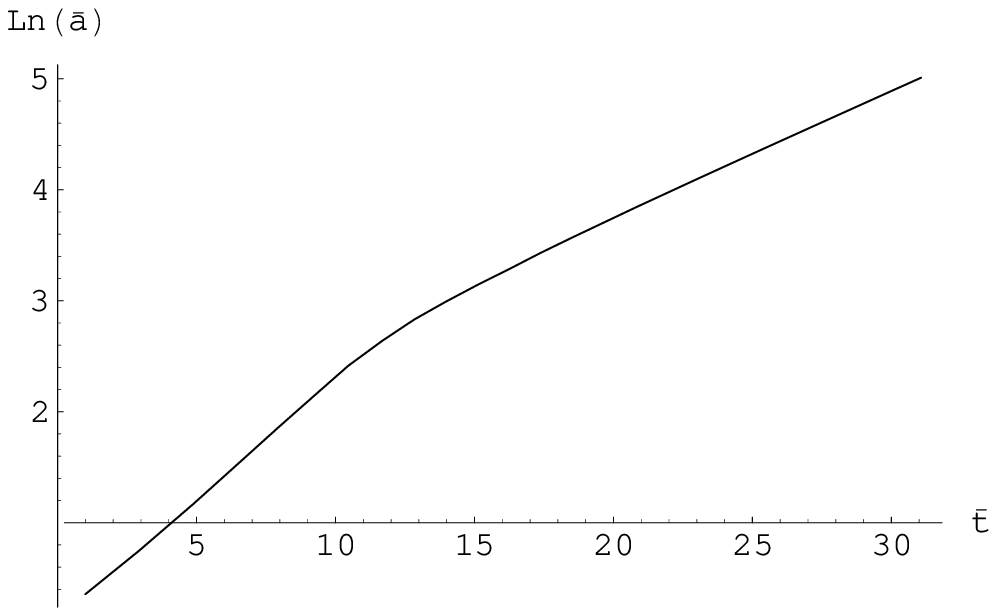}
\includegraphics[width=3.0in,height=2.0in]{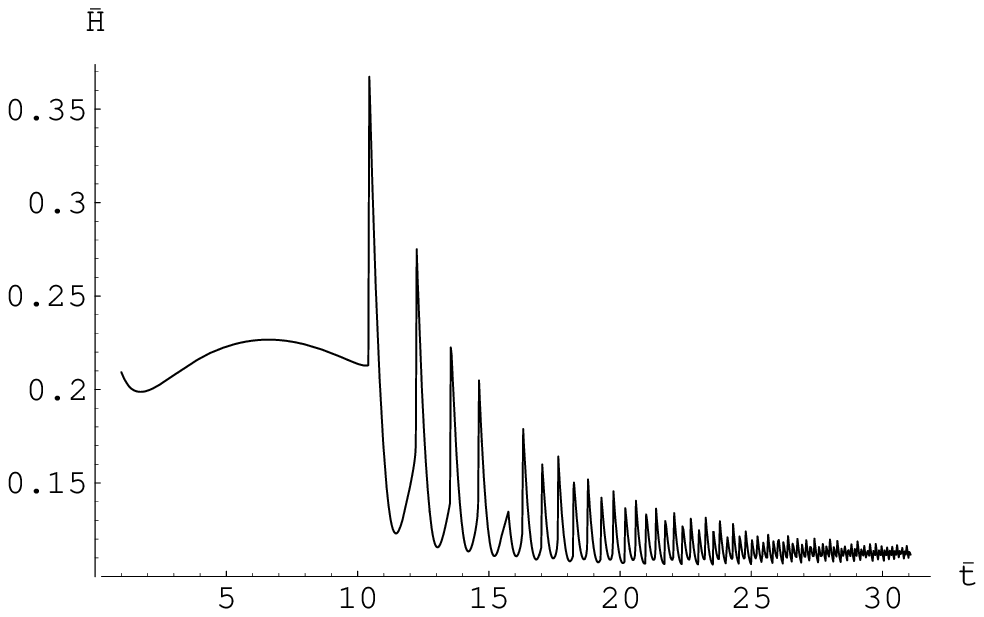}}
\caption {The time evolution of the logarithms of the scale factor $\bar a(\bar t)$ (left panel) and the Hubble
parameter $\bar H(\bar t)$ (right panel) for the trajectory drawn in Fig. \ref{theta} in the Brans-Dicke frame.
\label{BD}}
\end{figure*}
Figs. \ref{H,a3P} and \ref{q3P} demonstrate the behavior of the Universe before and after crossing $3\pi$. The vicinity of the branching point $2\pi$ is not shown because when $\theta$ approaches $2\pi$ the Universe  has the bouncing inflation described above. Hence, there are
3 phases sequentially: the short de Sitter-like stage during slow rolling in the vicinity of the
inflection point before crossing, then decelerating expansion just after the crossing with gradual transition to the accelerating stage again when $\theta$ approaches
the branching point $2\pi$.
Clearly that for another monotonic point
 $\theta=0$ the similar crossing behavior (without the bouncing stage when $\theta \to -\infty$) can be reached. Therefore, the monotonic
points $\theta=0$ and $\theta=3\pi$ are penetrable for the scalaron field $\theta$.

In appendix 4, it is shown that bouncing inflation in the vicinity of the branching point takes place
also in the Brans-Dicke frame.

 \medskip\medskip\medskip\medskip
\textbf{Summary}

 A possibility of inflation in
multidimensional cosmological models has been investigated. The main attention was paid to
nonlinear (in scalar curvature) models with quadratic $\bar{R}^2$ and
quartic $\bar{R}^4$ lagrangians. These models contain two scalar fields.
One of them corresponds to the scale factor of the internal space
and another one is related with the nonlinearity of the original
models. The effective four-dimensional potentials in these models
are fully determined  by the geometry and matter content of the
models. The geometry is defined by the direct product of the
Ricci-flat external and internal spaces. As a matter source, a monopole form field, D-dimensional bare cosmological
constant and tensions of branes located in fixed points have been included into consideration. The exact
form of the effective potentials depends on the relation between
parameters of the models and can take rather complicated view with a
number of extrema points.

First of all, it was found that a range of parameters which insures the
existence of zero minima of the effective potentials. These minima
provide sufficient condition to stabilize the internal space and,
consequently, to avoid the problem of the fundamental constant
variation. Zero minima correspond to the zero effective
four-dimensional cosmological constant. In general, one can also
consider positive effective cosmological constant which corresponds
to the observable now dark energy. However, it usually requires
extreme fine tuning of parameters of models.

Then, for corresponding effective potentials, the
possibility of the external space inflation was investigated. It have been shown that for
some initial conditions in the quadratic and quartic models can be
achieve up to 10 and 22 e-folds, respectively. An additionally bonus
of the considered model is that $R^4$ model can provide conditions
for the eternal topological inflation.

Obviously, 10 and 22 e-folds are not sufficient to solve the
homogeneity and isotropy problem but big enough to explain the
recent CMB data. To have the inflation which is long enough for
modes which contribute to the CMB, it is usually supposed that
$\triangle N \ge 15$ \cite{WMAP5}. Moreover, 22 e-folds is rather
big number to encourage the following investigations of the
nonlinear multidimensional models to find theories where this number
will approach  50-60. It can be seen that the increase of the
nonlinearity (from quadratic to quartic one) results in the increase
of $\triangle N$ in more that two times. So, there is a hope that
more complicated nonlinear models can provide necessary 50-60
e-folds.  Besides, this number is reduced in models where long
matter dominated (MD) stage followed inflation can subsequently
decay into radiation \cite{LiddleLyth,BV}. Precisely this scenario
takes place for given models. It also have been shown for quadratic and quartic
nonlinear models, that MD stage with the external scale factor
$a\sim t^{2/3}$ takes place after the stage of inflation. It happens
when scalar fields start to oscillate near the position of zero
minimum of the effective potential.

For the investigation of the dynamical behavior of the scalaron field $\phi$ and the Universe in nonlinear model
with curvature-squared and curvature-quartic correction terms: , the parameters $\alpha$ and $\gamma$ are chosen in such a way that the scalaron potential $U(\phi )$ is a multi-valued function consisting of a number of branches. These branches are fitted with each other either in the branching points (points $P_{2,3}$ in Fig. \ref{U(X)}) or in the monotonic points (points $P_{1,4}$ in Fig. \ref{U(X)}). The potential $U$ was optimized so that a way that it becomes the one-valued function of a new field variable $\theta =\theta(\phi)$ (see Fig. \ref{pot}). This has enabled to consider the dynamical behavior of the system in the vicinity of the branching and monotonic points in ($D=4$)-dimensional space-time. This investigation shows that the monotonic points are penetrable for scalaron field (see Figs. \ref{theta 3P}-\ref{q3P}) while in the vicinity of the branching points scalaron has the bouncing behavior and cannot cross these points. Moreover, there are branching points
where scalaron bounces  an infinite number of times with decreasing amplitude and the Universe asymptotically approaches the de Sitter stage (see Figs. \ref{theta}-\ref{q}). Such accelerating behavior is called bouncing inflation. It should be noted that for this type of inflation there is no need for original potential $U(\phi)$ to have a minimum or to check the slow-roll conditions. A necessary condition is the existence of the branching points. This is a new type of inflation. It is shown that this inflation takes place both in the Einstein and Brans-Dicke frames. It is found that this type of inflation for the model with the curvature-squared and curvature-quartic correction terms which play an important role during the early stages of the Universe evolution. However, the branching points take also place in models with $\bar R^{-1}$-type correction terms \cite{Frolov}. These terms play an important role at late times of the evolution of the Universe.

 There is no need for fine tuning of the initial conditions to get the bouncing inflation. In Figs. \rf{theta}-\rf{q}, for definiteness the initial conditions are chosen as $\theta =3.5$ and $E_{kin}=0$. However, the calculations show that these figures do not qualitatively change if one takes arbitrary $\theta \in (\pi,2\pi)$ and non-zero $E_{kin}$. Fig. \ref{force-mass} indicates that the minimum at $\theta =2\pi$ is stable with respect to tunneling through the barrier at this point. The situation is similar to the quantum mechanical problem with infinitely high barrier. It has been stressed already that the form of the potential Fig. \ref{pot} is not sufficient to predict the dynamical behavior of $\theta$. This field has very non-trivial behavior because of non-canonical kinetic term and singular (at the matching points) non-flat moduli space metric $G(\theta )$. Therefore, it is impossible to "jump" quantum mechanically from one branch to another. It is impossible to apply to given dynamical system the standard tunneling approach (e.g., in \cite{Coleman}). This problem needs a separate investigation. It is worth of noting that the Universe with a bounce preceding the inflationary period was considered in \cite{Yifu} where it was shown that due to a bounce the spectrum of primordial perturbations has the characteristic features. It indicates that the similar effect can take place in this model.


\newpage

 \section*{\center{CONCLUSION}}
 \addcontentsline{toc}{section}{CONCLUSION}\label{sec-summary}

\medskip\medskip
Several nonlinear multidimensional gravitational models are studied in this Thesis from the point of freezing stabilization and compactification of additional dimensions; and possibility of inflation. The main results may be constituted as follows.

Two separate branches of the
four-dimensional spaces are produced for multidimensional nonlinear models with the scalar
curvature of type $\bar R^{-1}$ after
dimensional reduction. For these type of models, simultaneous compactification and late-time acceleration is unreachable, due to branching of potential.

The pure gravitational model with curvature-quadratic and
curvature-quartic correction terms allows too the stable compactification of the internal space
for certain interval of the parameters (regions of
stability). Assuming the scale factor of the stabilized internal space to be in
order of the Fermi length: $b_{(0)1} \sim L_F \sim 10^{-17}$cm,
one arrive that $\alpha \sim L_F^2$ and for the effective cosmological
constant and masses folds: $-\Lambda_{eff}\sim m^{2}_{1}\sim
m^{2}_{\phi}\sim 1\mbox{TeV}^2$.

Adding forms to the model of $\bar R^{-1}$gives possibility to obtain positive minimum of the effective potential. In this model the freezing
stabilization of the internal spaces is provided. This allows to avoid the
problem of the fundamental constant variation in multidimensional
models \cite{Zhuk(IJMP),BZ}. At the same time, the stage
of the cosmic acceleration is achieved too. However, fine tuning is required to get parameters of the present-day
accelerating expansion.

Additionally the existence of domain walls was found for this model, which separates regions
with different vacua in the Universe without providing of inflation, because the effective potential is not flat enough around the saddle point. In case of parameter $z< 2$ the minimums of the potential is
metastable, i.e. the quantum tunneling is possible both in $\phi$ and in $\varphi$
directions. For $z\geq 2$, tunneling in $\varphi$ direction is
still possible due to $U_{eff}(\varphi ,\phi_0) \approx e^{b\varphi}
U(\phi_0) \rightarrow 0$ for $\varphi \rightarrow -\infty$ which is
less than any positive $\Lambda_{eff}$. This could result in the
materialization of bubbles of the new phase in the metastable one
(see e.g. \cite{Rubakov}). Hence, late-time acceleration is possible
only if characteristic lifetime of the metastable stage is greater
than the age of the Universe. For this model the global negative minimum
is also possible. This minimum is stable both in
classical and quantum limits. However, the acceleration is absent
because of its negativity.

Another feature of this model consists in multi-valued form of the effective potential, which was two branches for for each choice of parameter $\mu$. This gives possibility
for investigation of transitions from one branch to another one by analogy
with catastrophe theory or similar to the phase transitions in
statistical theory. In this
particular model the point $\phi =0$ corresponds to the singularity
$\bar R, R \to \pm \infty$. Indicating that the analog of the second order
smooth phase transition through the point $\phi =0$ is impossible in
this model. The analog
of the first order transition via quantum jumps is possible too.

The inflation has been investigated in the models with nonlinearities of type  $\bar{R}^2$ and $\bar{R}^4$ with a monopole form field, D-dimensional bare cosmological
constant and tensions of branes located in fixed points as source.
As result, the equivalent linear model has two scalar fields: one corresponds to the scale factor of the internal space and another the the nonlinearity of the original
models. Up to 10 and 22 e-folds can be achieved for  $\bar{R}^2$ and  $\bar{R}^4$ respectively, although it is not sufficient to solve the problems of
homogeneity and isotropy in
recent CMB data, where 50-60 e-folds are needed. This number is reduced in case if inflation, followed by long
matter dominated stage, can subsequently
decay into radiation \cite{LiddleLyth,BV}. This scenario takes
place for given models, as have been shown for $\bar{R}^2$ and  $\bar{R}^4$ models: matter dominated stage with the external scale factor
$a\sim t^{2/3}$ occurs after the stage of inflation. It happens
when scalar fields start to oscillate near the position of zero
minimum of the effective potential. Another significant problem for these models
consists in the spectral index. For example, in the case of $\bar{R}^4$
model, the spectral index: $n_s\approx 1+2\eta|_{\chi_{2(+)}}\approx 0.61$ which
is less than observable now $n_s\approx 1$. As possible solution of this problem, more complicated function
$f(\bar{R})$ can be used. For example, in \cite{Ellis} it is shown that quadratic and quartic nonlinearities
lead to increase of $n_s$.


As next step, the dynamical behavior of the scalaron field $\phi$ and the Universe in quadratic and quartic nonlinear model was investigated. In this case, the scalaron potential $U(\phi )$ can be a multi-valued function consisting of a number of branches, being connected in the branching points or in the monotonic points. Further reparametrization of the potential $U$ to the one-valued function of a new field variable $\theta =\theta(\phi)$, enables consideration of the dynamical behavior of the system in the vicinity of the branching and monotonic points. It appears that the monotonic points are penetrable for scalaron field while in the vicinity of the branching points scalaron has the bouncing behavior and cannot cross these points. Also branching points where found,
where scalaron bounces  an infinite number of times with decreasing amplitude, approaching asymptotically de Sitter stage. This king of accelerating behavior is called bouncing inflation. For this type of inflation there is no need for original potential $U(\phi)$ to have a minimum or to check the slow-roll conditions, i.e. no need for fine tuning. The existence of the branching points is only condition imposed. The minimum at $\theta =2\pi$ is stable for tunneling through the barrier at this point.

This represents a new type of inflation, which takes place both in the Einstein and Brans-Dicke frames and play an important role during the early stages of the Universe evolution. In the model of nonlinearity $\bar R^{-1}$ \cite{Frolov} the branching points also take place, playing an important role at late times of the evolution of the Universe. Therefore, bouncing inflation may be responsible for the late-time accelerating expansion of the Universe.



 \newpage
 \addcontentsline{toc}{section}{APPENDIXES}
 \section*{\center{APPENDIXES}}
 \setcounter{section}{0}
 \setcounter{equation}{0}

\medskip\medskip
 \addcontentsline{toc}{subsection}{Appendix 1. The critical dimensions in $f(\bar{R})$ theories\label{sec1}}
 \setcounter{subsection}{1}
 \medskip\medskip\medskip\medskip
 \textbf{Appendix 1. The critical dimensions in $f(\bar{R})$ theories }
  \medskip\medskip\medskip\medskip

\renewcommand{\theequation}{A1.\arabic{equation}}
\setcounter{equation}{0}

The existence of a critical dimension (in this case $D=8$) is a
rather general feature of gravitational theories with polynomial
scalar curvature terms (see, e.g., Refs. \cite{BC,paul,BMS}).
Following the paper \cite{GZBR}, it can be easily demonstrated for a
model with curvature nonlinearity of the type
\be{a6} f(\bar R)=\sum_{k=0}^N a_k \bar R^k \ee
for which the ansatz
\be{a7} e^{A\phi}=f'=\sum_{k=0}^N k a_k \bar R^{k-1} \ee
leads, similar like \rf{1.8a1}, to a potential
\be{a8} U(\phi)=\frac12 \left(f'\right)^{-D/(D-2)}\sum_{k=0}^N
(k-1)a_k \bar R^k. \ee
The condition of extremum \rf{1.11a} for this potential reads:
\be{a8-1} Df - 2\bar R f' = \sum_{k=0}^N \left(D-2k\right)a_k \bar
R^k =0\, . \ee
Thus, at the critical dimension $D=2N$ the degree of this equation
is reduced from $N$ to $N-1$. In this case the search of extrema is
considerably simplified.

In the limit $\phi\to +\infty$ the curvature will behave like $\bar
R\approx c e^{h \phi}$ where $h$ and $c$ can be defined from the
dominant term in \rf{a7}:
\be{a8-1a}e^{A\phi}\approx N a_N \bar R^{N-1}\approx N a_N c^{N-1}
e^{(N-1)h\phi}. \ee
Here the requirement $f'>0$ allows for the following sign
combinations of the coefficients $a_N$ and the curvature asymptotics
$\bar R (\phi\to \infty)$:
\ba{a8-2} N=2l: & \qquad & \sign [a_N ] = \sign [\bar R (\phi\to
\infty)]\nn\\
N=2l+1: & & a_N >0, \quad \sign [\bar R (\phi\to \infty)]=\pm 1. \ea
The other combinations, $N=2l: \ \ \sign [a_N ] =  -\sign [\bar R
(\phi\to \infty)]$, \  $N=2l+1: \ \ a_N <0, \ \sign [\bar R (\phi\to
\infty)]=\pm 1$, would necessarily correspond to the $f'<0$ sector,
so that the complete consideration should be performed  in terms of
the extended conformal transformation technique of Ref.
\cite{Maeda}. Such a consideration is out of the scope currently and the focus is restricted to the cases \rf{a8-2}.
The coefficients $h$ and $c$ are then easily derived as $h=A/(N-1)$
and $c=\sign(a_N)\left|N a_N\right|^{-\frac{1}{N-1}}$. Plugging this
into \rf{a8} one obtains
\be{a9} U(\phi\to +\infty)\approx
\sign(a_N)\frac{(N-1)}{2N}\left|Na_N\right|^{-\frac{1}{N-1}}
e^{-\frac{D}{D-2}A\phi}e^{\frac{N}{N-1}A\phi} \ee
and that the exponent \be{a10} \frac{D-2N}{(D-2)(N-1)}A \ee changes
its sign at the critical dimension $D=2N$:
\be{a11} U(\phi \to +\infty)\to
\sign(a_N)\frac{(N-1)}{2N}\left|Na_N\right|^{-\frac{1}{N-1}}\times
\left\{ \begin{array}{c}
  \infty \\
  1 \\
  0
\end{array}\right. \quad \begin{array}{c}
  \mbox{for}\ D>2N\, , \\
  \mbox{for}\ D=2N\, , \\
  \mbox{for}\ D<2N\, .
\end{array}
\ee

This critical dimension $D=2N$ is independent of the concrete
coefficient $a_N$ and is only defined by the degree $\deg_{\bar R}
(f)$ of the scalar curvature polynomial $f$. From the asymptotics
\rf{a11} follows that in the high curvature limit $\phi\to
+\infty$, within given oversimplified classical framework, the
potential $U(\phi)$ of the considered toy-model shows asymptotical
freedom for subcritical dimensions $D<2N$, a stable behavior for
$a_N>0$, $D>2N$ and a catastrophic instability for $a_N<0$, $D>2N$.
It should be noted that this general behavior suggests a way how to cure a
pathological (catastrophic) behavior of polynomial $\bar
R^{N_1}-$nonlinear theories in a fixed dimension $D>2N_1$: By
including higher order corrections up to order $N_2>D/2$ the theory
gets shifted into the non-pathological sector with asymptotical
freedom. More generally, one is even led to conjecture that the
partially pathological behavior of models in supercritical
dimensions could be an artifact of a polynomial truncation of an
(presently unknown) underlying non-polynomial $f(\bar R)$ structure
at high curvatures
--- which probably will find its resolution in a strong coupling
regime of $M-$theory or in loop quantum gravity.

 \addcontentsline{toc}{subsection}{Appendix 2. On the particular properties of potential $U_{eff}$\label{sec2}}
 \setcounter{subsection}{1}
 \medskip\medskip\medskip\medskip
 \textbf{Appendix 2. Self-similarity condition}
  \medskip\medskip\medskip\medskip

\renewcommand{\theequation}{A2.\arabic{equation}}
\setcounter{equation}{0}

Due to the zero minimum conditions $U(\phi_{min})=f_1^2=\lambda/2$,
the effective potential \rf{3.1} can be written in the form:
\ba{1b} &&U_{eff}(\varphi ,\phi )=U(\phi_{min})e^{-\,
\sqrt{\frac{2d_1}{d_1+2}}\;
\varphi}\nn\\
&&\times\left[\frac{U(\phi )}{U(\phi_{min})} + e^{-2\,
\sqrt{\frac{2d_1}{d_1+2}}\; \varphi} -2 e^{-\,
\sqrt{\frac{2d_1}{d_1+2}}\; \varphi} \right]\;.\qquad \ea
Exact expressions for $U(\phi)$ \rf{3.7} and \rf{4.2} indicate that
the ratio
\be{2b} \frac{U(\phi )}{U(\phi_{min})}=F(\phi,k,d_1) \ee
depends only on $\phi, k$ and $d_1$. Dimensionless parameter
$k=\xi\Lambda_D$ for the quadratic model and $k=\gamma \Lambda_D^3$
for the quartic model. In Eq. \rf{2b} one have to take into account that
$\phi_{min}$ is a function of $k$ and $d_1$:
$\phi_{min}=\phi_{min}(k,d_1)$.  Then, $U(\phi_{min})$ defined in
Eqs. \rf{3.7} and \rf{4.2} reads:
\be{3b} U(\phi_{min})=\Lambda_D\tilde F(\phi_{min}(k,d_1),k,d_1)\, .
\ee
Therefore, parameters $k$ and $d_1$ determine fully the shape of the
effective potential, and parameter $\Lambda_D$ serves for conformal
transformation of this shape. This conclusion is confirmed with the fact that all extrema  depend
only on $k$ and $d_1$. Thus, figures \ref{effpotr2}, and
\ref{effpotr4} for contour plots  are defined by $k$ and $d_1$ and
will not change with $\Lambda_D$. From the definition of the slow
roll parameters it is clear that they also do not depend on the
hight of potentials and in the given model depend only on $k$ and $d_1$
(see figures \ref{w1} and \ref{f1}). Similar dependence takes place
for difference $\Delta \phi =\phi_{max}-\phi_{min}$ drawn in Fig.
\ref{n}. Thus the conclusions concerning the slow roll and
topological inflations are fully determined by the choice of $k$ and
$d_1$ and do not depend on the hight of the effective potential, in
other words, on $\Lambda_D$. So, for fixed $k$ and $d_1$ parameter
$\Lambda_D$ can be arbitrary. For example, one can take $\Lambda_D$
in such a way that the hight of the saddle point $\chi_{2(+)}$ will
correspond to the restriction on the slow roll inflation potential
(see e.g. \cite{Lyth}) $U_{eff}\lesssim 2.2 \times
10^{-11}M_{Pl}^4$, or in the notations  $U_{eff}\lesssim 5.5\times
10^{-10}M_{Pl}^2$.

Above, the figure is indicated which (for given $k$ and $d_1$) do not
depend on the hight of the effective potential (on $\Lambda_D$).
What will happen with dynamical characteristics drawn in figures
\ref{fields4}, \ref{H4} and \ref{q4} (and analogous ones for the
quadratic model) if keeping fixed $k$ and $d_1$,  will change
$\Lambda_D$? In other words, the positions of the extrema
points (in $(\varphi,\phi)$-plane)are kept, but the hight of extrema experience change.
One can easily answer this question using the self-similarity
condition of the Friedmann equations. Let the potential $U$ in Eqs.
\rf{2} and \rf{3} be transformed conformally: $U \to c\, U$ where
$c$ is a constant. Next, let us introduce a new time variable $\tau
:= \sqrt{c}\,  t $. Then, from Eqs. \rf{2}-\rf{5} follows that the
Friedmann equations have the same form as for the model with
potential $U$ where time $t$ is replaced by time $\tau$. This
condition one may call the self-similarity. Thus, if in the model one
changes the parameter $\Lambda_D:\; \Lambda_D\to c\, \Lambda_D$, it
results (for fixed $k$ and $d_1$) in rescaling of all dynamical
graphics (e.g. Figures \ref{fields4} - \ref{q4}) along the time axis
in $1/\sqrt{c}$ times (the decrease of $\Lambda_D$ leads to the
stretch of these figures along the time axis and vice versa the
increase of $\Lambda_D$ results in the shrink of these graphics).
Numerical calculations confirm this conclusion. The property of the
conformal transformation of the shape of $U_{eff}$ with change of
$\Lambda_D$ for fixed $k$ and $d_1$ can be also called as the
self-similarity.

 \addcontentsline{toc}{subsection}{Appendix 3. Multi-component scalar field model\label{sec3}}
 \setcounter{subsection}{1}
 \medskip\medskip\medskip\medskip
 \textbf{Appendix 3. Multi-component scalar field model}
  \medskip\medskip\medskip\medskip

\renewcommand{\theequation}{A3.\arabic{equation}}
\setcounter{equation}{0}

Let us consider $n$ scalar fields minimally coupled to gravity in four
dimensions.  The effective action of this model reads \ba{1} &S& =
\frac{1}{16\pi G}\int d^4x \sqrt{|\tilde g^{(0)}|} \left(R[\tilde
g^{(0)}]\right.\nn\\&-&\left. G_{ij}\tilde g^{(0)\mu\nu}
\partial_{\mu}\varphi^i\partial_{\nu}\varphi^j -2 U(\varphi^1,\varphi^2,\ldots)\right)
\ea
where the kinetic term is taken in the canonical form:
$G_{ij}=\mbox{diag}(1,1,\ldots )$ (flat $\sigma$ model). Such
multi-component scalar fields originate naturally in
multidimensional cosmological models (with linear or nonlinear
gravitational actions) \cite{GZ1,GZ2,GMZ1}. Here, the the usual
conventions $c=\hbar =1$ is used, i.e. $L_{Pl}=t_{Pl}=1/M_{Pl}$ and $8\pi G
= 8\pi /M^2_{Pl}$. Here, scalar fields are dimensionless
and potential $U$ has dimension $[U] = \mbox{lehgth}^{-2}$.

Because the aim is to investigate dynamical behavior of our Universe in
the presence of scalar fields, let us suppose that scalar fields are
homogeneous: $\varphi^i=\varphi^i(t)$ and four-dimensional metric is
spatially-flat Friedmann-Robertson-Walker one: $\tilde g^{(0)}=
-dt\otimes dt + a^2(t)d\vec{x}\otimes d\vec{x}$.

For energy density and pressure follows:
\ba{2} \rho &=& \frac{1}{8\pi G}\left(\frac12 G_{ij}
\dot\varphi^i\dot\varphi^j +U\right)\,, \nn\\
P&=& \frac{1}{8\pi G}\left(\frac12 G_{ij} \dot\varphi^i\dot\varphi^j - U\right)\,;\\
&\Longrightarrow&
\left\{%
\begin{array}{ll}
  \frac12 G_{ij} \dot\varphi^i\dot\varphi^j= 4\pi
G(\rho+P)\;,\\\\
U= 4\pi G(\rho-P)\;.\label{3}\end{array}%
\right. \ea

The Friedmann equations  for considered model are
\be{4} 3 \left(\frac{\dot a}{a}\right)^2 \equiv 3 H^2 = 8\pi G \rho
= \frac12 G_{ij} \dot\varphi^i\dot\varphi^j +U\, , \ee
and
\be{5} \dot H = -4\pi G (\rho +P) = - \frac12 G_{ij}
\dot\varphi^i\dot\varphi^j\, . \ee
From these two equations, one obtains the following expression for the
acceleration parameter:
\ba{6} q &\equiv& \frac{\ddot a}{H^2 a} =  1 -\frac{4\pi G}{H^2}
(\rho + P) = -\frac{8\pi G}{6H^2} (\rho +3P)\nn\\ &=& \frac{1}{6H^2}
\left(-4 \times \frac12 G_{ij} \dot\varphi^i\dot\varphi^j +
2U\right)\, . \ea
It can be easily seen that the equation of state (EoS) parameter
$\omega = P/\rho$ and parameter $q$ are linearly connected:
\be{6a} q=-\frac12(1+3\omega)\, . \ee
From the definition of the acceleration parameter, it follows that
$q$ is constant in the case of the power-law and De Sitter-like
behavior:
\be{6b} q = \left\{\begin{array}{cc} (s-1)/s \;;\quad
a\propto t^s\, ,\\
1 \;;\quad
a\propto e^{H t}\, . \\
\end{array}\right.
\ee
For example, $q=-0.5$ during the matter dominated (MD) stage where
$s=2/3$.

Because the minisuperspace metric $G_{ij}$ is flat, the scalar field
equations are:
\be{7} \ddot\varphi^i +3H\dot \varphi^i + G^{ij}\frac{\partial
U}{\partial \varphi^j} = 0\, . \ee

For the action \rf{1}, the corresponding Hamiltonian is
\be{8a} \mathcal{H} = \frac{8\pi G}{2a^3}G^{ij}P_iP_j +
\frac{a^3}{8\pi G} U\, , \ee
where
\be{8b} P_{i} = \frac{a^3}{8\pi G} G_{ij} \dot \varphi^j \, \ee
are the canonical momenta and equations of motion have also the
canonical form
\be{8c} \dot \varphi^i = \frac{\partial \mathcal{H}}{\partial P_i}\,
,\quad \dot P_i = - \frac{\partial \mathcal{H}}{\partial
\varphi^i}\, . \ee
It can be easily seen that the latter equation (for $\dot P_i$) is
equivalent to the eq. \rf{7}.

Thus, the Friedmann equations together with the scalar field
equations can be replaced by the system of the first order ODEs:
\ba{11}
 &\dot \varphi^i& = \frac{8\pi G}{a^3} G^{ij}P_j\, ,\\
 &\dot P_i &= - \frac{a^3}{8\pi G}\frac{\partial U}{\partial
\varphi^i}\, ,
\label{12}\\
&\dot a &= a H\, ,\label{13}\\
&\dot H &= \frac{\ddot a}{a} - H^2 \nn\\
&&= \frac16 \left(-4 \times \frac12 G_{ij}
\dot\varphi^i\dot\varphi^j + 2U\right) - H^2 \label{14}\, \ea
with Eq. \rf{4} considered in the form of the  initial conditions:
\be{16} H(t=0) = \left.\sqrt{\frac13\left(\frac12 G_{ij}
\dot\varphi^i\dot\varphi^j +U\right)}\; \right|_{t=0}\, . \ee
One can make these equations dimensionless:
\ba{17} \frac{d \varphi^i}{M_{Pl}dt} &=& \frac{8\pi }{M^3_{Pl}a^3}
G^{ij}P_j,\nn\\
\Rightarrow  \frac{d \varphi^i}{dt} &=& \frac{8\pi }{a^3} G^{ij}P_j\,;\\
\frac{d P_i}{M_{Pl}dt} &=& - \frac{a^3 M^3_{Pl}}{8\pi}\frac{\partial
(U/M^2_{Pl})}{\partial \varphi^i}\,,\nn\\ \Rightarrow  \frac{d
P_i}{dt} &=& - \frac{a^3}{8\pi}\frac{\partial U}{\partial
\varphi^i}\label{18}\, . \ea
That is to say the time $t$ is measured in the Planck times
$t_{Pl}$, the scale factor $a$ is measured in the Planck lengths
$L_{Pl}$ and the potential $U$ is measured in the $M^2_{Pl}$ units.

The system of dimensionless first order ODEs together with
the initial condition \rf{16} is used for numerical calculation of the
dynamics of considered models with the help of a Mathematica package
\cite{KP}.


 \addcontentsline{toc}{subsection}{Appendix 4. Bouncing inflation in the Brans-Dicke frame\label{sec4}}
 \setcounter{subsection}{1}
 \medskip\medskip\medskip\medskip
 \textbf{Appendix 4. Bouncing inflation in the Brans-Dicke frame}
  \medskip\medskip\medskip\medskip

\renewcommand{\theequation}{A4.\arabic{equation}}
\setcounter{equation}{0}

According to Eq. \rf{1.9}, the four-dimensional FRW metrics in the Einstein frame \rf{5.1a1} and in the Brans-Dicke frame
are related as follows:
\be{a1}
-dt\otimes dt + a^2(t)d\vec{x}\otimes d\vec{x} =
f'\left[-d\bar t\otimes d\bar t + \bar a^2(\bar t)d\vec{x}\otimes d\vec{x}\right],
\ee
where $f'=X+1>0$ and $X$ is parameterized by Eq. \rf{5.3}.
Therefore, for the synchronous times and scale factors in both frames one obtains, correspondingly:
\ba{qwer3}
d \bar t &=& d t /\sqrt{f'(t)} \label{a2}\, ,\\
\bar a ({\bar t}) &=& a \left(t(\bar t)\right)/\sqrt{f'\left(t(\bar t)\right)}\label{a3} ,
\ea
which lead to the following equations:
\be{a4}
\bar t = \int^t_1\frac{d t}{\sqrt{X(t)+1}} +1\, ,
\ee
where the constant of integration is chosen in such a way that $\bar t (t=1)=1$, and
\ba{a61a}
\bar H (\bar t) &=& \left.\frac{d\bar a}{d\bar t} \frac{1}{\bar a}= \sqrt{X\left(t(\bar t)\right) +1}\right[H\left(t(\bar t)\right) \nn\\&-&\left.
\frac{1}{2(X\left(t(\bar t)\right)+1)}\frac{dX}{dt}\left(t(\bar t)\right)\right]\, .
\ea
From the latter equation follows the relation between the Hubble parameters in both frames.
In Fig. \ref{BD} the logarithms of the scale factor $\bar a(\bar t)$ and the Hubble parameter $\bar H (\bar t)$ ar depicted
for the trajectory drawn in Fig. \ref{theta}. These pictures clearly demonstrate that in the Brans-Dicke frame the Universe has also asymptotical de Sitter
stage when the scalaron field approaches the branching point $\theta = 2\pi$.
It is not difficult to verify that because $X(t\to +\infty) \to X_{max}$ and $dX/dt(t\to +\infty) \to 0$,
one obtains the following relation for the asymptotic values of the Hubble parameters in both frames:
\be{a7a1}
\bar H = H \sqrt {X_{max}+1}\, .
\ee

\newpage
\addcontentsline{toc}{section}{LITERATURE}
\bibliography{thesis_bib}

\newpage
\addcontentsline{toc}{section}{ACKNOWLEDGMENT}
 \setcounter{subsection}{1}
 \medskip\medskip\medskip\medskip
 \textbf{ACKNOWLEDGMENT}
  \medskip\medskip\medskip\medskip

First of all, I would like to thank my parents, especially my mother, who instilled in me a love for science.

With my greatest pleasure I would like to thank my supervisor Prof. Alexander Ivanovich Zhuk, who has opened for me the multidimensional cosmology, for his strong support in the research, fruitful discussions and many of very interesting walk tours we had around Sevastopol and Odessa. His capability for elegant solutions in very complex theoretical tasks will always amaze me.

I would like to address my warmest gratitude to professors of the Department of Theoretical Physics ONU: Prof. V.M. Adamyan, Prof. N.P. Malomuzh and Prof. \framebox[1.05\width]{A.V. Zatovskiy}$\:$, Dr. V.P. Olyiynik, Dr. M.Ya. Sushko and Dr. V.L. Kulinskiy; their lectures allowed me to enter the world of theoretical physics.

Then I like to thank my \textbf{T}eacher of physics (deserving capital T) Vadim Leonidovich Manakin, whose impact on my life is impossible to overestimate. There are people gifted with ability to kindle a passion to knowledge and science. I was lucky to meet such a person - Vadim Leonidovich.

And last, but definitely not least, I would like to acknowledge my close friend Volodymyr Volodymyrovych Nemertsalov for his great help in preparation of promotion ceremony.

Thank you all!

\newpage
\addcontentsline{toc}{section}{CURRICULUM VITAE}
 \setcounter{subsection}{1}
 \medskip\medskip\medskip\medskip
\textbf{CURRICULUM VITAE}
  \medskip\medskip\medskip\medskip

Tamerlan Adamovich Saidov was born in Grozny (Chechen Republic) in October 1, 1982. In 1999 he graduated with the certificate of merit the Richelieu Lyceum (Department of Chemistry and Biology) in Odessa, Ukraine.

In 2003 he received the Bachelor degree in Theoretical Physics with Excellence at the Faculty of Physics (Department of Theoretical Physics) of the Odessa I.I. Mechnikov  National University (Odessa, Ukraine). The study of the bachelor report was focused on Navier-Stokes equations in cylindrical coordinates.

In period of 2003 - 2004, Tamerlan Saidov has obtained the Master degree in Theoretical Physics with Excellence at the same department. The work of the master program was dedicated to the properties of fermions fields in models of charged leptons, and was performed under supervision of Dr. V.P. Olyiynik.

Simultaneously with studying of physics, in 2005 Tamerlan Saidov graduated the Faculty of Economy and Law (Department of Administrative Law) of Odessa I.I. Mechnikov  National University (Odessa, Ukraine) and received Master degree in Law. The master thesis was written on the administrative right and obligations with the focus on the specifics of their origin and prosecuting.

In the period of 2004 - 2008, Tamerlan Saidov was appointed as phd-student at the Department of Theoretical Physics of the Odessa I.I. Mechnikov  National University (Odessa, Ukraine), where under the supervision of Prof. A.I. Zhuk, he has undertaken the theoretical study on multidimensional cosmology: the compactification of extra dimensions and accelerated expansion of the Universe. The results of this work are described in this Thesis.

Starting from 2007 to 2011, he did his second phd-research, in the Department of Applied Physics of Eindhoven University of Technology (Eindhoven, The Netherlands). Where under the supervision of Dr. L. Pel, the crystallization and ion transport of salts was studied in porous media by Nuclear Magnetic Resonance technique.

From 2011, Tamerlan Saidov is working on the cancer detection by contrast ultrasound imaging in the Signals Processing Systems group at the Department of Electrical Engineering of Eindhoven University of Technology (Eindhoven, The Netherlands).

\end{document}